\documentclass[twoside,fontsize=12pt,BCOR=22.5mm,DIV=15,headsepline=true,headlines=1.5,headinclude=true,listof=totocnumbered,bibliography=totocnumbered,unicode,pdfa]{scrbook}  %
\pdfoutput=1
\immediate\pdfobj stream attr{/N 3}  file{PDFA/sRGBIEC1966-2.1.icm}
\pdfcatalog{%
/OutputIntents [ <<
/Type /OutputIntent
/S/GTS_PDFA1
/DestOutputProfile \the\pdflastobj\space 0 R
/OutputConditionIdentifier (sRGB IEC61966-2.1)
/Info(sRGB IEC61966-2.1)
>> ]
}

\usepackage[utf8]{inputenc}
\usepackage{lmodern}
\usepackage[T1]{fontenc}
\usepackage{csquotes}
\usepackage[ngerman,english]{babel}
\usepackage{setspace}
\AfterTOCHead{\singlespacing}
\KOMAoptions{DIV=last}

\usepackage{scrlayer-scrpage}

\usepackage{threeparttable}
\usepackage{dpfloat}
\usepackage[section]{placeins}

\usepackage{xcolor}
\usepackage{etoolbox}
\usepackage{amsfonts}
\usepackage{amssymb}
\usepackage[intlimits]{amsmath}
\usepackage{amsthm}
\usepackage{mathtools}
\usepackage[h]{esvect}
\usepackage{tensor}
\usepackage{booktabs}
\usepackage{bbm}
\pdfpkmode={supre}
\usepackage{nicefrac}
\usepackage{units}
\usepackage{nextpage}

\usepackage{graphicx}
\usepackage[export]{adjustbox}
\graphicspath{{graphics/}{plot/}{plot/omega_tau_1_transposed.make/}}

\usepackage{wrapfig}
\usepackage{hyperxmp}
\input glyphtounicode.tex
\input glyphtounicode-cmr.tex
\usepackage{bookmark}

\usepackage{calc}
\usepackage{suffix}
\usepackage{ifthen}
\usepackage{titling}
\usepackage{needspace}

\usepackage{makeidx}
\makeindex
\WithSuffix\newcommand\index*[2][]{%
  \ifthenelse{\equal{#1}{}}%
    {#2\index{#2}}%
    {#2\index{#1}}%
}

\usepackage[style=numeric-comp,natbib,sorting=none,autopunct=true,autocite=plain,firstinits=true,backend=biber]{biblatex}
\renewbibmacro{in:}{%
  \ifentrytype{article}{}{\printtext{\bibstring{in}\intitlepunct}}}

\bibliography{inspire,noinspire}

\newcommand{\hyp}{\hyphen{}}

\usepackage[]{hyperref}
\usepackage[all]{hypcap}

\makeatletter
\newcommand{\maketitlefront}{%
  \setparsizes{\z@}{\z@}{\z@\@plus 1fil}\par@updaterelative
  \begin{center}
    {\usekomafont{title}{\huge \@title\par}}%
    \vskip 2em
    {\ifx\@subtitle\@empty\else\usekomafont{subtitle}{\@subtitle\par}\fi}%
    \vskip 3em
    \ifx\@subject\@empty \else
      {\usekomafont{subject}{\@subject\par}}%
      \vskip 3em
    \fi
    {\usekomafont{date}{\@date \par}}%
    \vskip \z@ \@plus3fill
  \end{center}\par
  \vfill
  \begin{minipage}[b]{\linewidth}
    {\usekomafont{publishers}{\@publishers}}%
  \end{minipage}\par
  \@thanks\let\@thanks\@empty%
}

\newcommand{\maketitleback}{%
  \next@tpage
  \begin{minipage}[t]{\linewidth}
    \@uppertitleback
  \end{minipage}\par
  \vfill
  \begin{minipage}[b]{\linewidth}
    \@lowertitleback
  \end{minipage}\par
  \@thanks\let\@thanks\@empty%
}

\newcommand{\makededication}{%
  \ifx\@dedication\@empty
  \else
    \next@tdpage\null\vfill
    {\centering\usekomafont{dedication}{\@dedication \par}}%
    \vskip \z@ \@plus3fill
    \@thanks\let\@thanks\@empty
    \cleardoubleemptypage
  \fi%
}

\newcommand{\mytitlepage}{%
  \begin{titlepage}%
      \begin{spacing}{1}
    \maketitlefront%
    \maketitleback%
    \makededication%
  \end{spacing}
  \end{titlepage}%
}
\makeatother

\title{Electron Positron Pair Production in Strong Electric Fields}
\subject{%
  {\LARGE Dissertation}\medskip\\
  {\mdseries zur Erlangung des akademischen Grades\\
  \textit{doctor rerum naturalium} (Dr. rer. nat.)\bigskip\bigskip\\}
  \includegraphics[width=5cm]{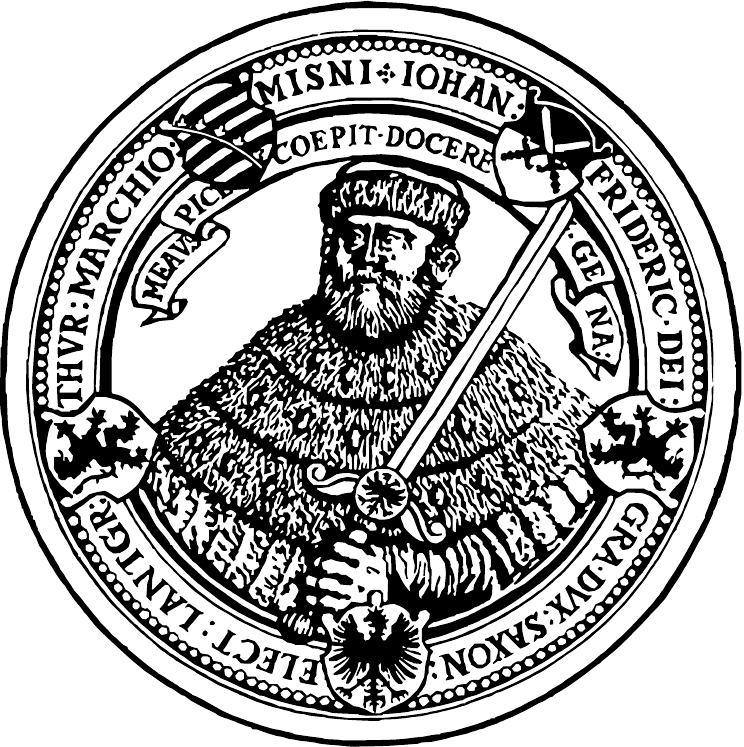}\vspace{-.5em}\\
  \rule{5cm}{2pt}\\
  seit 1558%
}
\author{Alexander Blinne}
\date{}
\publishers{\raggedright vorgelegt dem Rat der Physikalisch-Astronomischen
Fakultät der Friedrich-Schiller-Universität Jena\bigskip\bigskip\\
von M.\,Sc. Alexander Blinne\\
geboren am 19.10.1987 in Essen}
\uppertitleback{}
\lowertitleback{%
  {\usekomafont{sectioning}Gutachter}
  \begin{enumerate}
    \item Prof. Dr. Holger Gies\\
	Friedrich-Schiller-Universität Jena\\
	Theoretisch-Physikalisches-Institut
    \item Prof. Dr. Burkhard Kämpfer \\
     Helmholtz-Zentrum Dresden-Rossendorf\\
      Institut für Strahlenphysik
    \item Prof. Dr. Antony Ilderton \\
 Chalmers University of Technology\\
     Department of Physics
  \end{enumerate}
  Datum der Disputation: 24. November 2016%
}

\definecolor{rgbcyan}{rgb}{0,1,1}
\definecolor{rgbpurple}{rgb}{1,0,1}
\definecolor{rgbred}{rgb}{1,0,0}
\definecolor{rgbblue}{rgb}{0,0,1}

\hypersetup{
    pdflang = {en},
    unicode = {true},
    pdftitle = {\thetitle},
    pdfsubject = {Dissertation},
    pdfauthor = {\theauthor},
    pdfauthortitle = {Master of Science},
    pdfkeywords = {pair production, rotating fields, DHW},
    pdfstartview = {FitV},
    pdfview = {FitH},
    pdfpagelayout = {OneColumn},
    colorlinks = {true},
    urlcolor = {rgbcyan},
    citecolor = {rgbpurple},
    filecolor = {rgbred},
    linkcolor = {black},
}

\addtokomafont{title}{\Huge \bfseries}
\addtokomafont{publishers}{\bfseries \large}
\newcommand\numberthis{\addtocounter{equation}{1}\tag{\theequation}}
\newcommand{\commutator}[2]{\left[ #1 ,\, #2 \right]}
\newcommand{\Dt}{D_{\!t}\,}
\newcommand{\vDx}{\vv{D}_{\!\vv{x}}\,}

\newcommand{\hatbar}[1]{{\overline{\hat{#1}}}}
\newcommand{\hvec}[1]{{\hat{\vv{#1}}}}
\newcommand{\nablap}{\vv{\nabla}_{\!\vv{p}}\,}
\newcommand{\nablax}{\vv{\nabla}_{\!\vv{x}}\,}
\newcommand{\anticommutator}[2]{\left\lbrace #1 ,\, #2 \right\rbrace}
\newcommand{\vev}[2][0]{\left\langle#1\right| #2 \left|#1\right\rangle}

\newcommand{\Eqref}[1]{Eq.~\eqref{#1}}

\DeclareMathOperator{\diverg}{div}
\DeclareMathOperator{\rot}{rot}
\newcommand{\Eq}{{Eq.~}}
\newcommand{\Eqs}{{Eqs.~}}
\newcommand{\Sec}{{Sec.~}}
\newcommand{\Secs}{{Secs.~}}
\newcommand{\Cha}{{Cha.~}}
\newcommand{\Fig}{{Fig.~}}
\newcommand{\Figs}{{Figs.~}}
\newcommand{\App}{{App.~}}
\newcommand{\Tab}{{Tab.~}}
\newcommand{\Ref}{{Ref.~}}

\newcommand{\abs}[1]{\left\lvert#1\right\rvert}

\newlength{\colonwidth}
\newcommand{\definedby}{\hspace*{\colonwidth} &\mathrel{\hspace*{-\colonwidth}:=}}

\newcommand{\bbs}{\mathbbm{s}}
\newcommand{\bbp}{\mathbbm{p}}
\newcommand{\bbv}{\mathbbm{v}}
\newcommand{\bba}{\mathbbm{a}}
\newcommand{\bbt}{\mathbbm{t}}
\newcommand{\bbw}{\mathbbm{w}}
\newcommand{\bbW}{\mathbbm{W}}

\newcommand{\vac}{\mathrm{vac.}}
\newcommand{\crit}{\mathrm{cr.}}
\newcommand{\mx}{\mathrm{max}}

\renewcommand{\i}{{\mathrm{i}}}
\newcommand{\ii}{{\mathrm{i}}}
\newcommand{\e}{{\mathrm{e}}}
\newcommand{\cm}{{q}}

\newcommand{\id}{\mathbbm{1}}

\DeclareMathOperator\tr{tr}
\DeclareMathOperator\Rect{Rect}
\DeclareMathOperator\E{\text{\textbf{E}}}
\DeclareMathOperator\K{\text{\textbf{K}}}
\DeclareMathOperator\PiI{\boldsymbol{\Pi}}
\DeclareMathOperator\artanh{artanh}
\DeclareMathOperator{\sech}{sech}

\newcommand{\dd}[2][]{\ifstrempty{#1}{\mathop{\mathrm{d}#2}}{\mathop{\mathrm{d}^{#1}#2}}}

\newcommand{\tc}{t_\mathrm{c}}

\newenvironment{rarray}[1]
{%
  \setlength{\arraycolsep}{#1}
  \begin{array}
}{%
  \end{array}
}

\newcounter{framenum}
\DeclareLayer[{
    clone=scrheadings.head.even,
    area={3mm}{3mm}{\paperwidth-6mm}{\dimexpr\voffset+2.54cm-\headsep\relax},
    align=lt,
    contents={\setcounter{framenum}{(\thepage-1)/2}\hfill\includegraphics[scale=0.8]{thumb/f\theframenum}}
}]{scrheadings.head.bg.even}

\newpairofpagestyles[scrheadings]{headingsthumb}{}
\AddLayersAtBeginOfPageStyle{headingsthumb}{scrheadings.head.bg.even}
\AddLayersAtBeginOfPageStyle{plain.headingsthumb}{scrheadings.head.bg.even}

\begin{document}
  \setlength\footskip{26.83pt}
  \pagenumbering{alph}
  \mytitlepage

  \frontmatter
  \setcounter{page}{1}

  {
    \renewcommand*{\chapterheadstartvskip}{\vspace*{-1cm}}
    \begin{onehalfspacing}
\begin{otherlanguage}{ngerman}
\setlength{\parskip}{1ex} %
\setlength{\parindent}{0pt}
\cleardoublepage %
\phantomsection
\pdfbookmark[chapter]{Zusammenfassung}{zusammenfassung}
\addchap*{Zusammenfassung}
In dieser Arbeit wird Elektron-Positron Paarerzeugung in räumlich homogenen elektrischen (und magnetischen) Feldern behandelt.
Verschiedene Feldkonfigurationen werden untersucht um verschiedene Phänomene wie Multiphoton"=Paarproduktion, Sauter-Schwinger-Paarproduktion und dynamisch assistierte Paarproduktion zu studieren.
Der zentrale Fokus liegt auf gepulsten, rotierenden Feldern mit einer zentralen Frequenzkomponente, die rotierende Sauter-Pulse genannt werden.

Die Resultate werden mittels verschiedener numerischer Methoden gewonnen, die auf unterschiedlichen theoretischen Ansätzen basieren.
Eine generische Methode, welcher eine modifizierte quantenkinetische Gleichung zu Grunde liegt, wird aus dem Dirac-Heisenberg-Wigner (DHW) Formalismus abgeleitet.
Die numerische Lösung dieser Gleichung wird die Wigner-Methode genannt.
Andere Arten von Gleichungen werden ebenfalls aus dem DHW-Formalismus abgeleitet und mit dem Ziel, magnetische Felder einzubeziehen, numerisch gelöst.
Im Falle des rotierenden Sauter-Pulses wird zusätzlich eine komplett unabhängige numerische Methode entwickelt, welche auf einem semiklassischen Ansatz basiert und daher semiklassische Methode genannt wird.

Eine Reihe von Parameterstudien wird durchgeführt, um Paarproduktion in rotierenden Sauter-Pulsen zu verstehen.
In diesen Studien werden die Wigner-Methode und die semiklassische Methode ausschöpfend verglichen, wobei gefunden wird, dass sie sich ergänzen.
Dadurch ist es möglich, den kompletten Parameterbereich des rotierenden Sauter-Pulses abzudecken, wodurch Paarproduktionsraten für Laser-basierte Experimente mit gegenläufig propagierenden zirkular polarisiertem Licht berechnet werden können.
Die resultierenden Paarproduktionsspektren werden interpretiert.

Durch die generische Natur der Wigner-Methode ist es möglich, darüber hinaus weitere Feldkonfigurationen zu untersuchen, beispielsweise Laser-Pulse mit elliptischer Polarisation, Pulse mit chirp oder Überlagerungen zweier rotierender Sauter-Pulse, so genannte bichromatische Pulse.
Jede dieser Konfigurationen zeigt interessante Merkmale, unter anderem den dynamisch assistierten Schwinger-Effekt in bichromatischen Pulsen, was die Planung von hochintensitäts Laser-Experimenten positiv beeinflussen kann.
\end{otherlanguage}
\end{onehalfspacing}

    \begin{onehalfspacing}
\cleardoublepage %
\phantomsection
\pdfbookmark[chapter]{Abstract}{abstract}
\addchap*{Abstract}%
This work covers electron positron pair production in spatially homogeneous electric (and magnetic) fields.
Different field configurations are looked at in order to study various phenomena including multiphoton pair production, Sauter-Schwinger pair production and dynamically assisted pair production.
The main focus lies on pulsed, rotating fields with one main frequency component which are called rotating Sauter pulses.

The results are obtained via different numerical methods, that rest on different theoretical approaches.
A generic method is derived from the Dirac-Heisenberg-Wigner (DHW) formalism which entails a modified quantum kinetic equation.
We call the numerical solution of this equation the Wigner method.
Other types of equations are derived from the DHW formalism as well and numerically solved with the aim to include magnetic fields.
In the case of rotating Sauter pulses a completely different numerical method is developed, which is based on a semiclassical approach and therefore called the semiclassical method.

A number of parameter studies are conducted to understand pair production in these rotating Sauter pulses.
In those studies the Wigner method and the semiclassical method are compared exhaustively and found to complement each other.
This makes it possible to cover the complete range of parameters of the rotating Sauter pulse, which helps to calculate the pair production rates for experiments involving counter-propagating circularly polarized laser light.
An interpretation of the resulting pair production spectra is given.

Due to the general nature of the Wigner method it is possible to study more general field configurations which include pulses with elliptic polarization, chirped pulses or bichromatic rotating Sauter pulses.
Each of these exhibit interesting features, including the dynamically assisted Schwinger effect in bichromatic pulses, which could be useful in planning high-intensity laser experiments.
\end{onehalfspacing}

  }

  \mainmatter
  \pagestyle{headingsthumb}
    \phantomsection
    \renewcommand{\contentsname}{Table of Contents}
    \pdfbookmark[chapter]{\contentsname}{toc}
    \tableofcontents

  \chapter{Introduction}
In the first half of the 20th century the quantum theory of electromagnetic interactions has been developed.
The driving force was the urge to understand a very common bound state between a positively charged nucleus and negatively charged electrons, the atom.
When quantum mechanics, based upon Schrödinger's equation, turned out to be not sufficient to explain the exact structure of the spectrum, a relativistic kind of quantum mechanics was developed.
At the time this was called Dirac theory, as it involved the Dirac equation.
Dirac theory was problematic, because a one-particle interpretation of the Dirac equation in analogy to the nonrelativistic Schrödinger equation is not consistent.
This is due to the fact that the Dirac equation has eigenstates with negative energy and thus no stable ground state.
When the Dirac sea was postulated \autocite{Dirac:1930ek}, where almost all the eigenstates with negative energy should be occupied, the theory became useful.
Using Dirac's theory \autocite{Dirac1934} it was indeed possible to improve the understanding of hydrogen-like atoms, but the theory was not sufficient to explain, e.\,g., the Lamb shift or the anomalous magnetic moment of the electron.
Holes in this proposed Dirac sea were thought of as a possible explanation for the proton \autocite{Dirac:1930bga} at first, but when this did not work out, the existence of the positron was conjectured, which was experimentally confirmed later \autocite{Anderson:1932zz,Anderson:1933mb}.

These problems of Dirac's theory went away when the theory of quantum electrodynamics (QED) was developed.
The anomalous magnetic moment of the electron has been perturbatively calculated to high precision and also the Lamb shift has been explained.
This quantum theory led to a Nobel prize for J.\,S.\,Schwinger \autocite{schwinger}, R.\,P.\,Feynman \autocite{feynmanqed}, and S.\,Tomonaga in 1965.

In the classical limit it describes classical electrodynamics with special relativistic motion of classical charged particles.
The full quantum theory takes into account that there is no distinction between particles and fields and that particles can be viewed as excitation of the quantized fields.
The electrons and positrons are excitations of the Dirac field and the excitations of the electromagnetic field are the photons which convey the electromagnetic force.

Perturbative QED is well understood, but beyond perturbation theory there are still open questions.
Especially, a lot of the predicted effects of QED have yet to be observed in experiment, e.\,g., light\hyp{}light interaction via vacuum polarization \autocite{Heisenberg:1934pza,Schwinger:1951nm} or Sauter\hyp{}Schwinger pair production \autocite{Sauter:1931zz,Heisenberg:1935qt,Schwinger:1951nm}.
Pair production by interaction of high-energy photons with laser light has been observed in the famous SLAC-144 experiment \autocite{Burke:1997ew}.
This work will concentrate on electron positron pair production in strong fields which is a phenomenon that has its roots in the fact that the spectrum of the Dirac equation is not bounded from below.

If one thinks of slowly varying and very strong fields, pair production can be understood as a tunneling process.
In the extreme, for homogeneous and constant fields, the pair production rate per unit spacetime volume is given by the Schwinger formula
\begin{equation}
  \label{eqn:schwinger_formula}
  \mathcal{N} = \frac{\left( eE \right)^2}{4\pi^3} \e^{-\pi \frac{m^2}{eE}}\,,
\end{equation}
where $m$ is the electron mass and $e$ the (positive) elementary charge.
This is called Sauter\hyp{}Schwinger pair production, see \Ref\cite{Sauter:1931zz,Heisenberg:1935qt,Schwinger:1951nm}.
The result has an essential singularity in the perturbative limit $E\to0$ and can not be found using perturbation theory.
Because of this Sauter\hyp{}Schwinger pair production is thought of as a non\hyp{}perturbative effect.
It was shown recently that the essential singularity is a consequence of the electric field being constant for all times.
When the field is switched on adiabatically, it is possible to obtain analytic results which can be expanded into convergent power series.
Furthermore the series coefficients can be reproduced by perturbation theory\autocite{arXiv:1108.2615}.

\begin{figure}
  \hfil a)\raisebox{-\height+\baselineskip}{\includegraphics{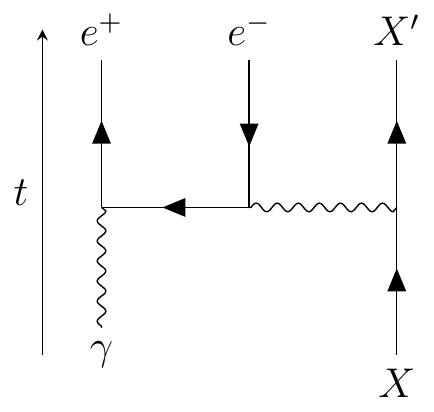}} \hfil b)~\raisebox{-\height+\baselineskip}{\includegraphics{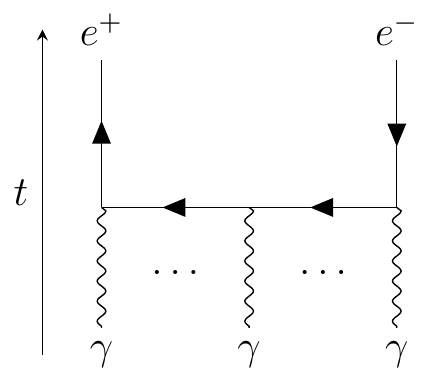}} \hfil
  \caption{Feynman diagrams for either a) Bethe\hyp{}Heitler or b) Breit\hyp{}Wheeler processes.}
  \label{fig:feynman_bethe_breit}
\end{figure}
If the fields are quickly changing, it is possible to use perturbation theory.
Often the varying electromagnetic fields are then understood in terms of propagating photons.
It is not possible to produce pairs from a single photon, as in the vacuum this would violate the conservation of energy and momentum.
This can be remedied in the presence of a nucleus $X$ which can absorb excess momentum and this effect is called Bethe-Heitler pair production \autocite{Bethe:1934za}, an example for a Feynman graph describing this process can be found in \Fig\ref{fig:feynman_bethe_breit}a).

Perturbative pair production in vacuum is only possible if there are at least two photons around, which is called Breit-Wheeler pair production \autocite{Breit:1934zz} and displayed in \Fig\ref{fig:feynman_bethe_breit}b).
If the number of photons is greater than two the process is called non-linear Breit-Wheeler process or just multiphoton pair production.
The methods described in this work cover both Schwinger pair production and multiphoton pair production and we apply them to pair production in absence of a nucleus.%

Unfortunately, any attempt at verifying this understanding in macroscopic electric fields with frequencies well below the threshold for resonant pair production $\Omega=m$, which are closer to Schwinger pair production than multiphoton pair production, is hampered by the exponentially small production rates due to \Eqref{eqn:schwinger_formula}.
The rapid development of optical or X-ray high-intensity lasers has lead to many suggestions for schemes for a first discovery \autocite{Ringwald:2001ib,Alkofer:2001ik,Ruf:2009zz,DiPiazza:2009py,Monin:2009aj,Baier:2009it,Heinzl:2010vg,Bulanov:2010ei,Gonoskov:2013ada}, also including the combination of laser light and strong Coulomb fields \autocite{Muller:2008zzd,Muller:2009zzf,DiPiazza:2009yi,DiPiazza:2010kg}.
As pair production is potentially a highly nonlinear process, the superposition of different fields is also a promising approach.
Lowering the effective tunneling threshold for Schwinger pair production by adding some medium energy photons is called dynamically assisted Schwinger pair production.
Previous research into pair production in such fields can be found in Refs.~\autocite{Schutzhold:2008pz,Dunne:2009gi,Fey:2011if,Jansen:2013dea,Akal:2014eua,Jansen:2015loa}.

However, while the highest field intensities in such systems may indeed gradually approach the critical intensity, $I_{\crit} = E_{\crit}^2 = \left( \frac{m^2}{e} \right)^2 \simeq 4.3\times 10^{29}\, \text{W/cm}^2$, further possible physical processes may set in that could partly or entirely swamp a pair production signal.
In particular QED cascades of successive radiation of accelerated charges and particle production from hard photons are expected to occur \autocite{Bell:2008zzb,Kirk:2009vk,Fedotov:2010ja,Bulanov:2010gb,Elkina:2010up,Nerush:2010fe,Nerush:2011xr,King:2013zw,Bulanov:2013cga,Bashmakov:2013iwa}, which may even fundamentally inhibit the generation of near critical intensities.
In view of a possible discovery of Schwinger pair production in strong laser fields, this gives rise to a crucial question: can a QED cascade seeded by an electron (sourced by impurities of an imperfect vacuum) be distinguished from a QED cascade seeded by Schwinger pair production?
Whereas the ensemble of electrons arising from impurities are likely to have an isotropic initial momentum distribution, the ensemble of Schwinger created pairs can be expected to carry information about the directionality of the electric field that lead to pair creation.
To quantify this difference for final observables, not only the QED cascade has to be computed, but first of all, the initial data from Schwinger pair production has to be determined.
For the cascade calculations performed so far, this is far from being trivial as spatio-temporal dependencies of the fields have to be accounted for.

The first extensions of Sauter's original work concentrated on one\hyp{}component electric fields.
In addition to unidirectional fields depending on either space \autocite{Kim:2000un,Nikishov:2003ig,Gies:2005bz,Dunne:2005sx,Dunne:2006st} or time \autocite{Dunne:2006st,Brezin:1970xf,Narozhnyi:1970uv,Popov:1971iga,Popov:1973az,Gavrilov:1996pz,Dunne:1998ni,Dietrich:2003qf,Piazza:2004sv}, exact solutions for specific classes of fields can be found in light cone variables \autocite{Tomaras:2001vs,Hebenstreit:2011cr}.
In \Ref\cite{Ilderton:2015qda} a connection between these three special cases was found using interpolating coordinates and the worldline instanton method.

Recently more involved fields have also been studied including electric fields that are not necessarily unidirectional and are spatially inhomogeneous in up to three dimensions \autocite{Dunne:2006ur,Dietrich:2007vw,Ruf:2009zz,Hebenstreit:2011wk} or depend on time \autocite{Hebenstreit:2010cc,Jansen:2013dea,Akal:2014eua,Otto:2015gla,Panferov:2015yda,Bell:2008zzb, Kirk:2009vk, Fedotov:2010ja, Bulanov:2010gb, Elkina:2010up, Nerush:2010fe, Nerush:2011xr, King:2013zw, Bulanov:2013cga, Bashmakov:2013iwa}.
Also unidirectional fields that depend on space as well as on time \autocite{Schneider:2014mla,Kohlfurst:2015niu} were studied.
A lot of studies concentrate on field configurations that could be found in counterpropagating laser light; this includes (nonlinear) Breit-Wheeler pair production \autocite{Nousch:2012hg,Nousch:2015pja} and pair production in pure electric fields near the antinodes of the magnetic field \autocite{Hebenstreit:2010cc,Jansen:2013dea,Akal:2014eua,Otto:2015gla,Panferov:2015yda}.

Different methods were developed including those that are exact on the mean-field level, e.\,g., the quantum kinetic theory (QKT) \autocite{Smolyansky:1997fc,Schmidt:1998vi} and the real-time Dirac-Heisenberg-Wigner (DHW) formalism \autocite{BialynickiBirula:1991tx,Hebenstreit:2011pm}.
A numerical scheme based upon the DHW formalism has been developed \autocite{Blinne:2013via}, which we will call the Wigner function method or, in short, the Wigner method.

While it is possible to obtain exact results using a scattering ansatz \autocite{Dumlu:2009rr,Dumlu:2011rr}, it is combined with some kind of a semiclassical approximation \autocite{Brezin:1970xf,Popov:1971ff,Popov:1971iga,Popov:1973uw,Marinov:1977gq,Kim:2000un,Popov:2001ak,Kim:2003qp,Piazza:2004sv,Kim:2007pm,Kleinert:2008sj,Kleinert:2012bu} in most of its applications.
This combination is sometimes referred to as the Wentzel-Kra\-mers-Brillouin (WKB) approach \autocite{Dumlu:2011rr,Strobel:2013vza,Dumlu:2015paa,Ilderton:2015qda} or even as the WKB approximation \autocite{Strobel:2013vza,Dumlu:2009rr}, while only the ansatz, but not the approximation, is taken from the original WKB method.
We will thus not refer to the method discussed here as the WKB method, but as the semiclassical scattering method.
Other semiclassical methods include the worldline instanton method \autocite{Dunne:2005sx,Dunne:2006st,Ilderton:2014mla,Ilderton:2015lsa}.

It was shown that in the case of linearly polarized, purely electric fields the QKT is equivalent to the DHW formalism \autocite{Hebenstreit:2010vz} as well as to the scattering approach \autocite{Dumlu:2009rr}.
The worldline instanton method and the semiclassical scattering method have been shown to agree for one-component fields \autocite{Dunne:2005sx,Dunne:2006st,Dumlu:2011cc} and for two-component fields \autocite{Strobel:2013vza}.

In this Thesis, I consider a number of the aforementioned topics.
For the most part, the Wigner method is used, which is based on the DHW formalism.
Some aspects of the DHW formalism are explained in \Cha\ref{cha:wigner_intro}.

Pair production in time-dependent, spatially homogeneous electric fields is the focus of \Cha\ref{cha:homogeneous}.
At first the modified quantum kinetic equations are derived from the DHW formalism and the Wigner method is explained in \Sec\ref{sec:wigner_method}.
After introducing a complementary semiclassical method in \Sec\ref{sec:semiclass}, the rotating Sauter pulse is introduced and studied in \Sec\ref{sec:simprot}, which may help bringing the quantum field theory studies a substantial step closer to QED cascade calculations.
The performance and accuracy of the Wigner method \autocite{Blinne:2013via} and the semiclassical method \autocite{Strobel:2014tha} are subject to a comparison in the context of the rotating Sauter pulse.
Additionally other kinds of time dependencies of the electric field are taken into account such as elliptically polarized fields or superpositions of fields with different carrier frequencies in the remaining sections of the chapter.

In \Cha\ref{chap:magnetic} an approach for the inclusion of spatially homogeneous magnetic fields in the context of the DHW formalism as an expansion of the Wigner method is explained and preliminary results are presented.

\chapter{The Wigner Function Formalism}
\label{cha:wigner_intro}
The Dirac-Heisenberg-Wigner formalism or Wigner function formalism is the basis for the Wigner method which will be used in \Cha\ref{cha:homogeneous} and explained in \Sec\ref{sec:wigner_method} as well as for the attempt to enhance it to include magnetic fields in \Cha\ref{chap:magnetic}.
The Wigner function itself is known since the 1930s \autocite{Wigner:1932eb} as a method to formulate quantum theory in the phase space.
Other functions are also used to do this, e.\,g., the Husimi function \autocite{Husimi1940} or the Glauber-Sudarshan function \autocite{Glauber:1963tx,Sudarshan:1963ts}.
In the context of electron positron pair production it was first introduced by Bialynicki-Birula, Górnicki and Rafelski \autocite{BialynickiBirula:1991tx} in 1991.
The \index*[DHW]{Dirac-Heisenberg-Wigner} formalism is also called the Wigner function formalism.
Comprehensive summaries for the present context as well as exact solutions for particular electric fields can be found in \Ref\cite{Hebenstreit:2010vz,Hebenstreit:2011pm}.
The details relevant to the work presented in this thesis will be given in this chapter.

The Wigner function formalism is based on the hermitian QED Lagrangian
\begin{align*}
  \mathcal{\hat{L}}(\hat{\Psi},\hatbar{\Psi},\hvec{A})=\frac12\left[\hatbar{\Psi}\gamma^\mu[\i\partial_\mu-eA_\mu]\hat{\Psi}-\left([\i\partial_\mu+eA_\mu]\hatbar{\Psi}\right)\gamma^\mu\hat{\Psi}\right]-m\hatbar{\Psi}\hat{\Psi}-\frac14F^{\mu\nu}F_{\mu\nu}
\end{align*}
from which the Dirac equation for the spinor $\hat{\Psi}$ and its adjoint form for the adjoint spinor $\hatbar{\Psi}=\hat{\Psi}^\dagger\gamma^0$
\begin{align*}
 \partial_t \hat{\Psi}(\vv{x}, t) &=
      -\gamma^0\vv{\gamma}
      \left[ \nablax - ie\hvec{A}(\vv{x}, t) \right]
      \hat{\Psi}(\vv{x}, t) - im\gamma^0\hat{\Psi}(\vv{x}, t) \\
      \partial_t\hatbar{\Psi}(\vv{x},t) &=
      -\left[
        \nabla_{\vv{x}} \hatbar{\Psi}(\vv{x}, t)
        +ie\hatbar{\Psi}(\vv{x}, t) \hvec{A}(\vv{x}, t)
      \right]
      \cdot \vv{\gamma}\gamma^0 + im\hatbar{\Psi}(\vv{x},t)\gamma^0
\end{align*}
follow.
Note that Einstein's sum convention is used.
The electromagnetic field will be treated as a classical external field and its equation of motion will be reduced to the requirement for the $\vv{E}$ and $\vv{B}$ to obey Maxwell's equations
\begin{align*}
  \diverg\vv{E}&=\rho & \rot\vv{E}&=-\dot{\vv{B}} \\
  \diverg\vv{B}&=0    & \rot\vv{B}&=\dot{\vv{E}}+\mu_0j\,.
\end{align*}

\pagebreak
Starting from the fermionic field operator $\hat{\Psi}$ and its conjugate $\hat{\Psi}^\dagger$, there are in principle two basic objects similar to two-point propagators,
\begin{align*}
    C^\pm_{ab}(t,\vv{x_1},\vv{x_2}) \definedby
    \vev{\hat{\Psi}_a(t,\vv{x_1})\hat{\Psi}^\dagger_b(t,\vv{x_2})}
    \pm
    \vev{\hat{\Psi}^\dagger_b(t,\vv{x_2})\hat{\Psi}_a(t,\vv{x_1})}\,,
\end{align*}
which could be used to build the Wigner function.
Since $C^+$ is the expectation value of the anticommutator of the field operators, which is by definition the delta function $\delta_{ab}\cdot\allowbreak\delta(\vv{x_2}-\vv{x_1})$, only $C^-$ contains non-trivial information.
Thus the commutator of the field operators will be the starting point in the definition of the Wigner function.
A center of mass coordinate $\vv{x}$ for the underlying two-point correlator is introduced while $\vv{s}$ denotes the separation vector
\begin{align*}
  \vv{x} \definedby \frac12\left( \vv{x_1} + \vv{x_2} \right) \\
  \vv{s} \definedby \vv{x_2} - \vv{x_1}\,.
\end{align*}
The Wigner operator $\hat{\mathcal{W}}$  will now be defined as the Fourier (Wigner) transform of the equal time density operator of two Dirac field operators in the Heisenberg picture w.\,r.\,t. the separation vector.
As we are interested in pair production from the vacuum, the Wigner function is defined by taking the vacuum expectation value $\vev{\Diamond}$ of the Wigner operator.
Note that the Dirac conjugate field operator $\hatbar{\Psi}=\hat{\Psi}^\dagger\gamma^0$ is used instead of the adjoint field operator such that the Wigner operator transforms homogeneously under Lorentz transformations.
\begin{align}
\label{eqn:WignerDef}
\begin{split}
  \mathcal{W}(t,\vv{x},\vv{p}) \definedby
  \vev{\hat{\mathcal{W}}(t,\vv{x},\vv{p})}
  \\
  \hat{\mathcal{W}}_{ab}(t,\vv{x},\vv{p}) \definedby
  -\frac12 \int d\vv{s}\,\e^{-\frac{\i}{\hbar}\vv{p}\cdot\vv{s}}\,\mathcal{\hat{C}}_{ab}(t,\vv{x},\vv{s})
  \\
  \mathcal{\hat{C}}_{ab}(t,\vv{x},\vv{s}) \definedby
  \hat{\Phi}(t,\vv{x},\vv{s})
  \commutator{\hat{\Psi}_a(t,\vv{x}+\nicefrac{\vv{s}}{2})}{\hatbar{\Psi}_b(t,\vv{x}-\nicefrac{\vv{s}}{2})}
  \\
  \hat{\Phi}(t,\vv{x},\vv{s}) \definedby
  \mathcal{P}\e^{-\i e\int_{\vv{x}+\nicefrac{\vv{s}}{2}}^{\vv{x}-\nicefrac{\vv{s}}{2}}\vv{\hat{A}}(t,\vv{x}')\cdot d\vv{x}'}
\end{split}
\end{align}
The Wilson line $\hat{\Phi}(t,\vv{x},\vv{s})$ was introduced to achieve gauge invariance \autocite{BialynickiBirula:1991tx,Hebenstreit:2011pm}, the path ordering will be dropped when the electromagnetic field is treated classically.
Choosing a straight line as integration path ensures the proper interpretation of $\vv{p}$ as kinetic momentum \autocite{Hebenstreit:2011pm}.
Please note that this is not the definition of the covariant Wigner function which depends on a spacetime point and a four-momentum, but the equal time Wigner function where an energy average has been taken \autocite{Hebenstreit:2011pm}.
The Fourier transform can be understood as measuring the plane wave content of the two-point correlator with respect to the origin $\vv{x}$.
The amplitudes of the plane waves in this correlator are then interpreted as quasi probability densities of particles with momentum $\vv{p}$ at position $\vv{x}$.

Consequently the Wigner function can be used to calculate the expectation values of any observable which is given in terms of the field operators.
If one of the variables $\vv{x}$ or $\vv{p}$ is integrated out, the remaining function is positive and can be interpreted as a particle density \autocite{BialynickiBirula:1991tx,Hebenstreit:2011pm}.

\section{Equation of Motion}
Using the Heisenberg equations of motion for every field operator in the definition \Eqref{eqn:WignerDef}, the equations of motion for the Wigner function can be derived.
The result is an infinite tower of coupled equations which describe not only the evolution of the Wigner function, but the evolution of the electromagnetic field as well.
This tower is called the BBGKY hierarchy after N.\,Bogoliubov, M.\,Born, H.\,Green, G.\,Kirkwood and J.\,Yvon \autocite{Hebenstreit:2011pm}.
In order to do practical numerical calculations, this tower can be truncated \autocite{Hebenstreit:2011pm}.
This Hartree-type or mean-field approximation, which will be explained below, is indeed the only approximation used in the Wigner method so far.

This approximation was already introduced by Bialynicki-Birula, Górnicki and Rafelski \autocite{BialynickiBirula:1991tx}.
We replace the expectation value of a product of operators by the product of the expectation values
\begin{align*}
  \vev{\hat{E}\,\hat{\mathcal{O}}} &\to\vev{\hat{E}}\cdot\vev{\mathcal{\hat{O}}}\,,\\
  \vev{\hat{B}\,\mathcal{\hat{O}}} &\to\vev{\hat{B}}\cdot\vev{\mathcal{\hat{O}}}\,.
\end{align*}

The infinite tower of equations of the BBGKY hierarchy is truncated at the first level for the Wigner function and at the zeroth level for the electromagnetic field, as the latter is treated as a classical background field.
The dynamical equation for the Wigner function can then be written as
\begin{equation}
  \label{eqn:Wigner-EoM}
  \Dt \mathcal{W}=-\frac12 \vDx \commutator{\gamma^0\vv{\gamma}}{\mathcal{W}}-\i m\commutator{\gamma^0}{\mathcal{W}}-\i \vv{P}\anticommutator{\gamma^0\vv{\gamma}}{\mathcal{W}}
\end{equation}
with the pseudo-differential operators
\begin{align}
  \label{eqn:diff-ops}
  \begin{split}
   \Dt &=\partial_t + e\int_{-\nicefrac12}^{\nicefrac12} d\lambda\,\vv{E}(t,\vv{x}+\i \lambda\nablap)\cdot\nablap\,,
\\
  \vDx &=\nablax + e\int_{-\nicefrac12}^{\nicefrac12} d\lambda\,\vv{B}(t,\vv{x}+\i \lambda\nablap)\times\nablap\,,
\\
  \vv{P} &=\vv{p}-\i e\int_{-\nicefrac12}^{\nicefrac12} d\lambda\,\lambda\vv{B}(t,\vv{x}+\i \lambda\nablap)\times\nablap\,.
  \end{split}
\end{align}
Here, we use the conventions $\anticommutator{\gamma^\mu}{\gamma^\nu}=+2\eta^{\mu\nu}=+2\operatorname{diag}(1,\allowbreak-1,\allowbreak-1,\allowbreak-1)$ and work in temporal gauge $A_0=0$.
The electric field $\vv{E}$ and magnetic field $\vv{B}$ are given by
\begin{align}
  \label{eqn:temporalgauge}
  \begin{split}
   \vv{E}&=-\partial_t\vv{A}
   \\
   \vv{B}&=\nablax\times\vv{A}\,.
  \end{split}
\end{align}

In the language of Feynman diagrams, the mean-field approximation corresponds to neglecting radiative corrections, which is justified by the smallness of the fine-structure constant $\alpha$.
The Wigner function can be decomposed in terms of a complete basis of the Dirac bilinears, ($\mathbbm{1},\allowbreak\gamma^5,\allowbreak\gamma^\mu,\allowbreak\gamma^\mu\gamma^5,\allowbreak\sigma^{\mu\nu}:=\frac{\i}{2}\commutator{\gamma^\mu}{\gamma^\nu}$),
\begin{align}
  \label{eqn:wigner_fierz}
 \mathcal{W} = \frac14
      \left(
        \mathbbm{1} \bbs + \i \gamma_5 \, \bbp
        +\gamma^\mu \, \bbv_\mu + \gamma^\mu \gamma_5 \, \bba_\mu
        +\sigma^{\mu\nu} \, \bbt_{\mu\nu}
      \right)
\end{align}
with correspondingly transformed coefficient functions, $\bbs,\bbp,\bbv_\mu,\bba_\mu,\bbt_{\mu\nu}$.
By inserting \Eq\eqref{eqn:wigner_fierz} into \Eq\eqref{eqn:Wigner-EoM} the latter can be decomposed, yielding
\begin{equation}
  \label{eqn:Wigner-EoM-Fierz}
  \begin{rarray}{1.5pt}[c]{rlll}
      \qquad\Dt \bbs
      &=
      &+2\vv{P}\cdot\vv{\bbt}^{1}
    \\
      \Dt \bbp
      &=
      &-2\vv{P}\cdot \vv{\bbt}^{2}
      &-{2m \, \bba^0 }
    \\
      \Dt \bbv^0
      &=-{\vDx\vv{\bbv}}
    \\
      \Dt \bba^0
      &=-{\vDx\vv{\bba}}
      &
      &+{2m\,\bbp}
    \\
      \Dt \vv{\bbv}
      &=-{\vDx \bbv^0}
      &-{2\vv{P}\times \vv{\bba} }
      &-{2 m\, \vv{\bbt}^{1}}
    \\
      \Dt \vv{\bba}
      &=-{\vDx \bba^0}
      &-{2\vv{P}\times \vv{\bbv} }
    \\
      \Dt \vv{\bbt}^{1}
      &=-{  \vDx \!\times  \vv{\bbt}^{2} }
      &-{2\vv{P}\, \bbs}
      &+{2m\, \vv{\bbv}}
    \\
      \Dt \vv{\bbt}^{2}
      &=+{ \vDx\! \times \vv{\bbt}^{1}}
      &+{2\vv{P}\, \bbp}
  \end{rarray}
\end{equation}
with two vectors $\vv{\bbt}^{1/2}$ containing the components of the antisymmetric tensor $\bbt_{\mu\nu}$
\begin{align}
  \label{eqn:wigner_comp_t}
  \vv{\bbt}^{1} \definedby 2\bbt^{i0}\vv{e_i},   &
  \vv{\bbt}^{2} \definedby \bbt_{ij}\epsilon^{ijk}\vv{e_k}\,.
\end{align}
For $\vv{E}$ and $\vv{B}$ that vanish at asymptotically early times, $t\to -\infty$, initial conditions are given by the vacuum solution with non-vanishing components
\begin{align}
  \label{eqn:wigner_vac}
   \bbs_\vac
  &=  \frac{-2m}{\omega(\vv{p}\,)}\,,
  &
   \vv{\bbv}_\vac
  &=   \frac{-2\vv{p}}{\omega(\vv{p}\,)}\,,
\end{align}
where
\begin{align}
 \omega^2\definedby \vv{p}^2+m^2.  \label{eq:scE}
\end{align}
The vacuum solution can also be written in matrix form as
\begin{align*}
  \mathcal{W}_\mathrm{vac}&=\frac14(\mathbbm{1}\mathbbm{s}_\mathrm{vac}+\gamma^\mu\mathbbm{v}_{\mathrm{vac}\mu})  = -\frac{m}{2\omega}\mathbbm{1} + \frac{\vv{\gamma}\cdot\vv{p}}{2\omega}\,.
\end{align*}
The equation of motion \Eqref{eqn:Wigner-EoM-Fierz} combined with the initial condition \Eqref{eqn:wigner_vac} defines the initial value problem.

\section{Observables} %
Using the definition of the Wigner function and the symmetries of QED, the conserved quantities charge
\begin{align}
  \label{eqn:obs_charge}
  \mathcal{Q} &= e \int \dd\Gamma\,\bbv_0(t,\vv{x},\vv{p})\,,
 \intertext{energy}
 \label{eqn:obs_energy}
\begin{split}
  \mathcal{E} &= \int \dd\Gamma\,(\underbrace{\vv{p}\cdot\vv{\bbv}(t,\vv{x},\vv{p})+m\,\bbs(t,\vv{x},\vv{p})}_{\epsilon(t,\vv{x},\vv{p})}) \\
              &\qquad\qquad\qquad\qquad + \frac12 \int \dd[3]x \left( \abs{\vv{E}(t,\vv{x})}^2 + \abs{\vv{B}(t,\vv{x})}^2 \right)\,,
\end{split}
\intertext{momentum}
\nonumber
\mathcal{\vv{P}} &= \int \dd\Gamma\,\vv{p}\,\bbv_0(t,\vv{x},\vv{p}) + \int \dd[3]x \vv{E}(t,\vv{x})\times\vv{B}(t,\vv{x})\,,
 \intertext{angular momentum}
 \label{eqn:angular_momentum}
\begin{split}
 \mathcal{\vv{S}} &= \int \dd\Gamma\,\left( \vv{x}\times\vv{p}\,\bbv_0(t,\vv{x},\vv{p})-\frac12\vv{\bba}(t,\vv{x},\vv{p}) \right) \\
                  &\qquad\qquad\qquad\qquad + \int \dd[3]x \vv{x}\times\left(\vv{E}(t,\vv{x})\times\vv{B}(t,\vv{x})\right)
\end{split}
 \intertext{and Lorentz boost operator}
 \nonumber
 \begin{split}
 \vv{K} &= t\mathcal{\vv{P}} -  \int  \dd\Gamma \vv{x} \bigl( \vv{p}\cdot\vv{\bbv}(t,\vv{x},\vv{p})+m\,\bbs(t,\vv{x},\vv{p}) \bigr) \\
                  &\qquad\qquad\qquad\qquad - \frac12 \int \dd[3]x \left( \abs{\vv{E}(t,\vv{x})}^2 + \abs{\vv{B}(t,\vv{x})}^2 \right)
 \end{split}
\end{align}
can be derived \autocite{BialynickiBirula:1991tx}.
The phase space measure $\dd\Gamma$ is given by $\dd\Gamma=\dd[3]{\vv{x}}\frac{\dd[3]{\vv{p}}}{\left( 2\pi \right)^3}$.
The quantity
\begin{equation}
 \label{eqn:energdens}
 \epsilon(t,\vv{x},\vv{p}) = \vv{p}\cdot \vv{\bbv}(t,\vv{x},\vv{p})+m\,\bbs(t,\vv{x},\vv{p})
\end{equation}
in the integrand in \Eqref{eqn:obs_energy} is of special interest as it can be interpreted as a (phase space) energy density of the fermionic fields.
From it the particle density is calculated by subtracting the vacuum solution \( \epsilon_\mathrm{vac} =  m \mathbbm{s}_\mathrm{vac} +\vv{p}\cdot\vv{\mathbbm{v}}_\mathrm{vac} = -2\omega \) and normalizing the result to the energy of a particle pair
\begin{align*}
  f &\mathrel{:=} \frac{1}{2\omega} \left( \epsilon - \epsilon_\mathrm{vac} \right) = \frac{1}{2\omega}\epsilon + 1\,.  \numberthis\label{eqn:oneparticledist}
\end{align*}
In spatially homogeneous, unidirectional, purely electric fields this distribution function $f$ was shown to be identical to the definition of the distribution function from QKT \autocite{Hebenstreit:2010vz}.
Observe that \Eqref{eqn:energdens} can be written in a more generic way like
\begin{align}
  \label{eqn:EpsilonProjectorFormula}
  \epsilon[\mathcal{W}] = \tr[\mathcal{W}(m\mathbbm{1}+\vv{p}\cdot\vv{\gamma})]\,.
\end{align}
As a consequence, \Eqref{eqn:oneparticledist} can be written in terms of a projection of the Wigner function
\begin{align}
  \label{eqn:general_f}
  f[\mathcal{W} - \mathcal{W}_\mathrm{vac.}] &= \frac{1}{2\omega}\tr\left[
	  \left( \mathcal{W} - \mathcal{W}_\mathrm{vac.} \right)
	  \left( m\mathbbm{1}+\vv{p}\cdot\vv{\gamma} \right)
	  \right]\,.
\end{align}
It is important to stress that the particle interpretation is only valid when $\vv{E}=\vv{B}=0$.
For the cases discussed in this work this is true at asymptotically large times.
For instance, the total number of particles ($=$ number of anti-particles $=$ number of pairs) produced out of the vacuum is given by
\begin{equation*}
 N=\lim_{t\to\infty}\int \dd\Gamma\, f(t,\vv{x}, \vv{p}).
\end{equation*}
Possible definitions for particle numbers at intermediate times are discussed in \Ref\cite{Dabrowski:2014ica,Dabrowski:2016tsx}.
When spatial homogeneity is assumed, the distribution function does not depend on position and it does not make sense to integrate over space, as this would yield infinity.
It is then sufficient to integrate over the momentum to find the spatial pair production density which will be called the particle yield
\begin{equation}
 \label{eqn:numberofparticles}
 \mathcal{N}:=\lim_{t\to\infty}\int \frac{\dd[3]{\vv{p}}}{\left( 2\pi \right)^3} f(t,\vv{p})\,.
\end{equation}
In some of the numerical calculations the asymptotic distribution function has not been calculated for all $\vv{p}$, but only for a slice with $p_z=0$.
In those cases the particle yield in the $p_x$-$p_y$-plane is called $\mathcal{N}_{xy}$ and defined as
\begin{equation}
 \label{eqn:numberofparticles2d}
 \mathcal{N}_{xy}:=\lim_{t\to\infty}\int \frac{\dd{p_x}}{2\pi}\int\frac{\dd{p_y}}{2\pi} \left.f(t,\vv{p})\right|_{p_z=0}\,.
\end{equation}

\section{Quantum States}
\label{sec:quantumstates}
In addition to the full one-particle distribution function, the Wigner function gives access to the spinor space of the Dirac field and hence the various quantum states of electrons.
As such the information about spin and chirality of the produced pairs can be extracted from the Wigner function.
An important property of the Wigner function is that it behaves under conjugation like a Dirac matrix
\begin{align}
  \label{eqn:conjugation_property}
  \mathcal{W}^\dagger = \gamma^0\mathcal{W}\gamma^0\,.
\end{align}
When a projector $P$ is applied to the Wigner function, this property might not hold for $P\mathcal{W}$.
That is because the Wigner function is build from the commutator of two field operators (see \Eqref{eqn:WignerDef}) and the Projector $P$ does not apply to both field operators in a symmetric way.
But how should the projector be applied to the spinor and/or the adjoint spinor in order to get a result, that fulfills \Eqref{eqn:conjugation_property}?
It is in general possible to regain the property described in \Eqref{eqn:conjugation_property} for any matrix $A$ by replacing it with
\begin{align*}
 B=\frac12\left( A+\gamma^0A^\dagger\gamma^0 \right)\,.
\end{align*}
Now \Eqref{eqn:conjugation_property} holds for $B$
\begin{align*}
 \gamma^0B\gamma^0=\frac12\left( \gamma^0A\gamma^0 + A^\dagger \right)=B^\dagger\,.
\end{align*}
Applying this procedure to $P\mathcal{W}$ for any projector $P$ results in
\begin{align}
  \label{eqn:atdef}
  P@\mathcal{W} \definedby \frac12\left( P\mathcal{W} + \gamma^0(P\mathcal{W})^\dagger\gamma^0 \right) \\
  \nonumber
               &= \frac12\left( P\mathcal{W} + \mathcal{W}\gamma^0P^\dagger\gamma^0 \right)\,.
\end{align}
The operator $@$ is now defined as a shorthand for applying an operator to a matrix in the way defined in \Eqref{eqn:atdef}.
From $\sum_i P_i=\id$ follows $\sum_i P_i@\mathcal{W}=\mathcal{W}$ for any $\mathcal{W}$.
It should however be noted, that applying the operation $P@$ to $\mathcal{W}$ is not idempotent, because in general $P@(P@\mathcal{W})\neq P@\mathcal{W}$.

What do $P@\mathcal{W}$ and specifically the two terms $P\mathcal{W}$ and $\mathcal{W}\gamma^0P^\dagger\gamma^0$ mean?
Remember that the Wigner function is basically build from a commutator
\begin{align*}
  \mathcal{W}_{ab}\sim[\hat{\Psi}_a(\vv{x}_+,t),\hatbar{\Psi}_b(\vv{x}_-,t)]\,.
\end{align*}
A term $P\mathcal{W}$ results, when the spinor in the first argument of the commutator $\hat{\Psi}$ is replaced by the same spinor after the projection $P$ was applied,
\begin{align*}
  \hat{\Psi} \to P\hat{\Psi}\,.
\end{align*}
Because the entries of the projection matrix are just numbers and can be removed from the commutator we can deduce
\begin{align*}
 \mathcal{W}_{ab} &\sim[\hat{\Psi}_a,\hatbar{\Psi}_b]
 \\
 &\to [P_{ac}\hat{\Psi}_c,\hatbar{\Psi}_b]
 \\
 &= P_{ac}[\hat{\Psi}_c,\hatbar{\Psi}_b]
 \\
 &\sim P_{ac}\mathcal{W}_{cb} = (P\mathcal{W})_{ab}\,.
\end{align*}
Analogously a term $\mathcal{W}\gamma^0P^\dagger\gamma^0$ results from applying the projection operator to the adjoint spinor and replacing
\begin{align*}
 \hatbar{\Psi} &\to \overline{P\hat{\Psi}}\,.
\end{align*}
This is identical to applying the adjoint projector from the right
\begin{align*}
 \mathcal{W}_{ab} &\sim[\hat{\Psi}_a,\hatbar{\Psi}_b]
 \\
 &\to [\hat{\Psi}_a,(\overline{P\hat{\Psi}})_b]
 \\
 &= [\hat{\Psi}_a,\left( \hat{\Psi}^\dagger P^\dagger \gamma^0 \right)_b]
 \\
 &= [\hat{\Psi}_a,\hatbar{\Psi}_d]  \left( \gamma^0 P^\dagger \gamma^0 \right)_{db}
 \\
 &\sim \left( \mathcal{W}\gamma^0P^\dagger\gamma^0 \right)_{ab}\,.
\end{align*}

By these results the question of how the projector should be applied to both of the spinor operators in the Wigner function is answered.
We can now apply the generic definitions for the energy density in the Dirac field $\epsilon$ or the one-particle distribution function $f$ from \Eqs\eqref{eqn:EpsilonProjectorFormula} and \eqref{eqn:general_f} to the projected Wigner function $P_s@\mathcal{W}$ for any state $s$ and calculate the particle distribution for specific quantum states.
\begin{align}
  \nonumber
  \epsilon_s(\mathcal{W}) \definedby
  \tr[(P_s@\mathcal{W})(m\mathbbm{1}+\vv{p}\cdot\vv{\gamma})]
  \\
  \label{eqn:general_f_projected}
  f_s[\mathcal{W}]
    \definedby
    \frac{1}{2\omega}\epsilon_s(\mathcal{W}-\mathcal{W}_\vac)\,.
\end{align}
Useful projectors $P$ could correspond to various spinor eigenstates, e.\,g., spin eigenstates, charge eigenstates, chiral eigenstates or combinations thereof.
Please note that these definitions are different from those in \Ref\cite{Blinne:2015zpa}, but the relevant results are unchanged.

We will now look at some possible projections.
Some well known projection operators are the following \autocite{opac-b1131978}
\begin{align*}
  &\text{Chiral projection} &    P_{\mathrm{r/l}} &= \frac12(\mathbbm{1}\pm\gamma_5)  \\
  &\text{Charge projection} &    P_{\mathrm{p/e}} &= \frac12(\mathbbm{1}\pm Q) = \frac12(\mathbbm{1}\mp \gamma^0)  \\
  &\text{Spin projection} &    P_{(a,b,c),\pm} &= \frac12(\mathbbm{1}\pm (\i a\gamma^2\gamma^3 + \i b\gamma^3\gamma^1 + \i c\gamma^1\gamma^2))
  \\ &&&= \frac12(\mathbbm{1}\pm (a \sigma^{23} + b \sigma^{31} + c \sigma^{12})\,.
\end{align*}
The spin projector requires $a^2+b^2+c^2=1$ to ensure $P_{(a,b,c),\pm}^2=P_{(a,b,c),\pm}$.

\subsubsection{Chiral Projection}
The results for the energy density for the chiral states $\mathrm{r/l}$ are
\begin{align*}
  \epsilon_\mathrm{r/l}(\mathcal{W}) &= \frac{m}{2}\mathbbm{s} + \frac{1}{2}(\vv{p}\cdot(\vv{\mathbbm{v}}\pm\vv{\mathbbm{a}}))
  \\
  &= \frac12\epsilon(\mathcal{W}) \pm\frac12\vv{p}\cdot\vv{\mathbbm{a}}\,.
\end{align*}
For the distribution function $f$ this turns into
\begin{align}
  \nonumber
  f_\mathrm{r/l} &= \frac{1}{2\omega}\epsilon_\mathrm{r/l}(\mathcal{W}-\mathcal{W}_\vac)
  \\
  \nonumber
  &= \frac{1}{2\omega}\left( \epsilon_\mathrm{r/l}(\mathcal{W})-\epsilon_\mathrm{r/l}(\mathcal{W}_\vac) \right)
  \\
  \nonumber
  &= \frac{1}{2\omega}\left( \frac12\underbrace{\epsilon(\mathcal{W})}_{2\omega (f-1)}\pm \frac12 \vv{p}\cdot\vv{\bba}-\frac12\underbrace{\epsilon(\mathcal{W}_\vac)}_{-2\omega} \right)
  \\
  \nonumber
  &=\frac12\left( f \pm \delta f_\mathrm{c} \right)
  \intertext{with the chiral asymmetry of the particle density}
  \label{eqn:asymm_chiral}
  \delta f_\mathrm{c} \definedby \frac{1}{2\omega}\vv{p}\cdot\vv{\bba}\,.
\end{align}

\subsubsection{Charge Projection}
When the charge projections $P_\mathrm{p/e}$ are applied the energy density and one-particle distribution function are
\begin{align}
 \nonumber
 \epsilon_\mathrm{p/e} &= \frac{m}{2}\left( \mathbbm{s} \mp \bbv^0 \right) + \frac{1}{2}(\vv{p}\cdot\vv{\mathbbm{v}})
 \intertext{and}
 \nonumber
 f_\mathrm{p/e} &= \frac12\left( f \mp \delta f_\mathrm{Q} \right)
 \intertext{with the charge asymmetry of the particle density}
 \label{eqn:asymm_charge}
 \delta f_\mathrm{Q} &= \frac{1}{2\omega}m\bbv^0\,.
\end{align}
This result is no surprise given that we already interpret $\bbv^0$ as the charge density due to \Eqref{eqn:obs_charge}.

\subsubsection{Spin Projection}
In case of the spin projections the result for any direction $\vv{n}$ with $\vv{n}^2=1$ is
\begin{align}
  \nonumber
  \epsilon_{\vv{n},\pm} &= \frac{1}{2}\left( m\bbs + \vv{p} \cdot\vv{\bbv}  \right)
  \pm \frac{1}{2}\left( - \vv{p}\cdot\vv{n}\,\bba_0 + m\,\vv{n} \cdot \vv{\bbt}^2 \right)
  \intertext{and}
  \nonumber
  f_{\vv{n},\pm} &= \frac12\left( f \pm \delta f_{\vv{n}} \right)
  \intertext{with the spin asymmetry of the particle density}
  \label{eqn:asymm_spin}
  \delta f_{\vv{n}} &= \frac{1}{2\omega}\left(- \vv{p}\cdot\vv{n}\,\bba_0 + m\,\vv{n} \cdot \vv{\bbt}^2 \right)\,.
\end{align}
On the one hand this result does not seem to correspond to \Eqref{eqn:angular_momentum}.
On the other hand this result is perfectly understandable, because the spin operators $\sigma^{ij}$ for $i\neq j$ select the $\vv{\bbt}^2$ component, as is evident from the decomposition of the Wigner function \Eqref{eqn:wigner_fierz}.

\subsubsection{Magnetic Moment}
It is however possible to construct observables that do correspond to \Eqref{eqn:angular_momentum}, by combining a spin projection with a charge projection.
For simplicity let
\begin{align*}
  P_\mathrm{u/d} \definedby P_{(0,0,1),\pm}\,.
\end{align*}
We can now construct projectors that correspond to magnetic moment by combining the 4 states with specific spin and charge into two groups as given by
\begin{align*}
  P_{\mu_z^+} &= P_\mathrm{p}P_\mathrm{u}+P_\mathrm{e}P_\mathrm{d} = \frac12(\id+\gamma_5\gamma^3) \\
  P_{\mu_z^-} &= P_\mathrm{e}P_\mathrm{u}+P_\mathrm{p}P_\mathrm{d} = \frac12(\id-\gamma_5\gamma^3)\,.
\end{align*}
We now obtain
\begin{align*}
  \epsilon_{\mu_z^\pm} &:= \frac{1}{2}(m(\mathbbm{s}\mp\mathbbm{a}_z)+\vv{p}\cdot\vv{\mathbbm{v}}\pm(\vv{p}\times\vv{\mathbbm{t}})_z)\,.
\end{align*}
The particle density is given by
\begin{align}
 \nonumber
 f_{\mu_z^\pm} &= \frac12\left( f \pm \delta f_{\mu_z} \right)
 \intertext{with its asymmetry}
 \label{eqn:asymm_magnetic}
 \delta f_{\mu_z} & = \frac{1}{2\omega}\left( -m\mathbbm{a}_z+(\vv{p}\times\vv{\mathbbm{t}})_z \right).
 \intertext{The analogue results for magnetic moment in the $x$ and $y$ direction are}
 \nonumber
 \delta f_{\mu_x} & = \frac{1}{2\omega}\left( -m\mathbbm{a}_x+(\vv{p}\times\vv{\mathbbm{t}})_x \right),
 \\
 \nonumber
 \delta f_{\mu_y} & = \frac{1}{2\omega}\left( -m\mathbbm{a}_y+(\vv{p}\times\vv{\mathbbm{t}})_y \right).
\end{align}
These results contain the spin density $\vv{\bba}$ identified in \Eqref{eqn:angular_momentum}.

  \chapter{Homogeneous Electric Fields}
\label{cha:homogeneous}
This chapter will cover electron positron pair production in time\hyp{}dependent homogeneous electric fields with $\vv{B}=0$.
In \Sec\ref{sec:wigner_method} the Wigner function formalism is applied to this special case and the Wigner method is derived.
Afterwards a semiclassical method for two-component electric fields is introduced \Sec\ref{sec:semiclass}.
The rotating Sauter pulse is introduced in \Sec\ref{sec:simprot}, which makes up a large part of this thesis.
The semiclassical method has so far been applied only to the rotating Sauter pulse and will be discussed for this special case in \Sec\ref{sec:numsemi}.
Afterwards the phenomenology of the rotating Sauter pulse regarding pair production is explained in \Sec\ref{sec:phenomenology} as a preliminary consideration, before it is used as an example to compare the available methods with each other in \Sec\ref{sec:cmp}.
Afterwards the results of the studies conducted on the rotating Sauter pulse are presented in \Secs\ref{sec:simprotTY} to \ref{sec:spin_states}.
Towards the end of this chapter (\Secs\ref{sec:general_polarization} to \ref{sec:Bichromatic}) a number of different electric field pulses are explored.

\section{The Wigner Method}
\label{sec:wigner_method}
In a spatially homogeneous setup ($\nablax=0$) with a pure electric field ($\vv{B}=0$) the pseudo differential operators from \Eq\eqref{eqn:diff-ops} simplify to
\begin{align}
  \label{eqn:diff-ops-simp}
  \vDx &= 0
  \,,\,
  \vv{P} = \vv{p}
  \text{ and }
  \Dt = \partial_t + e \vv{E}(t) \cdot\nablap\,.
\end{align}
As a result the system of 16 homogeneous differential equations of motion decouples into two independent sets of 10 and 6 equations each.
Only the set of 10 equations for the functions
\begin{align*}
  \bbw&=\begin{pmatrix}
    \bbs, & \vv{\bbv}, & \vv{\bba}, & \vv{\bbt}\,
  \end{pmatrix}^\intercal\,, &
  \left(\vv{\bbt}\right)_i \definedby  \bbt_{0i}-\bbt_{i0}
\end{align*}
has a non-vanishing initial condition, while the remaining functions are zero for all times
\begin{align}
  \label{eqn:zero_components}
  \bbp=\bba^0=\bbv^0&=0\,, & \vv{\bbt}^2=\vv{0}.
\end{align}
What remains of the equations of motion can be written as
\begin{align}
 \nonumber
 \left(
    \partial_t
    +e \vv{E}(t) \cdot \nablap
  \right)
  \begin{pmatrix}
    \bbs \\ \vv{\bbv} \\ \vv{\bba} \\ \vv{\bbt}
  \end{pmatrix}
  &=  \begin{pmatrix}
        0 & 0 & 0 & 2\vv{p}^{\,\intercal} \\
        0 & 0 & -2\vv{p}\times &-2m \\
        0 & -2\vv{p}\times & 0 &0 \\
        -2\vv{p} & 2m & 0 & 0
      \end{pmatrix}
      \cdot
      \begin{pmatrix}
        \bbs \\ \vv{\bbv} \\ \vv{\bba} \\ \vv{\bbt}
      \end{pmatrix}
  \\
  \label{eqn:Wigner-EoM-Homogen}
 \Leftrightarrow\quad
\left(
    \partial_t
    +e \vv{E}(t) \cdot \nablap
  \right) \, \bbw
  &=  \mathcal{M}\cdot\bbw\,.
\end{align}

In the numerical calculations a slightly different representation of the Wigner Matrix is used. The Wigner function is written as
\begin{align*}
 \begin{split}
  4\mathcal{W}&=\left( 2(f-1)\frac{m}{\omega}-\frac{\vv{p}\cdot\vv{v}}{m} \right)\mathbbm{1}
  +\i\gamma_5\mathbbm{p}+\gamma^0\mathbbm{v}_0-\vv{\gamma}\cdot\left( \vv{v}+2(f-1)\frac{\vv{p}}{\omega} \right)
  \\
  &\qquad\qquad+\gamma^\mu\gamma_5\mathbbm{a}_\mu+\sigma^{\mu\nu}\mathbbm{t}_{\mu\nu}\,.
 \end{split}\numberthis\label{eq:f2}
\end{align*}
By comparing the two representations we see
\begin{align}
  \label{eqn:wigner_subst}
  \mathbbm{s} &=  \frac{2m}{\omega}(f-1)-\frac{\vv{p}\cdot\vv{v}}{m}  &    \vv{\mathbbm{v}}=\vv{v}+\frac{2\vv{p}}{\omega} (f-1)\,.
\end{align}
Obviously this substitution does not change the degrees of freedom, as it is a linear system of equations with determinant $\det\frac{\partial\left( \bbs, \vv{\bbv} \right)}{\partial \left( f-1,\vv{\bbv}_\mathrm{n} \right)}=\frac{2\omega}{m}>0$ in each point in time.
Additionally we need to make sure that the variable $f$ as introduced by \Eqref{eq:f2} is indeed the same object as the one-particle distribution function introduced in \Eqref{eqn:oneparticledist}. To this end we will start with \Eqref{eqn:energdens} and reproduce \Eqref{eqn:oneparticledist}.
\pushQED{\qed}
\begin{align*}
  \epsilon &= m\mathbbm{s}+\vv{p}\cdot\vv{\mathbbm{v}} \\
  &=  m\left( \frac{2m}{\omega}(f-1)-\frac{\vv{p}\cdot\vv{v}}{m} \right) + \vv{p}\cdot \left( \vv{v}+\frac{2\vv{p}}{\omega} (f-1) \right) \\
  &= 2\frac{m^2}{\omega}(f-1)-\vv{p}\cdot\vv{v} + \vv{p}\cdot\vv{v} + 2\frac{\vv{p}^2}{\omega}(f-1) \\
  &= 2\frac{m^2+\vv{p}^2}{\omega}(f-1) \\
  &= 2\omega(f-1)\hfill \qedhere
\end{align*}

Applying this substitution to the equations of motion \Eqref{eqn:Wigner-EoM-Homogen} results in
\begin{align}
  \label{eqn:Wigner-EoM-NoB}
  \begin{split}
  \left( \partial_t + e\vv{E}\cdot\nablap \right) f(t,\vv{p})       &= \frac{1}{2\omega}e\vv{E}\cdot\vv{v} \,,
  \\
  \left( \partial_t + e\vv{E}\cdot\nablap \right) \vv{v}(t,\vv{p}) &= \frac{1}{2\omega^3}\left( \vv{p} (e\vv{E}\cdot\vv{p})-\omega^2e\vv{E} \right)(f-1)\\
                &\hspace{1cm} -\frac{1}{\omega^2}\vv{p}(e\vv{E}\cdot\vv{v})
                              -\vv{p}\times\vv{a}
                              -2\vv{t} \,,
  \\
  \left( \partial_t + e\vv{E}\cdot\nablap \right) \vv{a}(t,\vv{p}) &= -\vv{p}\times\vv{v} \,,
  \\
  \left( \partial_t + e\vv{E}\cdot\nablap \right) \vv{t}(t,\vv{p}) &= 2\left(\vv{v}+\vv{p}(\vv{p}\cdot\vv{v})\right)\,,
  \end{split}
\end{align}
which is a system of 10 inhomogeneous partial differential equations for $f$ and 9 auxiliary quantities $\vv{v}, \vv{a}=\vv{\bba}, \vv{t}=\vv{\bbt}$.
The initial conditions at $t\to-\infty$ for all $\vv{p}$ are given by
\[
  f = 0,\ \vv{v}=\vv{a}=\vv{t}=\vv{0}\,.
\]

Due to the nature of the substitution in \Eq\eqref{eqn:wigner_subst}, the one-particle distribution function is readily accessible from the numerical results.
The formulae for the various asymmetries of the one-particle distribution function $f$ given in \Sec\ref{sec:quantumstates} are already written in terms of the one-particle distribution function and the remaining Wigner function components.
Applying what we know in the spatially homogeneous, purely electric case (\Eqref{eqn:zero_components}) and using the same substitution as for the Wigner method \Eqref{eqn:wigner_subst}, we arrive at formulae to calculate these asymmetries for the numerical results.

The chiral and magnetic moment asymmetries, \Eqs\eqref{eqn:asymm_chiral} and \eqref{eqn:asymm_magnetic} are not modified besides using a different symbol for the Wigner components
\begin{align}
  \label{eqn:asymm_chiral_homog}
  \delta f_\mathrm{c} &= \frac{1}{2\omega}\vv{p}\cdot\vv{a}\,,
  \\
  \label{eqn:asymm_magnetic_homog}
  \delta f_{\mu_z} &= \frac{1}{2\omega}\left( -ma_z+(\vv{p}\times\vv{t})_z \right).
\end{align}
In contrast the spin and charge asymmetries from \Eqs\eqref{eqn:asymm_spin} and \eqref{eqn:asymm_charge} vanish due to \Eqref{eqn:zero_components}.

\subsection{The Modified Quantum Kinetic Equation} %
In order to transform the system \Eqref{eqn:Wigner-EoM-NoB} of PDEs into ODEs we apply the method of characteristics.
This is based on the requirement that the kinetic momentum $\vv{p}$ follows the solution of the classical equation of motion for a particle with charge $e$ in the external field with canonical momentum $\vv{q}$
\begin{align}
  \label{eqn:kinetic_momentum}
  \vv{p}_{\hspace{-2pt}\vv{q}}(t) &=\vv{q} -e\vv{A}(t) \,.
\end{align}
This allows the partial derivatives w.\,r.\,t. $\vv{p}$ to be absorbed into the total temporal derivative according to
\begin{align*}
  \frac{\dd{}}{\dd{t}}g(t,\vv{p}_{\hspace{-2pt}\vv{q}}(t)) &= \left[\bigl( \partial_t + (\partial_t p_i)\partial_{p_i}\bigr)g(t,\vv{p})\right]_{\vv{p}=\vv{p}_{\hspace{-2pt}\vv{q}}(t)} \\
  &= \left[\bigl( \partial_t + e\vv{E}(t)\nablap\bigr)g(t,\vv{p})\right]_{\vv{p}=\vv{p}_{\hspace{-2pt}\vv{q}}(t)}\,.
\end{align*}
Additionally any function $g(t,\vv{p}_{\hspace{-2pt}\vv{q}}(t))$ can now be reinterpreted as a function $\tilde{g}(t,\vv{q})$, where the canonical momentum $\vv{q}$ is now merely a parameter enumerating different trajectories.
Along any of those trajectories the functions can be calculated by solving a set of ordinary differential equations without interchanging information with other trajectories.
Since results are always being taken for $t\to\infty$, it is convenient to gauge the vector potential $\vv{A}$ in such a way that $\vv{p}_{\hspace{-2pt}\vv{q}}(t)\to\vv{q}$ for $t\to\infty$\,.

The result of this procedure is a modified quantum kinetic equation \autocite{Blinne:2013via}, which can be solved numerically to directly calculate the one-particle distribution function $f(t,\vv{p})$ at $t\to\infty$.
The kinetic momentum along the classical trajectory $\vv{p}_{\hspace{-2pt}\vv{q}}(t)$ will be denoted $\vv{p}$ in the context of the Wigner method.
The modified quantum kinetic equations read
\begin{align}
  \label{eqn:Wigner-EoM-ModQKT}
  \begin{split}
  \dot{f}       &=\frac{1}{2\omega}e\vv{E}\cdot\vv{v} \,,
  \\
   \dot{\vv{v}} &=\frac{1}{2\omega^3}\left( \vv{p} (e\vv{E}\cdot\vv{p})-\omega^2e\vv{E} \right)(f-1)\\
                &\hspace{4cm} -\frac{1}{\omega^2}\vv{p}(e\vv{E}\cdot\vv{v})
			      -\vv{p}\times\vv{a}
			      -2\vv{t} \,,
  \\
  \dot{\vv{a}} &=-\vv{p}\times\vv{v} \,,
  \\
  \dot{\vv{t}} &=2\left(\vv{v}+\vv{p}(\vv{p}\cdot\vv{v})\right)\,,
  \end{split}
\end{align}
where the dot above any function denotes the total temporal derivative, e.\,g., $\dot{f}=\frac{\dd{}}{\dd{t}}f$.
Combined with the initial conditions
\begin{align}
  \label{eqn:wigner_method_initial}
  f = 0,\ \vv{v}=\vv{a}=\vv{t}=\vv{0}
\end{align}
at $t\to-\infty$\,, the initial value problem is well defined.
The modified quantum kinetic equations turn into a homogeneous ODE when $E(t)$ is set to $\vv{0}$.
From this we can deduce that the initial condition \Eqref{eqn:wigner_method_initial} is also true at a finite time $t_0$ when $E(t)=0$ for $t<t_0$.

\subsection{Numerical Calculations}
The electron mass is the base unit used in the numerical calculations, while $\hbar=c=1$.
This means that time and length are measured in inverse electron masses, while the unit of energy and momentum is the electron mass.
The elementary charge vanishes from the numerical calculations when all electric fields are given in units of the critical field strength as in
\begin{align*}
  e\vv{E}(t)= e\,E_\crit\,\frac{\vv{E}(t)}{E_\crit} = m^2 \frac{\vv{E}(t)}{E_\crit}\,.
\end{align*}
As a benchmark test, it is useful to consider the Sauter pulse, $E(t)=E_0\sech^2  \frac{t}{\tau}  $.
Due to the exponential suppression of the electric field for $|t|\gg\tau$, it is sufficient to start solving the ODE at a time $t_0=-10\tau$ and stop at a time $t_1=10\tau$.
Outside of this region the electric field is sufficiently small such that $\dot{f}\approx0$.

In order to be able to solve the modified quantum kinetic equations \eqref{eqn:Wigner-EoM-ModQKT} numerically, the kinetic momentum \Eqref{eqn:kinetic_momentum} must be known.
Either one is able to calculate the vector potential $\vv{A}(t)$ corresponding to the input field $\vv{E}(t)$ or one has to solve the same classical equation of motion over and over along each trajectory.
It is in principle, for a lot of different $\vv{E}(t)$, possible to calculate $\vv{A}(t)$ analytically.
In a numerical code it would be impractical to have to change these analytical results each time a new configuration for the electric field is used.
Additionally, if no analytic solution can be found or can not be written in terms of available numerical functions, the vector potential must be calculated numerically anyway.
For these reasons the decision was made to calculate $\vv{A}(t)$ numerically with sufficient precision, for all needed $t$, once at the beginning of the program.

Our first numerical solutions have been found using \textit{Wolfram Mathematica} \autocite{mathematica}, starting with the set of homogeneous differential equations for the Wigner function from \Eqref{eqn:Wigner-EoM-Homogen} with the method of characteristics applied to obtain ordinary differential equations.
In contrast to all later approaches, these calculations have been carried out using an analytic solution for the vector potential.
These calculations quickly showed numerical problems due to the necessity to calculate the one-particle density function $f$
after finishing the integration using \Eqs\eqref{eqn:energdens} and \eqref{eqn:oneparticledist}.
The first problem is, that for small pair production rates, the solution would only differ very little from the vacuum solution, which leads to loss in precision from the difference in \Eqref{eqn:oneparticledist}.
On the other hand the execution of \Eqref{eqn:energdens} also includes large cancellations.
This realization led to introducing the substitutions \Eqref{eqn:wigner_subst}, which were then incorporated into any later numerical implementation.

Because the integration processes for every single data point of the pair production spectrum are independent, parallelization of the calculation is trivial.
Various technical approaches to split up the work and collect the results later have been tested.

\subsubsection{Ppsolve}
When the high-level nature of \textit{Wolfram Mathematica} \autocite{mathematica} hampered the numerical performance, a new implementation was created written in the \textit{Python} programming language using the \textit{Scipy} \autocite{scipy2001,numpyscipy2011} package for scientific computation.

\textit{Scipy} contains an implementation of a generic ODE solver called \textit{LSODA}, part of \textit{ODEPACK} \autocite{hindmarsh1982odepack}.
The advantage of the \textit{LSODA} algorithm is that it switches between stiff and non-stiff methods adaptively.
The implementation accepts two parameters, \texttt{rtol} (relative tolerance) and \texttt{atol} (absolute tolerance), and then chooses the step size such that it ensures the approximate integration error $\varepsilon$ to be smaller than
\[
\varepsilon = \texttt{atol} + |\mathbf{x}|\,\texttt{rtol}\,,
\]
where $|\mathbf{x}|$ is the Euclidean norm of the solution vector.
Previous experience \autocite{Blinne:2013via} showed, that $\texttt{rtol}$ should be set to 0, because of big intermediate function values, which would spoil the overall precision.
As a result the only external parameter to the numerical calculations is the absolute error tolerance $\texttt{atol}=10^{-k}$.

The evaluation of the right-hand sides of the equations was carried out using a compiled \textit{C++} component in order to improve performance.
This component was built using a specialized part of \textit{Scipy}, called \textit{weave}.
It used the \textit{Blitz++} library \autocite{veldhuizen1998arrays} to generate optimized code.

For parallelization a work queue was implemented using \textit{MPI4Python} such that it was possible to use the computer cluster of the university.
This work queue would distribute all the single ODE solver runs across all the processors of all the nodes assigned to a job, while a single job could easily contain a multi-dimensional parameter scan.
The workload distribution using this model turned out to be quite uniform, maximizing the computation load.
The total cpu time for a parameter scan was however difficult to predict such that jobs were often cancelled when they exceeded their wall time limit.

\subsubsection{Charwigner}
While looking into other available ODE solver implementations, the \textit{odeint} library \autocite{Ahnert2011} came into focus, which is part of the \textit{boost} \autocite{boost} library project.
It offers a lot of different algorithms through a unified interface.
The wish to try out some of these implementations lead to a new implementation of the Wigner method using the \textit{C++} language.
The \textit{Blitz++} library \autocite{veldhuizen1998arrays} was continued to be used.

Most of the ODE solvers in \textit{boost.odeint} allow adaptive step size control that can be controlled by giving absolute and relative error tolerances similar to \textit{LSODA}, called \texttt{abserr} and \texttt{relerr} in this case.
A notable trick was the use of the g++ compiler switch \texttt{-ffast-math}, which disables strict IEEE 754 compliance in order to do computations faster.
The switch has the side-effect of changing \texttt{-fexcess-precision=} \texttt{standard} to \texttt{-fexcess-precision=fast}, which enables the computation to benefit from the excess precision of cpu registers in floating point calculations.
This enabled the code to integrate the equations of motion with settings for the absolute error tolerances as small as $10^{-14}$ with \texttt{double} precision arithmetic.

\section{The Semiclassical Method for Two-Component Electric Fields}\label{sec:semiclass}
Besides the Wigner method, pair production by two-component homogeneous electric fields can also be computed by a semiclassical method \autocite{Strobel:2014tha,Blinne:2015zpa}.
This is a generalization of similar methods for unidirectional electric fields \autocite{Brezin:1970xf,Popov:1971ff,Popov:1971iga,Popov:1973az,Marinov:1977gq,Popov:2001ak,Kleinert:2008sj,Kleinert:2012bu} which are based on a scattering ansatz.
The pair production rate for a constant rotating field can be calculated analytically \autocite{Strobel:2014tha}, but for other choices of the electric field numerical calculations might be required.
An example of a numerical application of this method to a specific field pulse is given in \Sec\ref{sec:numsemi}.

The leading semiclassical order (often referred to as exponential factor) of Sauter\hyp{}Schwinger pair production for constant rotating fields has been studied using the scattering method in \Ref\cite{Popov:1973uw,Strobel:2013vza} and using the worldline instanton method in \Ref\cite{Xie2012}.
This was extended to include the next order (often referred to as prefactor) in \Ref\cite{Strobel:2014tha} for the scattering method and in \Ref\cite{Ilderton:2015qda} in interpolating coordinates for the worldline instanton method.

Note that the scattering ansatz presented in the following is exact until the approximation in \Eqref{eq:APPROX} is performed.
Indeed, it is possible to construct a Riccati equation for the reflection coefficient and to solve it numerically as was done for one-component fields in \Ref\cite{Dumlu:2011rr,Dumlu:2009rr}.
However for one-component electric fields the Riccati approach has been shown to be equivalent to the QKT \autocite{Dumlu:2009rr}, which in turn is equivalent to the Wigner method \autocite{Hebenstreit:2010vz}.
While this is not necessarily true for the two-component case we still expect the Riccati approach to have a numerical behavior comparable to the one of the Wigner method.

\subsection{Solution of the Dirac Equation}
 We start from the Dirac equation
 \begin{align*}
   \left(\left[\ii  \partial_\mu-e A_\mu(x)\right]\gamma^\mu-m\right)\hat{\Psi}(\vv{x},t)=0
\end{align*}
 and decompose the Dirac field as
\begin{align*}
   \hat{\Psi}(x,t)=\int\frac{\dd[3]{\cm}}{(2\pi)^3}\e^{\ii\vv{\cm}\vv{x}}\sum_{s=\pm1}&\left(\psi_{\vv{\cm},s}(t)\hat{a}_{\vv{\cm},s}
 +\tilde{\psi}_{\vv{\cm},s}(t)\hat{b}^\dagger_{-\vv{\cm},s}\right),%
\end{align*}
where
\begin{align*}
\tilde{\psi}_{\vv{\cm},s}(t):=\mathcal{C}\psi_{\vv{\cm},s}(t)^*\,,
\quad
\mathcal{C}=\ii\gamma^2\gamma^0.
\end{align*}
The Dirac field satisfies the canonical equal-time anticommutation relations
\begin{align*}
  \left\{\hat{\Psi}_a(\vv{x},t),\hat{\Psi}^\dagger_b(\vv{y},t)\right\}=\delta_{ab}\cdot \delta^3(\vv{x}-\vv{y})\,,
\end{align*}
provided the mode operators obey the corresponding relations
\begin{align*}
 &\left\{\hat{a}_{\vv{\cm},s},\hat{a}_{\vv{k},r}^\dagger\right\}=(2\pi)^3 \delta^3(\vv{k}-\vv{\cm})\delta_{rs}\,,\\
 & \left\{\hat{b}_{\vv{\cm},s},\hat{b}_{\vv{k},r}^\dagger\right\}=(2\pi)^3 \delta^3(\vv{k}-\vv{\cm})\delta_{rs}\,,\\
 &\left\{\hat{a}_{\vv{\cm},s},\hat{b}_{\vv{k},r}^\dagger\right\}=0
\end{align*}
and the modes satisfy the Wronskian condition
\begin{align}
 \sum_{s=\pm1}\left(\psi_{\vv{\cm},s}(t)\psi_{\vv{\cm},s}(t)^\dagger+\tilde\psi_{\vv{\cm},s}(t)\tilde\psi_{\vv{\cm},s}(t)^\dagger\right)=\mathbbm{1}. \label{eq:flatWronskian}
\end{align}
For convenience we choose to work in the Weyl representation, i.e.
\begin{align*}
 \gamma^{j}=\begin{pmatrix}
             0 &\sigma^j\\
             -\sigma^j & 0
            \end{pmatrix},
&&
  \gamma^{0}=\begin{pmatrix}
             0 &\mathbbm{1}\\
             \mathbbm{1} & 0
            \end{pmatrix}\,,
\end{align*}
where \(\sigma^j\) are the Pauli matrices.

For two-component fields solely depending on time [\(A_\mu(x)=(0,\allowbreak A_x(t),\allowbreak A_y(t),\allowbreak 0)\)] one can make the ansatz
\begin{align}
 \psi_{\vv{\cm},s}(t)=C_s\begin{pmatrix}
                      s\, m\, \psi_1^s(t)\\
                       \,m \, \psi_2^s(t)\\
                      -s( \cm_z+s\epsilon_\perp)\, \psi_1^s(t)\\
                      ( \cm_z+s\epsilon_\perp)\, \psi_2^s(t)\\
                       \end{pmatrix}\label{eq:firstAnsatz}
\end{align}
for \(s=\pm1\), with a fixed normalization constant $C_s$.
Here the transverse energy
\begin{align*}
\epsilon_\perp^2\definedby \cm_z^2+m^2
\end{align*}
is introduced.
This ansatz for the solution of the Dirac equation can be derived from the ansatz that is used in \Ref\cite{Strobel:2014tha} to solve the squared Dirac equation.
Due to $(\cm_z+s\epsilon_\perp)(\cm_z-s\epsilon_\perp)=-m^2$ we observe that \(\psi_{\vv{\cm},s}(t)\) and \(\psi_{\vv{\cm},-s}(t)\) are independent, i.\,e.,
\begin{align*}
 \psi_{\vv{\cm},s}(t)^\dagger\cdot \psi_{\vv{\cm},-s}(t)=0\,.
\end{align*}
The solutions we will find below for \(s=\pm1\) thus represent two independent solutions.
Inserting this ansatz into the Dirac equation leads to the defining equations for the unknown mode functions $\psi_{1/2}^s$
\begin{align}
  \label{eqn:mode_equations}
  \begin{split}
 \i\, \dot{\psi}_1^s(t)+s\, \epsilon_\perp\psi_1^s(t)-s\, p_{x-y}(t)\psi_2^s(t)&=0\,,\\
 \i\, \dot{\psi}_2^s(t)+s\, \epsilon_\perp\psi_1^s(t)-s\, p_{x+y}(t)\psi_2^s(t)&=0\,,
  \end{split}
\end{align}
where we have defined
\begin{align*}
p_{x\pm y}(t)\definedby p_x(t)\pm\i p_y(t)\,.
\end{align*}
Please note that the vector potential from the minimal coupling in the Dirac equation is of course present in \Eqs\eqref{eqn:mode_equations} via $\vv{p}=\vv{q}-e\vv{A}(t)$.
The Wronskian condition in \Eqref{eq:flatWronskian} holds if
\begin{align}
\label{eq:PsiNorm}
\left|\psi_1^s(t)\right|^2+\left|\psi_2^s(t)\right|^2=1
\end{align}
and
\begin{align*}
 C_s=\frac{1}{\sqrt{2\epsilon_\perp(q_z+\epsilon_\perp)}}\,.
\end{align*}

The unknown mode functions $\psi_{1/2}^s$ are now replaced with a WKB ansatz
\begin{align}
 \psi^{s}_{1}(t)&=\frac{\sqrt{cp_{x-y}(t)}}{\sqrt{2\omega(t)}}\sqrt{cp_\parallel(t)}\left(\alpha_s(t)\frac{\e^{-\frac{\i}{2}K_s(t)}}{\sqrt{\omega(t)+s\epsilon_\perp}} +\i \beta_s(t)\frac{\e^{\frac{\i}{2}K_s(t)}}{\sqrt{\omega(t)-s\epsilon_\perp}}\right),\label{eq:ansatz1} \\
 \label{eq:ansatz2}
 \psi^{s}_{2}(t)&=s\frac{\sqrt{cp_{x+y}(t)}}{\sqrt{2\omega(t)}}\sqrt{cp_\parallel(t)}\left(\alpha_s(t)\frac{\e^{-\frac{\i}{2}K_s(t)}}{\sqrt{\omega(t)-s\epsilon_\perp}} -\i \beta_s(t)\frac{\e^{\frac{\i}{2}K_s(t)}}{\sqrt{\omega(t)+s\epsilon_\perp}}\right)\,,
\end{align}
introducing the Bogoliubov coefficients $\alpha_s(t)$ and $\beta_s(t)$.
See \Ref\cite{Strobel:2014tha} for a motivation.
The integrals are given by
\begin{align}
K_s(t)\definedby K_0(t)-s K_{xy}(t)\,,\label{eq:K_s}\\
K_0(t)\definedby 2\int_{-\infty}^{t} \omega(t') \dd{t'}, \label{eq:Kint}\\
K_{xy}(t)\definedby \epsilon_\perp\int_{-\infty}^{t} \frac{\dot{p}_x(t')p_y(t')-\dot{p}_y(t')p_x(t')}{\omega(t')p_\parallel(t')^2}\dd{t'} \label{eq:K_xy}
\end{align}
 with
\begin{align*}
 p_\parallel(t)^2\definedby p_x(t)^2+p_y(t)^2\,,
 \\
 \omega(t)^2 &= (\vv{\cm}-e\vv{A}(t))^2+m^2\,.
\end{align*}
From this ansatz, using \Eqref{eqn:mode_equations}, the evolution equations for the Bogoliubov coefficients
\begin{align}
 \dot{\alpha}_s(t)&=\frac{\dot{\omega}(t)}{2 \omega(t)}G^s_+(t) \e^{\i K_{s}(t)}\beta_s(t)\,,\label{eq:2CompAlpha}\\
 \dot{\beta}_s(t)&=\frac{\dot{\omega}(t)}{2 \omega(t)}G^s_-(t) \e^{-\i K_{s}(t)}\alpha_s(t)\label{eq:2CompBeta}
\end{align}
are found, where
\begin{align*}
 G^s_\pm(t)=\i s\frac{\epsilon_\perp}{ p_\parallel(t)}\pm \frac{\dot{p}_x(t)p_y(t)-\dot{p}_y(t)p_x(t)}{\dot{p}_x(t)p_x(t)+\dot{p}_y(t)p_y(t)}\frac{\omega(t)}{ p_\parallel(t)}\,.
\end{align*}
Using \Eqs\eqref{eq:ansatz1} and \eqref{eq:ansatz2} in the normalization condition \Eqref{eq:PsiNorm} we find
\begin{align*}
 \left|\alpha_s(t)\right|^2+\left|\beta_s(t)\right|^2=1\,.
\end{align*}

\subsection{Momentum Spectrum of Produced Pairs} %
The transmission probability
\begin{align}
 W^s(\vv{\cm}):=\lim_{t\rightarrow\infty} \left|\beta_s(t)\right|^2 \label{eq:trans}
\end{align}
can be interpreted as the number of produced electron positron pairs as a function of the canonical momentum \(\vv{\cm}\).
Using appropriate boundary conditions \autocite{Dumlu:2011rr}
\begin{align*}
 \beta_s(-\infty)=0, && \alpha_s(-\infty)=1\,,
\end{align*}
one can find a multiple-integral description for \(\dot{\beta}^\pm(t)\) by iteratively using \Eqs\eqref{eq:2CompAlpha} and \eqref{eq:2CompBeta} following the ideas introduced in \Ref\cite{Berry1982}. We now use the fact that the integrals are dominated by regions around the classical turning points
\begin{align}
  \omega(t_p^{\pm}):=0 \label{eq:turningpoints}\,.
\end{align}
According to \Eqref{eq:scE} the \(t_p^{\pm}\) are found in complex conjugate pairs.
By deforming the contour we extract the turning points for which
\begin{align}
 \Im[K_0(t_p)]<0\,. \label{eq:constraint}
\end{align}
If in the following \(t_p\) is used without the superscript \(\pm\)  it will always refer to the turning point of the pair \(t_p^\pm\) which fulfills \Eqref{eq:constraint}.
Assuming that the turning points represent poles of order \(\nu_{t_p}\), one finds \autocite{Berry1982,Strobel:2014tha}
\begin{align}
\frac{\dot{\omega}(t)}{\omega(t)}&\approx\frac{dK_0(t)}{dt}\frac{\nu_{t_p}}{\nu_{t_p}+2}\frac{1}{K_0(t)-K_0(t_p)} \label{eq:APPROX}\,.
\end{align}
One can now approximate the preexponential factor in each integrand in the multiple\hyp{}integral series by its behavior around the poles  \(t_p\) given by \Eqref{eq:APPROX} to find
\begin{align}
\beta_s(\infty)\approx-2\sum_{t_p}\e^{-\i K_s(t_p)} \sin\left(\frac{\pi\nu_{t_p}}{2(\nu_{t_p}+2)}\right).\label{eq:approx2}
\end{align}
This approximation is semiclassical in the sense that the exponential factor, which is not approximated, presents the leading semiclassical order.
The approximation in \Eqref{eq:approx2} breaks down if the turning points get too close to each other in the complex plane, such that \Eqref{eq:APPROX} is no longer valid, see Sec.~\ref{sec:cmp}.

Since the examples covered in the present work have simple turning points, i.e. \(\nu_{t_p}=1\), the semiclassical momentum spectrum of \Eqref{eq:trans} takes the form
\begin{align}
 W^s(\vv{\cm})=\left|\sum_{t_p}\e^{-\i K_s(t_p)}\right|^2 \label{eq:MomentumSpectrum}\,.
\end{align}
The total particle yield is then given by summing over the independent solutions
\begin{align}
  \label{eq:MomentumSpectrumTotal}
  W(\vv{\cm})=W^+(\vv{\cm}) + W^-(\vv{\cm})\,.
\end{align}

\section{The Rotating Sauter Pulse}\label{sec:simprot}
Now that two methods are available for pair production in homogeneous electric fields, let us define an electric field which we will study and use as an example to compare the performance and accuracy of the methods.
A spatially homogeneous, monochromatic rotating electric field under a $\cosh^2$\hyp{}envelope in the absence of a magnetic field will be referred to as a rotating Sauter pulse.
It is given by
\begin{align}
  \label{eqn:puls-sauter-rot}
  \vv{E}(t)=\frac{E_0}{\cosh^2\left(\nicefrac{t}{\tau}\right)} \begin{pmatrix}
                    \cos(\Omega t) \\
                    \sin(\Omega t) \\
                    0
                  \end{pmatrix}\,,
\end{align}
characterized by a maximum field strength $E_0$, an angular rotation
frequency $\Omega$ and a pulse duration $\tau$. This field
configuration can be viewed as a model for the field in an anti-node
of a standing wave mode with appropriate circular polarization.
This field has been at the center for a lot of the studies presented in this thesis.
Later also superpositions of those fields will be taken into account, see \Sec\ref{sec:Bichromatic}.

\begin{wrapfigure}{r}{\widthof{\includegraphics{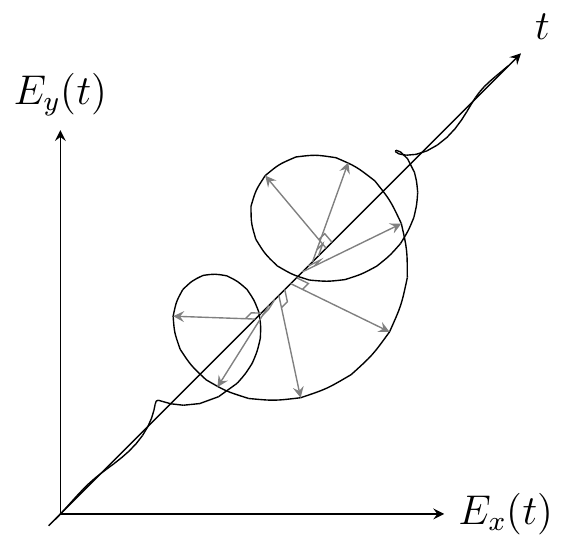}}+2mm}
  \centering
  \includegraphics{rotatingfield}
  \caption{Time dependency of the rotating Sauter pulse.}
  \label{fig:field}
\end{wrapfigure}
For the discussion, it is useful to introduce the dimensionless parameters
\begin{equation}
  \varepsilon= \frac{E_0}{E_\crit}\,, \quad \sigma=\Omega\tau\,,, \label{eq:dimless}
\end{equation}
where $\varepsilon$ measures the maximum field strength in units of the critical field strength $E_\crit=\frac{m^2c^3}{e\hbar}\approx1.3\cdot10^{18}\,\unitfrac{V}{m}$ and $\sigma$ is a measure for the number of full rotation cycles within the pulse duration.
The dimensionful
parameters will be given in units of the QED scale, i.\,e., the electron
mass $m$. For instance, the pulse duration is measured in units of
the Compton time $\tc=1/m$ in units where $\hbar=c=1$.

The time evolution of this field is illustrated in
Fig.~\ref{fig:field}. In the limit $\Omega=0$, the rotating field
collapses to a non-rotating Sauter\hyp{}type field, which is one of the few
examples where the Wigner function can be calculated analytically \autocite{Hebenstreit:2010vz}.
Note that a carrier envelope
phase $\phi$ with the replacement $\Omega t \to \Omega t+\phi$ would
have no effect, as it can be transformed to zero by a rotation of the
coordinate system in the $(x,y)$-plane.

\subsection{The Semiclassical Method for the Rotating Sauter Pulse}\label{sec:numsemi}
\begin{figure}
 \centering
 \includegraphics{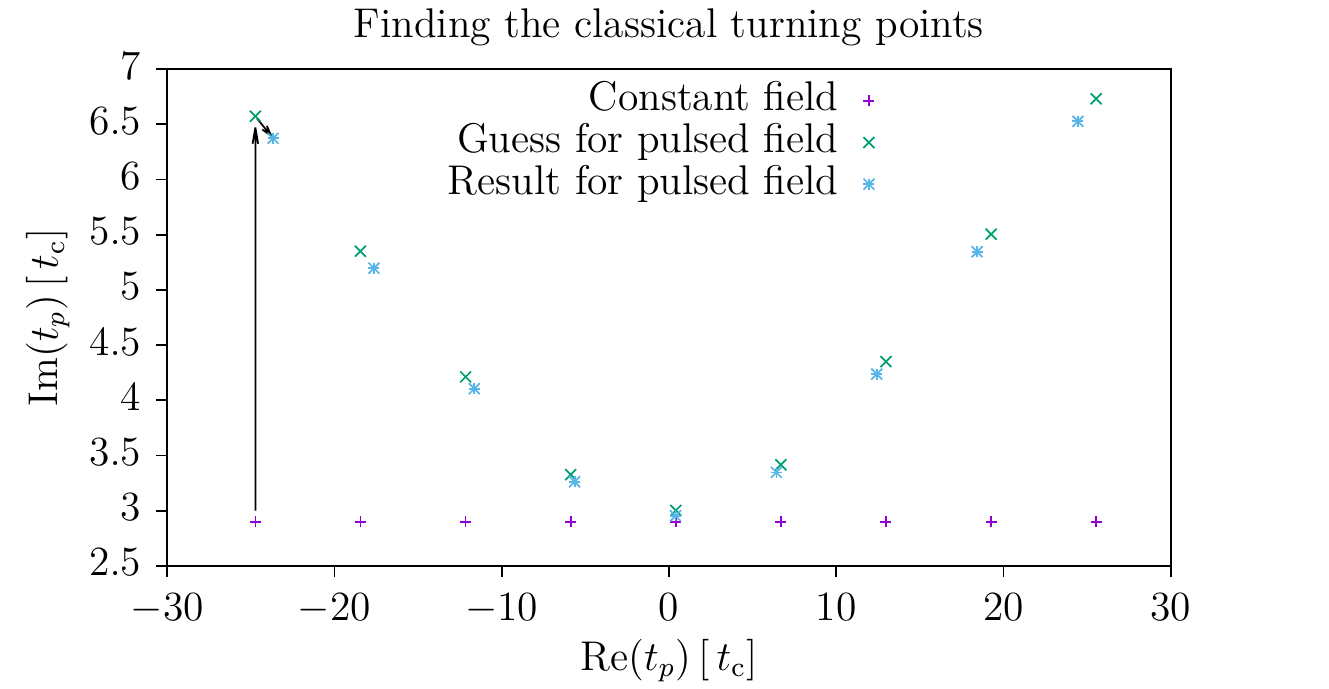}
 \caption[Positions of the turning points of the rotating Sauter pulse.]
 {
  This figure shows how the turning points for the rotating Sauter pulse are found.
  The turning points of the constant rotating field are shifted away from the real axis by replacing the field strength parameter by the pulse shape (upward arrow).
  Afterwards the correct turning points can be found by a numerical search (smaller arrow).%
 }
 \label{fig:tp_numeric}%
\end{figure}
The semiclassical method was discussed for a generic two-component electric field in \Sec\ref{sec:semiclass}.
The pair production rate for a constant rotating field can be calculated analytically \autocite{Strobel:2014tha}, but numerical calculations are required to solve for the rotating Sauter pulse.
In order to calculate the semiclassical pair production rates, first the classical turning points of the given potential need to be found.
This is done by numerically solving $\omega(t_p)=0$ for complex $t_p$ using a Newton-Raphson method \autocite{Atkinson1989}, which needs an initial guess that is in some sense close to the desired solution.
The known turning points for the constant rotating field discussed in appendix  \ref{app:constrot} and given in \Eqref{eq:rotatingcompltp} may be used as a starting point.
Unfortunately, these points are too far away from the desired solutions for a reliable numerical search.
If, however, the field strength parameter $\varepsilon$ in \Eqref{eq:rotatingcompltp} is replaced by the pulse shape $\frac{E(t_j)}{E_\crit}$
\begin{align*}
  \varepsilon\rightarrow \frac{E(t_j)}{E_\crit}=\frac{\varepsilon}{\cosh^2\left(\frac{\Re(t_j)}{\tau}\right)}\,, %
\end{align*}
the result is a sufficient guess for the starting point
(see \Fig\ref{fig:tp_numeric} for a depiction of this behavior).
In this way we also get a nomenclature for the turning points, by giving them the same name as the corresponding ones of the constant rotating field.

For the computation the momentum grid is divided into several parts for parallelization.
For each of these parts the number of used pairs of turning points is chosen adaptively.
To this end, the turning point \(t_0\) is considered first.
Afterwards, for increasing integer $j$, \(t_j\) and \(t_{-j}\) are added in pairs until their contribution to \(W_s(\vv{\cm})\) is less than \(0.1\%\).

The semiclassical method relies heavily on integrals in the complex plane.
These are expressed in terms of multiple real integrals by parameterization of the integration paths.
Afterwards the \textit{GNU Scientific Library} \autocite{Gough2009} is used to carry out the real\hyp{}valued integrals, specifically with the use of adaptive Gauss-Kronrod \autocite{Kronrod1965} and Clenshaw-Curtis \autocite{ClenshawCurtis} rules.
The adaptive algorithms are also tuned by an absolute and a relative error tolerance.
Still, it is necessary to evaluate the vector potential $\vv{A}$ with $\vv{E}=-\frac{\dd{}}{\dd t}\vv{A}$ for complex times.
The indefinite integral of the field given by \Eq\eqref{eqn:puls-sauter-rot} can be given as
\begin{align}
  \label{eqn:field-integral}
  \int \vv{E}(t)\,\dd{t}&= \frac{\varepsilon E_\crit\tau}{2} \cdot
   \begin{pmatrix}
    2\Re\left[  \e^{-\i t\Omega} H_1-\e^{\frac{t}{\tau}(2-\i\tau\Omega)}\frac{\tau\Omega}{2+\i\tau\Omega}H_2 \right]\\
      \qquad+2\cos(t\Omega)\tanh(\frac{t}{\tau})
      \\[10pt]
      2\Re\left[ \i\e^{-\i t\Omega} H_1+\e^{\frac{t}{\tau}(2-\i\tau\Omega)}\frac{\tau\Omega}{-2+\i\tau\Omega}H_2 \right]\\
      \qquad+2\sin(t\Omega)\tanh(\frac{t}{\tau})
      \\[10pt]
      0
   \end{pmatrix}
\end{align}
with
\[
\begin{array}{rrrl}
 H_1 = \tensor*[_2]{F}{_1}\big( 1,&  -\frac{\i}2 \tau\Omega,& 1-\frac{\i}{2}\tau\Omega,& -\e^{\frac{2t}{\tau}} \big), \\
 H_2 = \tensor*[_2]{F}{_1}\big( 1,& 1-\frac{\i}2 \tau\Omega,& 2-\frac{\i}{2}\tau\Omega,& -\e^{\frac{2t}{\tau}} \big),
\end{array}
\]
where $\tensor*[_2]{F}{_1}$ denotes the Gaussian hypergeometric function \autocite{gradshteyn2007}.
Due to the singularities at solutions of $\cosh^2\left(\nicefrac{t}{\tau}\right)=0$, the vector potential must have branch cut discontinuities.
These singularities are found on the imaginary axis at
\[
t_\mathrm{k} = \i\, \frac{2k+1}{2}\pi\tau\,.
\]
The form given in \Eqref{eqn:field-integral} has discontinuities on straight lines that start at the singularities on the imaginary axis and continue parallel to the real axis towards positive real infinity, while it is continuous everywhere else.
By exploiting the symmetries of the electric field $E_\mathrm{x/y}(t)\to\pm E_\mathrm{x/y}(-t)$, the continuous region with negative real part can be carefully mirrored towards the right-hand side of the imaginary axis leaving all the discontinuities strictly on the imaginary axis.
Finally the evaluation of the Gaussian hypergeometric function with complex arguments is left to a code named \textit{AEAE} which is described in \Ref\cite{Michel2008}.

For large values of $\sigma$ the arguments to the hypergeometric function become large and the \textit{AEAE} library fails to compute its values reliably.
In these cases different methods for evaluating the complex vector potential are available in the implementation.
One of these methods, mainly used for testing, computes each requested value of the complex vector potential from scratch, by performing a line integral over the complex electric field.
Another method precomputes the complex vector potential numerically on a grid in order to use spline interpolation.

\subsubsection{The Locally Constant Rotating Field Approximation} %
It is possible to approximate the momentum spectrum of the rotating Sauter pulse using the result for the rotating rectangular pulse field
\begin{align*}
 \vv{E}=\varepsilon E_\crit \Rect\left(\frac{t}{\tau}\right)\begin{pmatrix}
                    \cos(\Omega t) \\
                    \sin(\Omega t) \\
                    0
                  \end{pmatrix},
\end{align*}
with the rectangular box function
\begin{align*}
 \Rect(x)=\Theta(x)-\Theta(x-1)\,,
\end{align*}
where \(\Theta(x)\) is the Heaviside step function.
Fields of this form can be treated analytically as shown in \Ref\cite{Strobel:2014tha}, in a way similar to the constant rotating field
\begin{align}
  \label{eqn:rotating_const}
  \vv{E}(t) = \varepsilon E_\crit \begin{pmatrix}
    \cos(\Omega t) \\ \sin(\Omega t) \\ 0
                                  \end{pmatrix}\,.
\end{align}

The idea is to replace the field by a sum of rectangular pulses with pulse length \(\tau_0\) and a different constant field strength given by the form of the pulse \(E(t)\), i.e.~replace \(E(t)\) by
\begin{align*}
 E(t)\approx \sum_{j=0}^\infty \left[ E\left(\left(\frac12-j\right)\tau_0\right)\Rect
 \left(\frac{t}{\tau_0}+j\right) + E\left(\left(\frac12+j\right)\tau_0\right)\Rect
 \left(\frac{t}{\tau_0}-j\right)  \right].
\end{align*}
Now one can compute the momentum spectrum of the pair creation rate using the analytic result for the pair creation rate for each of these pulses.
The shorter the length \(\tau_0\) the better becomes this approximation which we call the locally constant rotating field approximation, or LCRFA.
Using that for the rectangle pulse the only turning points which contribute are those whose real part lies within the pulse range, it is possible to perform the limit \(\tau_0\rightarrow0\) which leads to
\begin{align}
  W^s_\text{approx}(\vv{\cm})=\left|\sum_{j=0}^\infty\left(  \e^{K_s\left(\vv{\cm},{E\left(\Re[t_j]\right)}\right)}+\e^{K_s\left(\vv{\cm},{E\left(\Re[t_{-j}]\right)}\right)} \right)\right|^2 .\label{eq:approx}
\end{align}
Here \(K_s(\vv{\cm},E)\) is the integral from \Eqref{eq:K_s} which is given by \Eqs\eqref{eq:Krot} and \eqref{eq:K_xyrot} and \(t_j\), represent the turning points given in \Eqref{eq:rotatingcompltp}.

\pagebreak
The LCRFA approximates the field by a constant rotating field at every time.
Therefore effects from the time variation caused by the shape of the pulse are neglected, while leaving the effects of the rotation intact.
Accordingly the approximation is reasonable for long enough pulses in which the time scale of the rotation \(1/\Omega\) is smaller than the time scale of the pulse \(\tau\), i.e.~\(\sigma:=\Omega\tau\gg1\).

\subsection{Phenomenology of the Rotating Sauter Pulse}
\label{sec:phenomenology}
As mentioned in the introduction, pair production in strong fields is provided by at least two quite distinct processes, the Sauter-Schwinger pair production in slowly varying, very strong fields and the multi-photon pair production in quickly varying fields.
This duality of regimes is known for example in atomic ionization in a sinusoidal electric field.
There the Keldysh adiabaticity parameter $\gamma$ is used to compare the characteristic frequency of the field $\Omega$ with the intrinsic tunneling frequency $\Omega_\mathrm{T}$ of the bound electron according to
\begin{align*}
  \gamma\definedby \frac{\Omega}{\Omega_\mathrm{T}}\,, & \Omega_\mathrm{T} &= \frac{eE_0}{\sqrt{2m E_\mathrm{b}}}\,,
\end{align*}
given the field strength $E_0$ and binding energy $E_\mathrm{b}$.
This has been adopted in strong-field QED as
\begin{align}
  \label{eq:keldysh_omega}
  \gamma_\Omega \definedby %
  \frac{\Omega}{\varepsilon m}\,,
\end{align}
where the dimensionless field\hyp{}strength parameter defined in \Eqref{eq:dimless} was used.
For the rotating Sauter pulse described here this parameter is not good enough, because it disregards another important parameter: the pulse duration $\tau$.
When analyzing the non-rotating Sauter pulse, the characteristic frequency is replaced by the inverse pulse duration yielding
\begin{align}
 \label{eq:keldysh_tau}
  \gamma_\tau \definedby %
  \frac{1}{\tau\varepsilon m}\,.
\end{align}
The question boils down to asking what frequency can be thought of as a characteristic frequency in a laser pulse, that has two independent time scales.

One possibility would be a combination of both scales of the form
\begin{align*}
  \tilde{\Omega} &= \sqrt{\left( a\frac{1}{\tau}  \right)^2 + \left( \Omega \right)^2}\,.
\end{align*}
A somewhat less general idea of a combined Keldysh parameter $\gamma^*$ has already been introduced in \Ref\cite{Blinne:2013via}, which used $a=1$.
When a much larger set of data was analyzed, it turned out that the parameter $a$ should be varied.
We propose a new definition with $a=\frac{\pi}{2}$
\begin{align}
 \label{eq:keldysh_combined}
  \gamma^* \definedby \frac{1}{\varepsilon m}\sqrt{\left( \frac{\pi}{2}\frac{1}{\tau}  \right)^2 + \Omega^2} \\
  \nonumber
  &= \frac{1}{\tau\varepsilon m}\sqrt{\left( \frac{\pi}{2} \right)^2 + \sigma^2}
\end{align}
which far better fits the numerical data, as can be seen in \Fig\ref{fig:total_yield_unified}.
Obviously the limits
\begin{align}
  \label{eqn:keldysh_combined_sigmatoinf}
  \gamma^*&\xrightarrow{\sigma\to\infty} \gamma_\Omega \\
  \nonumber
  \gamma^*&\xrightarrow{\makebox[\widthof{$\scriptstyle\sigma\to\infty$}]{$\scriptstyle\sigma\to0$}} \frac{\pi}{2}\gamma_\tau
\end{align}
hold.
The combined Keldysh parameter $\gamma^*$ can also be written in terms of the Keldysh parameters $\gamma_\Omega$ and $\gamma_\tau$ as
\begin{align}
  \label{eqn:keldysh_combined_intermsof}
  \gamma^* &= \sqrt{\left( \frac{\pi}{2}\gamma_\tau \right)^2 + \gamma_\Omega^2}\,.
\end{align}
While the combined parameter $\gamma^*$ might be well suited for a collective quantitative description, when used to separate parameter regimes its meaning is just
\begin{align}
  \label{eqn:gammastar_ll_1}
 \gamma^* \ll 1 &\Leftrightarrow \left( \gamma_\tau \ll 1 \right) \wedge \left( \gamma_\Omega \ll 1 \right)\\
 \label{eqn:gammastar_gg_1}
 \gamma^* \gg 1 &\Leftrightarrow \left( \gamma_\tau \gg 1 \right) \vee \left( \gamma_\Omega \gg 1 \right)\,.
\end{align}

In different parameter regimes, different expectations arise for the resulting spectra from both Schwinger and multiphoton pair production.
In order to be able to discuss the results in the later section, let us first talk about these processes.

\subsubsection{Multiphoton Pair Production and Effective Mass}
Multiphoton pair production rests on the idea that the electric field can be described in terms of photons which correspond to Fourier modes.
If the field contains photons with high enough frequencies, some of them can combine to produce an electron-positron pair.
For simplicity polarization is neglected and a complex plane wave
\begin{align*}
  E(t)=\varepsilon\,E_\crit\, \frac{\e^{-\i\Omega t}}{\cosh\left( \frac{t}{\tau} \right)^2}
\end{align*}
under a $\cosh^2$ envelope is considered.
Its Fourier transform is given by
\begin{align*}
  \tilde{E}(\omega)\definedby \int_{-\infty}^{\infty}\dd t \e^{\i\omega t}E(t)
  \\
  &=\varepsilon E_\crit \pi\tau^2\left( \omega-\Omega \right)\frac{1}{\sinh\left( \frac{\pi}{2}\tau\left( \omega-\Omega \right) \right)}
  \\
  &=2\varepsilon E_\crit \tau \,h\left( \tau\left( \Omega-\omega \right)  \right)
\end{align*}
with a symmetric, peaked function
\begin{align}
  \label{eqn:function_h}
  h(x):=\frac{\tfrac{\pi}{2} x}{\sinh\left( \tfrac{\pi}{2} x \right)}=h(-x)\,,\quad h(0)=1\,.
\end{align}
It is peaked at the pulse frequency $\Omega$ and has a maximum value
\begin{align*}
  \tilde{E}_\mathrm{max.} &= \tilde{E}(\Omega) = 2 \varepsilon E_\crit\tau\,.
\end{align*}
The static component, compared to the peak spectral component, is given by
\begin{align*}
  \frac{\tilde{E}(0)}{\tilde{E}_\mathrm{max.}} = h\left( \sigma \right)\,,
\end{align*}
which is exponentially suppressed for large $\sigma$, because
\begin{align*}
  h\left( \sigma \right)\xrightarrow{\sigma\to\infty} \frac{\pi}{2} \sigma\,\e^{-\frac{\pi}{2} \sigma}\,.
\end{align*}
That means that for large enough $\sigma$, the spectral density peak at $\omega=\Omega$ dominates the spectrum and we can expect clear multiphoton pair production.
This does not mean that multiphoton effects should not be expected for smaller $\sigma$ at all, it only means that they will only become pure and clearly separated from Schwinger pair production effects for higher $\sigma$.

What does multiphoton pair production mean?
The idea is that $n$ photons interact to create an electron-positron pair.
The electron and positron need to be on-shell, that means
  $\vv{p}^2 = \omega^2-m^2$
with the total energy $\omega$.
This total energy comes from $n$ photons of frequency $\Omega$, but only half of the total energy goes into either the positron or the electron.
The total momentum for any of the particles is thus given by
\begin{align}
  \nonumber
  \abs{\vv{p}} &= \sqrt{\left( \frac{n\Omega}{2} \right)^2-m^2} \\
  \label{eqn:multiphoton_circle_baremass}
               &= \frac12\sqrt{\left( n\Omega \right)^2-\left( 2m \right)^2}\,.
\end{align}

Charged particles in oscillating fields do behave as if their mass is slightly higher due to the kinetic energy of their periodic movement (ponderomotive potential).
This has been shown to be the case also for pair production in pulsed, oscillating fields \autocite{Kohlfurst:2013ura}.
For linearly polarized fields a good approximation for the effective mass is given by
\begin{equation*}
  m_*=m\sqrt{1+\frac{e^2}{m^2}\frac{\varepsilon^2}{2\Omega^2}}\,.
\end{equation*}
For the rotating field, which has the same $x$-component and an additional $y$\hyp{}component, the analogue result is
\begin{equation*}
  m_*^\text{rot.}=m\sqrt{1+\frac{e^2}{m^2}\frac{\varepsilon^2}{\Omega^2}}\,.
\end{equation*}

Exchanging the mass in \Eqref{eqn:multiphoton_circle_baremass} for the effective mass we expect pairs to be created by multiphoton pair production to have momenta of
\begin{align}
  \label{eqn:multiphoton_circle_effmass}
 \abs{\vv{p}} &= \frac12\sqrt{\left( n\Omega \right)^2-\left( 2m^* \right)^2}
\end{align}
for $n>n_\mathrm{min.}$, where $n_\mathrm{min.}$ is given by the requirement of a positive radicand
\begin{align}
  \label{eqn:multiphoton_nmin}
  n_\mathrm{min.} = \left\lceil\frac{2m^*}{\Omega}\right\rceil\,,
\end{align}
where $\lceil x\rceil$ is the smallest positive integer $k$ with $k\geq x$.

For short pulses the spectrum of the pulse has a high bandwidth which offers the possibility of having multiphoton pair production with different momenta, because then not only photons with the characteristic frequency are in the pulse.
The components of the spectrum where $2m^*=k\omega$ for some integer $k$ will produce pairs at rest.
Depending on the interplay between the time scales of multiphoton pair production and the pulse, the produced pairs are subject to acceleration by the electric field and end up with $\vv{p}=\vv{A}(0)$.

\subsubsection{Schwinger Pair Production}
If Schwinger pair production is dominant, there should be a pair production peak when the field strength $\lvert\vv{E}\rvert$ has a (local) maximum.
Those particles would have a kinetic momentum in the final distribution which is dictated by the vector potential at their creation time.

The rotating Sauter pulse will always have the global maximum of the field strength at $t=0$.
The vector potential at $t=0$ can be analytically calculated in this case, resulting in
\begin{align}
  \nonumber
  \vv{A}(0)&=\int_0^\infty \vv{E}(t)\,\dd t   \\
  \nonumber
          &=\int_0^\infty\frac{E_\crit\,\varepsilon}{\cosh^2(\nicefrac{t}{\tau})}\begin{pmatrix}
                                                                                            \cos(\Omega t) \\ \sin(\Omega t)
                                                                                          \end{pmatrix} \,\dd t
                                                                                        \\
  \label{eqn:vector_potential_t0}
                                                                                        &= E_\crit\,\varepsilon\,\tau \begin{pmatrix}
                                                                                          \nicefrac{\tfrac{\pi}{2} \sigma}{\sinh\left( \frac{\pi}{2} \sigma\right)}
                                                                                          \\
                                                                                          \frac{\sigma}{4}\left(  H_{\i\frac{\sigma}{4}}  - H_{-\frac12+\i\frac{\sigma}{4}} + \mathrm{c.c.}\right)
                                                                                           \end{pmatrix}\,,
\end{align}
with $H_n$ being the analytic continuation of the Harmonic Numbers $H_n=\sum_{k=1}^n \frac{1}{k}=\int_0^1\dd x\frac{1-x^n}{1-x}$.

While the result for $A_x$ is quite simple and can be written with the previously discussed function $h(\sigma)$ (see \Eqref{eqn:function_h}) by
\begin{align*}
 A_x(0)&=E_\crit\,\varepsilon\,\tau  h(\sigma)\,,
\end{align*}
the result for $A_y$ is not as straightforward to understand.
We can separate the dependencies from the parameters $\tau$ and $\sigma$ by defining a function $g(\sigma)$ according to
\begin{align*}
  A_{y}(0)&=E_\crit\,\varepsilon\,\tau \,g(\sigma) \\
  g(\sigma) \definedby \frac{\sigma}{4}\left(  H_{\i\frac{\sigma}{4}}  - H_{-\frac12+\i\frac{\sigma}{4}} + \mathrm{c.c.}\right)
\end{align*}
and discussing its properties.
\begin{figure}
 \centering
 \includegraphics{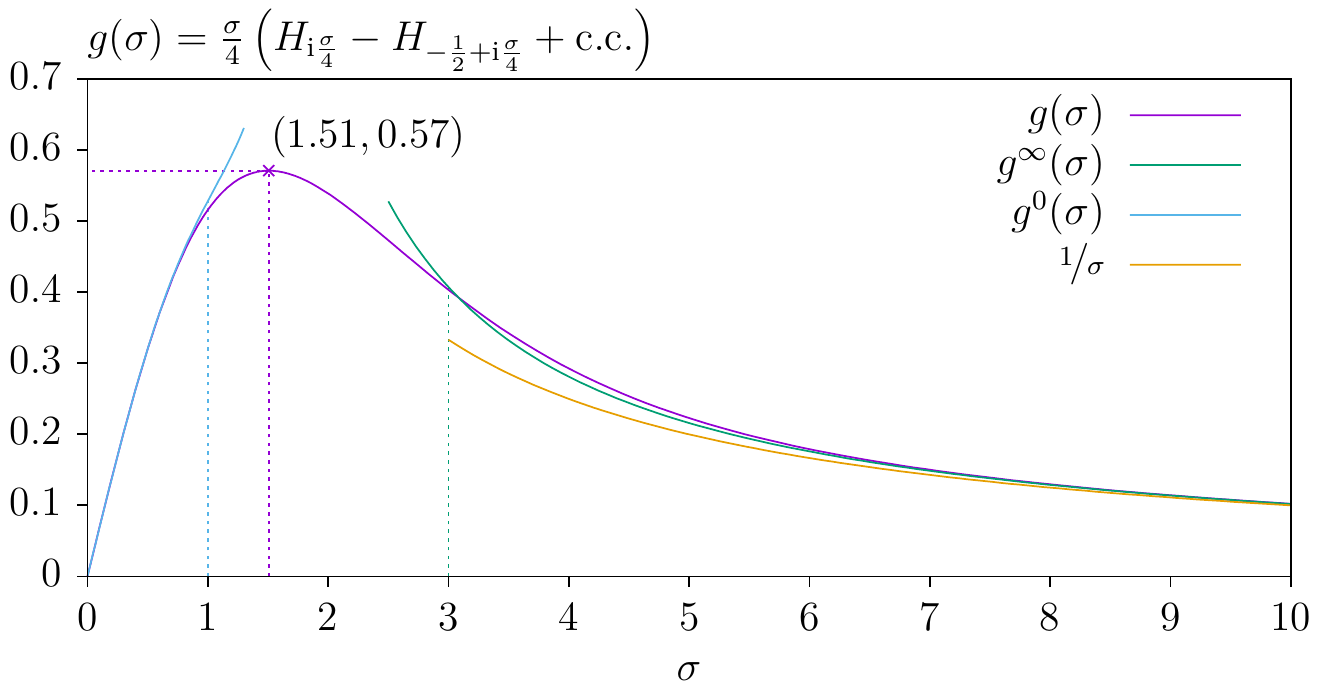}
 \caption[Asymptotics of an auxiliary function used to calculate the vector potential of the rotating Sauter pulse.]{
   This figure shows the asymptotics of the function $g(\sigma)$ which is used to calculate the $y$\hyp{}component of the vector potential at $t=0$.
 }
 \label{fig:vector_potential_asymp}
\end{figure}
\begin{figure}
 \centering
 \includegraphics{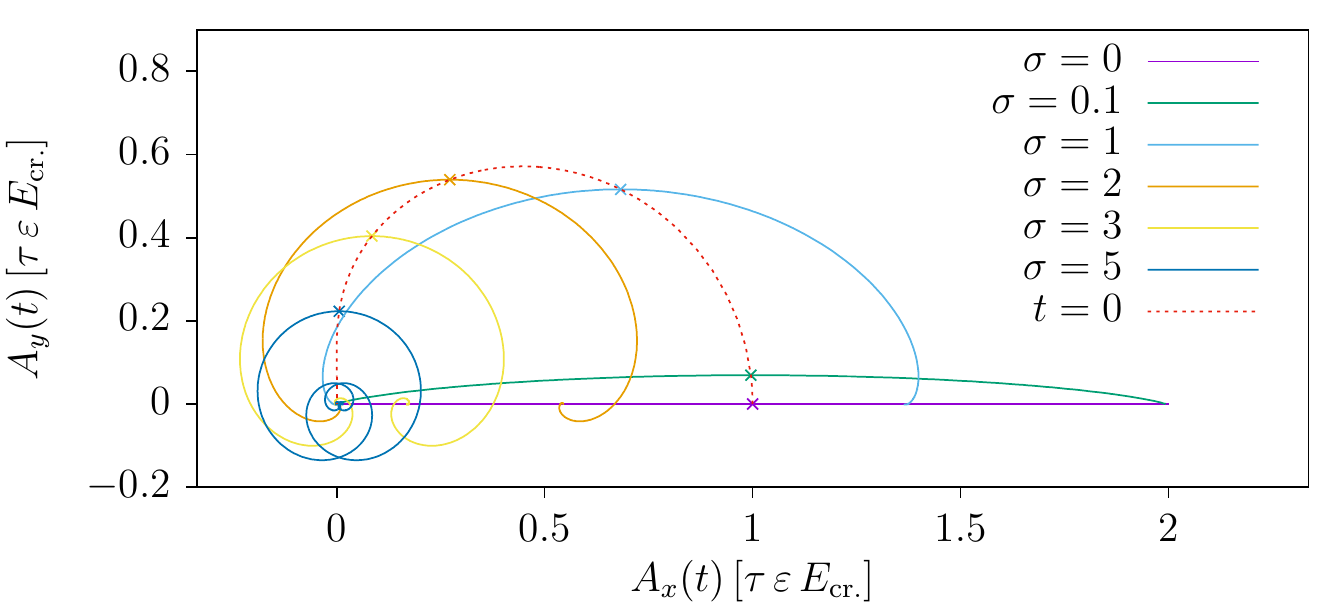}\\[4mm]%
 \includegraphics{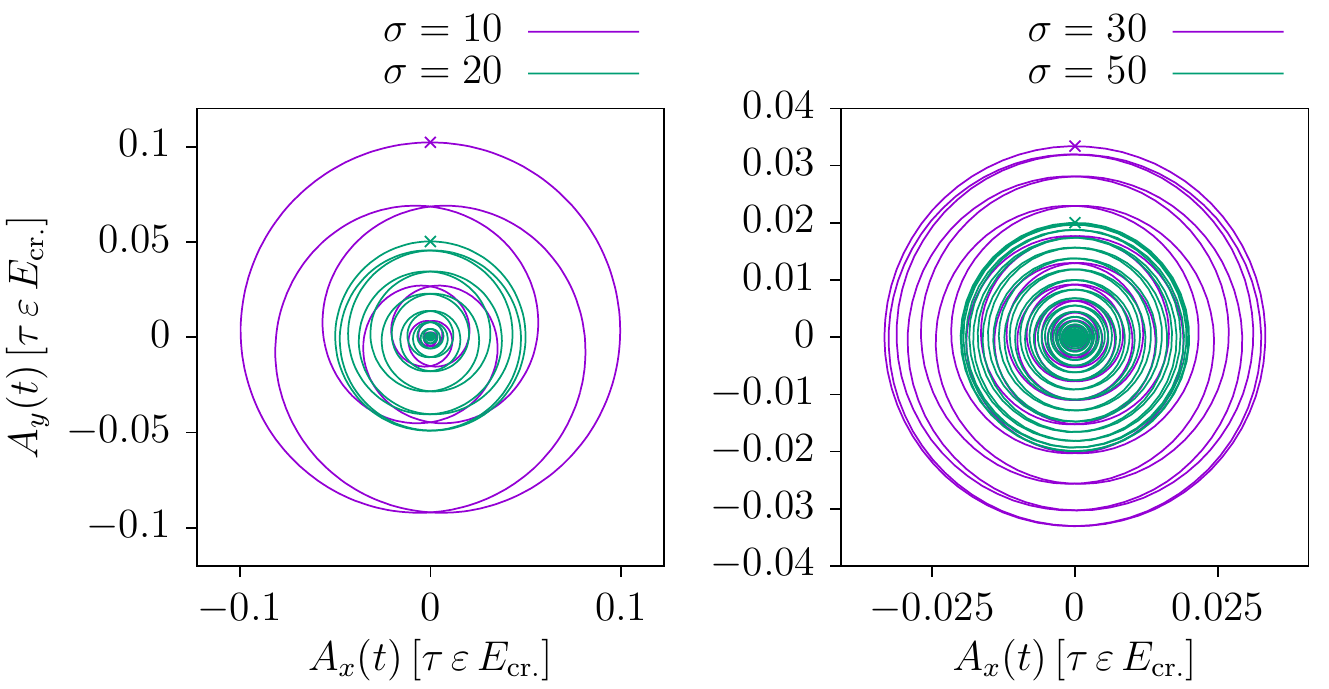}
 \caption[Loci of the vector potential of the rotating Sauter pulse]{
   Loci of the vector potential in units of $\tau\,\varepsilon\,E_\crit$ for different values of $\sigma$.
   The curves for higher values of $\sigma$ are shown in separate plots, note the different axes ranges.
   Crosses mark the vector potential at $t=0$ where maximal field strength occurs.
   The dashed line marks this point for all $\sigma$.
 }
 \label{fig:vector_potential_loci}
\end{figure}
The function $g(\sigma)$ has a simple series expansion at $\sigma\to\infty$ given by
\begin{align*}
  g(\sigma)&=\sum_{n=1}^\infty \frac{(-4)^n\,(4^n - 1)\,\zeta(1 - 2n)}{\sigma^{2n-1}}
  \intertext{with the Riemann $\zeta$ function. The first few terms are}
  g(\sigma)&=\frac{1}{\sigma}+\frac{2}{\sigma^3}+\frac{16}{\sigma^5}+\frac{272}{\sigma^7}+\mathcal{O}\left( \frac{1}{\sigma^9} \right)\,.
\end{align*}
At $\sigma=0$ another series expansion is given by
\begin{align*}
  g(\sigma)&= \ln(2) \,\sigma + \sum_{n=1}^\infty  \frac{(-1)^n(4^n - 1) \zeta(2 n + 1) }{16^n}\sigma^{2 n + 1}\,.
\end{align*}

A useful approximation of $g(\sigma)$ for $\sigma>3$ or $\sigma<1$ is given by taking these expansions to the orders $\sigma^{-3}$ and $\sigma^{5}$, respectively
\begin{align*}
 g^\infty(\sigma)\definedby\frac{1}{\sigma}+\frac{2}{\sigma^3} \\
 g^0(\sigma)\definedby \ln(2) \,\sigma -\frac{3}{16} \zeta(3) \sigma^3 + \frac{15}{256} \zeta(5) \sigma^5 \\
 &\approx 0.693 \sigma- 0.225 \sigma^3 + 0.0608 \sigma^5\,,
\end{align*}
which is displayed in \Fig\ref{fig:vector_potential_asymp}.
The maximum of $g(\sigma)$ is located at approximately $\sigma_\mathrm{max.}\approx1.506$ with $g(\sigma_\mathrm{max.})\approx0.571295$.
\pagebreak

Some interesting relations to the Keldysh adiabaticity parameters can be found by taking the limits $\sigma\to0$ and $\sigma\to\infty$ for the result given in \Eqref{eqn:vector_potential_t0}.
For the $y$\hyp{}component the result in the leading order in $\sigma\to\infty$ is
\begin{align}
  \label{eqn:vector_potential_t0_y_sigmatoinf}
  A_y(0)\xrightarrow{\sigma\to\infty} E_\crit\,\varepsilon\,\tau\,\frac{1}{\Omega\tau}=E_\crit\,\frac{\varepsilon}{\Omega} =E_\crit\,\frac{\hbar}{mc^2}\frac{1}{\gamma_\Omega}=E_\crit\,\tc\,\frac{1}{\gamma_\Omega}
\end{align}
in terms of the frequency based Keldysh parameter $\gamma_\Omega$.

In the limit of $\sigma\to0$, taking only the linear term into account, the result is
\begin{align*}
 A_y(0)\xrightarrow{\sigma\to0}  \ln(2)\,E_\crit\,\varepsilon\,\tau \,\sigma=\ln(2)\,E_\crit\,\tc\,\frac{\sigma}{\gamma_\tau}
\end{align*}
with the time scale based Keldysh parameter $\gamma_\tau$.
The limits of the $x$\hyp{}component are given by
\begin{align}
  \label{eqn:vector_potential_t0_x_sigmatoinf}
  A_x(0)&\xrightarrow{\sigma\to\infty} E_\crit\,\varepsilon\,\tau\,\pi\,\sigma\,\e^{-\frac{\pi\,\sigma}{2}} \to 0\\
  \nonumber
  A_x(0)&\xrightarrow{\makebox[\widthof{$\scriptstyle\sigma\to\infty$}]{$\scriptstyle\sigma\to0$}} E_\crit\,\varepsilon\,\tau\left( 1-\frac{\pi^2\sigma^2}{24} \right)
  =E_\crit\,\tc\,\frac{1}{\gamma_\tau}\,.
\end{align}
The complete result for the vector potential at $t=0$ is visualized as the dashed line in \Fig\ref{fig:vector_potential_loci}.
For $\sigma\to0$ the vector potential at $t=0$ is dominated by $A_x$, for large $\sigma$ the $A_y$\hyp{}component is dominating.

\subsubsection{Consequences for Pair Production Spectra}
\label{sec:expectations}
As is evident from QKT and the DHW formalism, the pair production spectrum in unidirectional electric fields have cylindrical symmetry with respect to the direction of the field.
In the language of the rotating Sauter pulse that means that for $\sigma=0$, where the field is in fact a unidirectional oscillating field as investigated in \Ref\cite{Kohlfurst:2013ura}, the pair production probability depends on $p_\shortparallel=p_x$ and $p_\perp=\sqrt{p_y^2+p_z^2}$.
In the rotating field for large $\sigma$, another symmetry arises.
In the case of a constant rotating field (see \Eqref{eqn:rotating_const}), the pair production probability depends on $p_\shortparallel=\sqrt{p_x^2+p_y^2}$ and $p_\perp=p_z$.
This is because a translation of the constant rotating field in time can be canceled by a rotation around the $p_z$\hyp{}axis.
As the pair production probability should not depend on time it must also be the same for every angle of rotation around the $p_z$\hyp{}axis.
As will be shown through a number of examples in \Sec\ref{sec:typical_spectra}, this is true also in the pulsed case for $\sigma\gtrsim20$, independent of which pair production process is dominant.
In both of these cases the subscripts $\shortparallel$ and $\perp$ refer to the momentum direction being parallel or perpendicular to the electric field, however with linear fields the $p_y$ and $p_z$ direction are both perpendicular to the field while in rotating fields only the $p_z$ direction is perpendicular to the field.

\pagebreak
The total particle yield as defined in \Eqref{eqn:numberofparticles} can be calculated by exploiting this symmetry according to
\begin{align}
 \nonumber
 \mathcal{N} &=\iiint \frac{\dd[3]{\vv{p}}}{\left( 2\pi \right)^3} f(\vv{p})\\
 \nonumber
 &=\frac{1}{\left( 2\pi \right)^3}\int_{-\infty}^{\infty} \dd{p_2} \int_0^{\infty} \dd{p_1} \int_0^{2\pi} \dd{\varphi} p_1\,f(p_1,p_2) \\
 \label{eqn:numberofparticles_isotropic}
 &=\frac{1}{\left( 2\pi \right)^2}\int_{-\infty}^{\infty} \dd{p_2} \int_0^{\infty} \dd{p_1}  p_1\,f(p_1,p_2)\
\end{align}
with $p_1=p_\perp=\sqrt{p_y+p_z}$ and $p_2=p_\shortparallel=p_x$ for $\sigma=0$ (linear fields) or with $p_1=p_\shortparallel=\sqrt{p_x+p_y}$ and $p_2=p_\perp=p_z$ for rotating pulses with $\sigma\gtrsim20$.
For small, non\hyp{}vanishing values of $\sigma$ the full three\hyp{}dimensional integral has to be calculated.

If multiphoton pair production is dominant, we expect, within the $p_x$-$p_y$ plane, rings with radii given by \Eqref{eqn:multiphoton_circle_effmass} with $n=n_\mathrm{min.}, n_\mathrm{min.}+1, n_\mathrm{min.}+2,\dotsc$ with $n_\mathrm{min.}$ given by \Eqref{eqn:multiphoton_nmin}.
These rings are present with linear oscillating fields, but have gaps that are the result of interference.
When the $p_z$ axis is taken into account, the rings should be completed to spheres, that may or may not have gaps.

If Schwinger pair production is dominant, the produced pairs should lie in the vicinity of the locus of the vector potential, with a peak at the point $\vv{p}=\vv{A}(0)$.
Due to $E_z=0$ for all times we also have $A_z=0$ for all times, hence the vector potential is always in the $p_z=0$ plane.
Hence pairs should be produced only with small momenta in $z$ direction, $|p_z|<1\,m$, and the pair production probability should decay quickly when moving away from $p_z=0$.

If the characteristic frequency scale is large enough, $\gamma^*\gg1$ (see \Eqref{eqn:gammastar_gg_1}), photons with large enough energies should exist to allow for multiphoton pair production.
If the vector potential at $t=0$ is large (e.\,g. greater than $E_\crit\,\tc$), this implies that the field is strong and slowly varying, which would favor Schwinger pair production.
For large sigma, according to \Eqref{eqn:vector_potential_t0_y_sigmatoinf}, this corresponds to $\gamma_\Omega\ll1$, and implies $\gamma_\tau\ll1$.
For small sigma, according to \Eqref{eqn:vector_potential_t0_x_sigmatoinf}, this corresponds to $\gamma^*=\gamma_\tau\ll1$, and in turn implies $\gamma_\Omega\ll1$.
In both of these cases, according to \Eqref{eqn:gammastar_ll_1}, the conditions correspond exactly to the inverse of the previously mentioned condition for multiphoton pair production.
In conclusion one can say that $\gamma^*\ll1$ and $\abs{\vv{A}(0)}\gg1$ are phenomenologically equivalent, as are $\gamma^*\gg1$ and $\abs{\vv{A}(0)}\ll1$.

\subsection{Comparison of the Wigner and Semiclassical Methods}\label{sec:cmp}  %
\begin{figure}[p]
 \centering
 \includegraphics{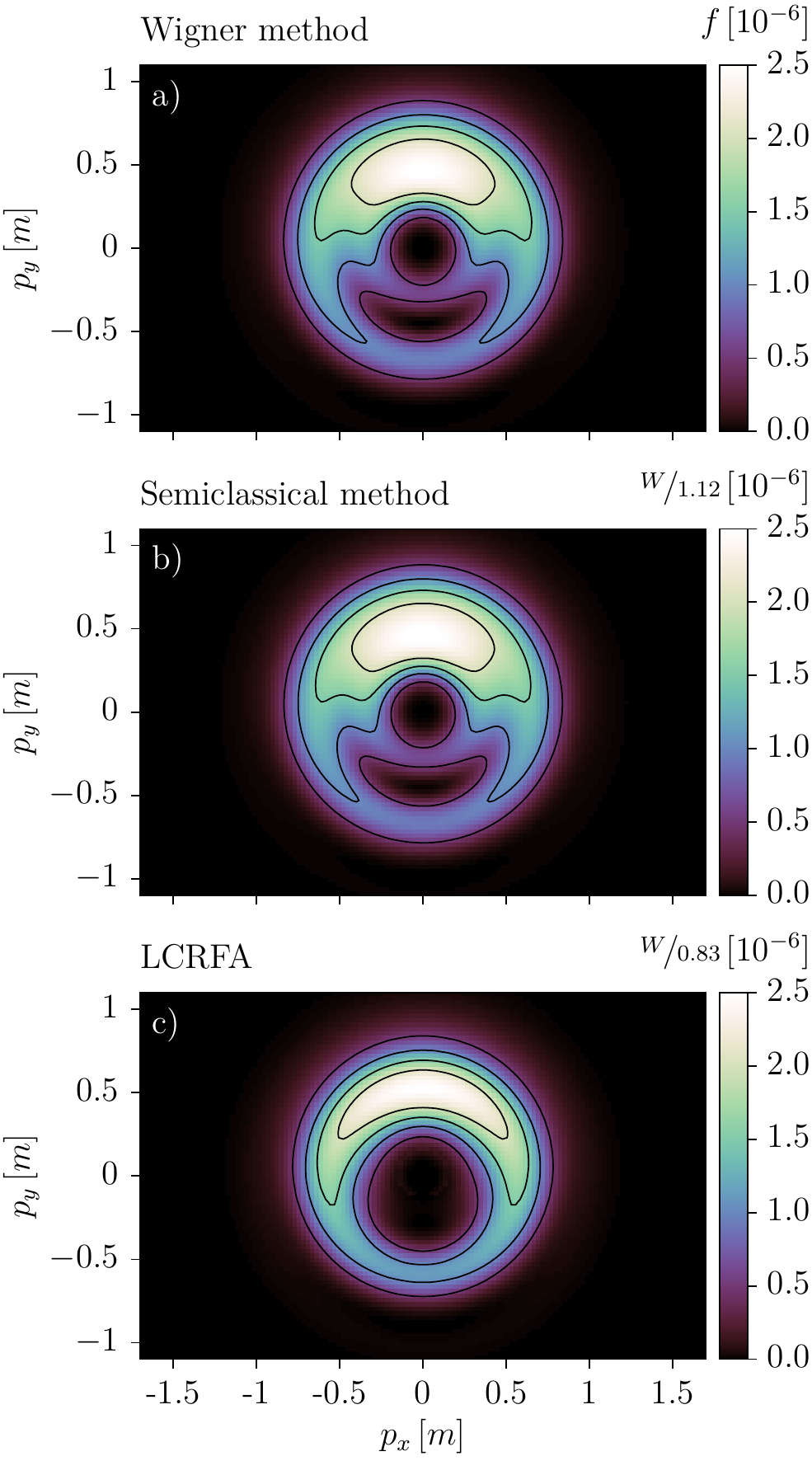}
 \caption[Momentum spectrum of a specific rotating Sauter pulse as given by three different methods.]
 {
   Momentum spectrum of the Sauter pulse for \(\tau=10/m\), \(\sigma=6\) and \(\epsilon=0.1\).
   The levels of the contour lines are indicated by the marks of the color box.
   Panel a): The Wigner result.
   Panel b): The semiclassical result divided by 1.12.
   Panel c): The result of the LCRFA divided by 0.83.
 }
\label{fig:spectra_methods}
\end{figure}
As we have three different methods for calculating pair production spectra at our disposal (the Wigner method, the full semiclassical method and the LCRFA), we should start with a comparison of these methods.
We do so for the example of the rotating Sauter pulse in \Eqref{eqn:puls-sauter-rot}.
The limit to the non-rotating pulse \(\Omega\rightarrow0\) can be treated analytically with both the Wigner method and the scattering approach (see Appendix~\ref{app:Sauter} for more details).

As for the constant rotating field discussed in \Ref\cite{Strobel:2014tha} we find that there is an infinite number of turning points for the rotating Sauter pulse.
But in contrast to the constant field case the turning points in the general case have different real and imaginary parts (see \Fig\ref{fig:tp_numeric} for a plot of the turning points).
This would in principle require a separate treatment of all of them.
However, the closer a pair of turning points is to the real axis, the bigger is its influence on the pair creation rate \autocite{Dumlu:2011rr}, such that it is sufficient to study a finite number of turning points in order to have a good approximation for the pair creation rate (see Appendix~\ref{sec:numsemi} for details).
Note that this holds true also within the LCRFA where it is sufficient to evaluate the sum in \Eqref{eq:approx} up to a finite \(|j|\).

\begin{figure}
 \centering
 \includegraphics{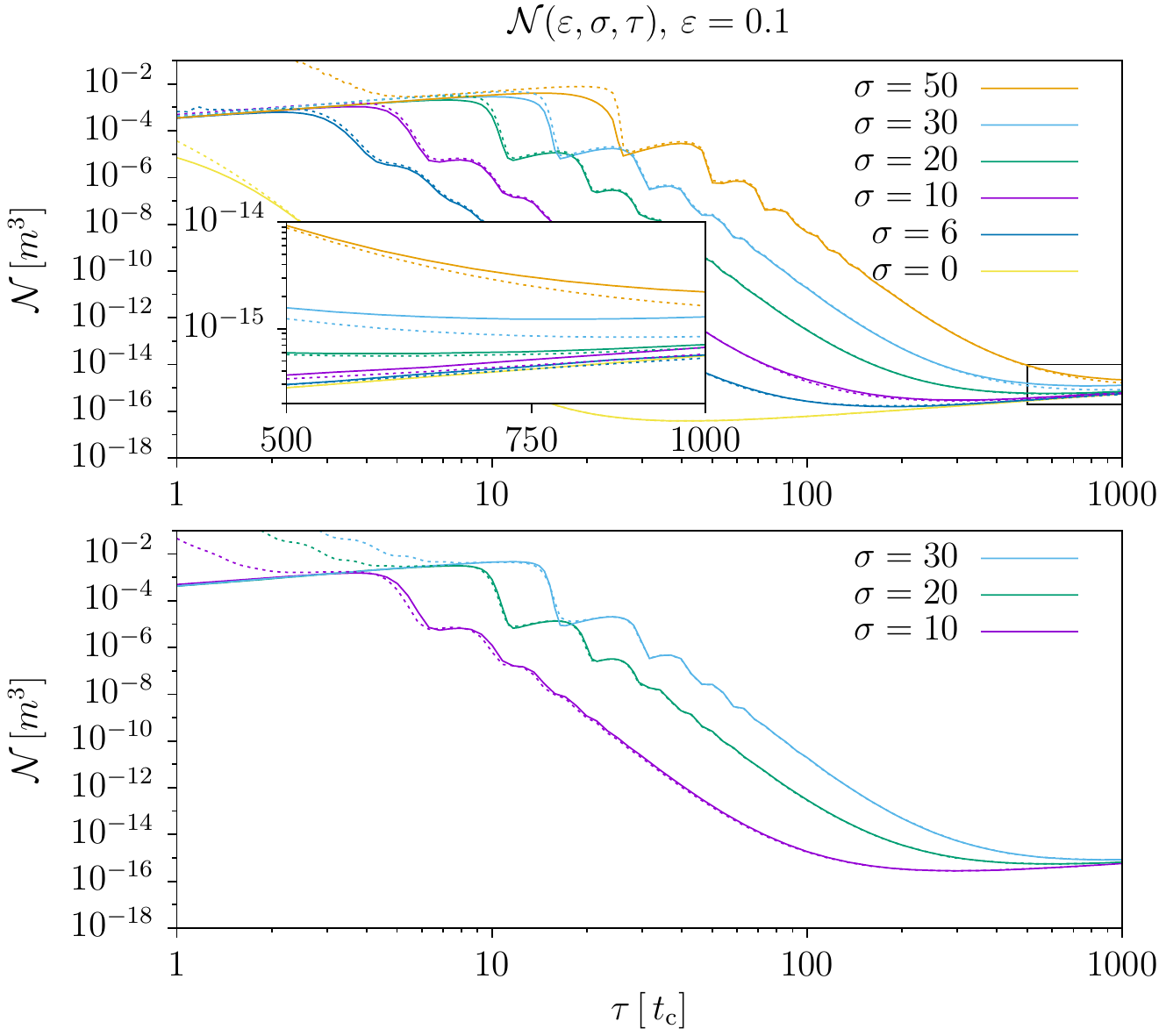}
 \caption[Comparison of the total particle number per Compton volume of rotating Sauter pulses as a function of the pulse length $\tau$.]
 {
   Comparison of the total particle number per Compton volume of the rotating Sauter pulse for $\varepsilon=0.1$ as a function of the pulse length $\tau$.
   \\
   Top panel:
   Solid lines show particle yield as calculated using the Wigner method, dashed lines show particle yield as calculated using the semiclassical method.
   In the cases $\sigma\in\{6,10\}$ a noise suppression method has been used when integrating over the spectra of the Wigner method to obtain the three dimensional totals.
   \\
   Bottom panel:
   Solid lines show the particle yield as calculated using the numerical semiclassical method, dashed lines show particle yield as calculated using the LCRFA.
   One finds that for long enough pulses the approximation agrees with the numerical results.
 }
 \label{fig:total_yield}
\end{figure}
\begin{figure}
 \centering
 \includegraphics{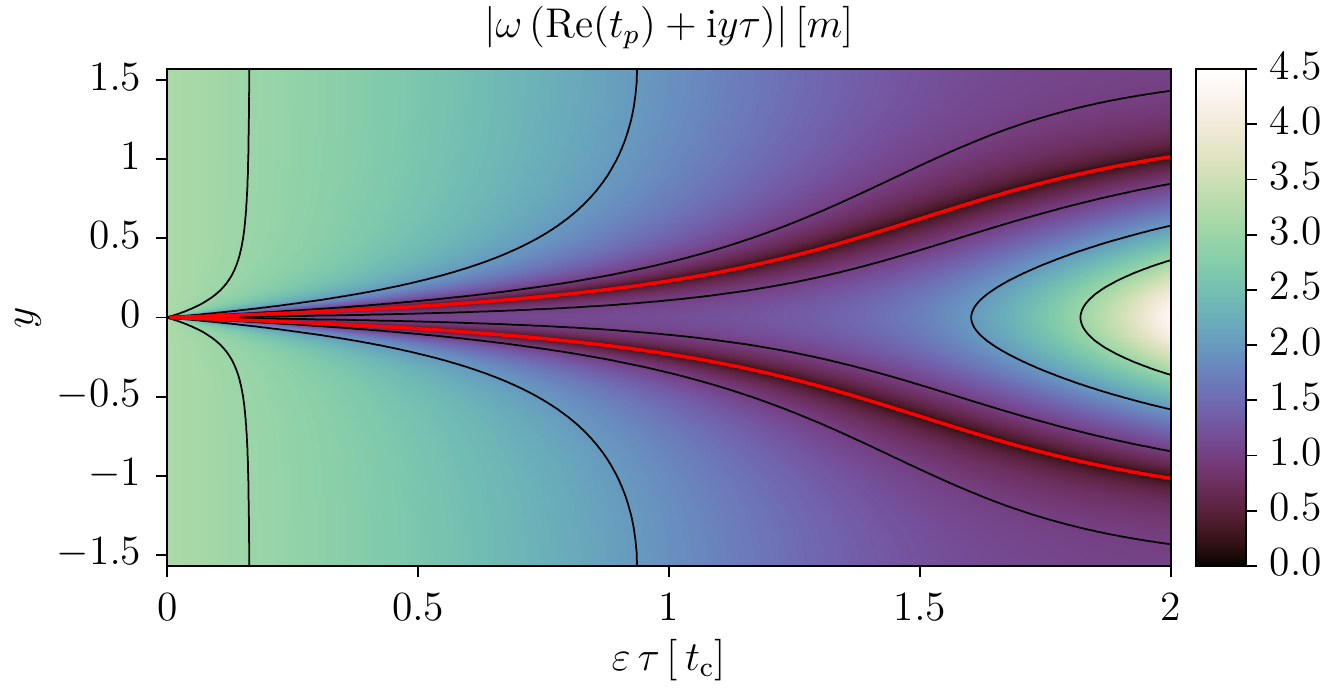}
 \caption[Energy of particles with specific kinetic momentum in the constant rotating field with different parameters at complex times.]
 {
   Value of \(|\omega(t)|\) for \(t=\Re[t_p]+\ii y \tau\) depending on  \(\varepsilon\,\tau\) for \(\cm_x=3m,\,\cm_y=\cm_z=0\). We see that for small \(\varepsilon\,\tau\) the turning points (red line) get closer and the assumption that \Eqref{eq:APPROX} holds for every turning point is not satisfied anymore.
 }
 \label{fig:TPSauter}
\end{figure}
\begin{figure}
 \centering
 \includegraphics{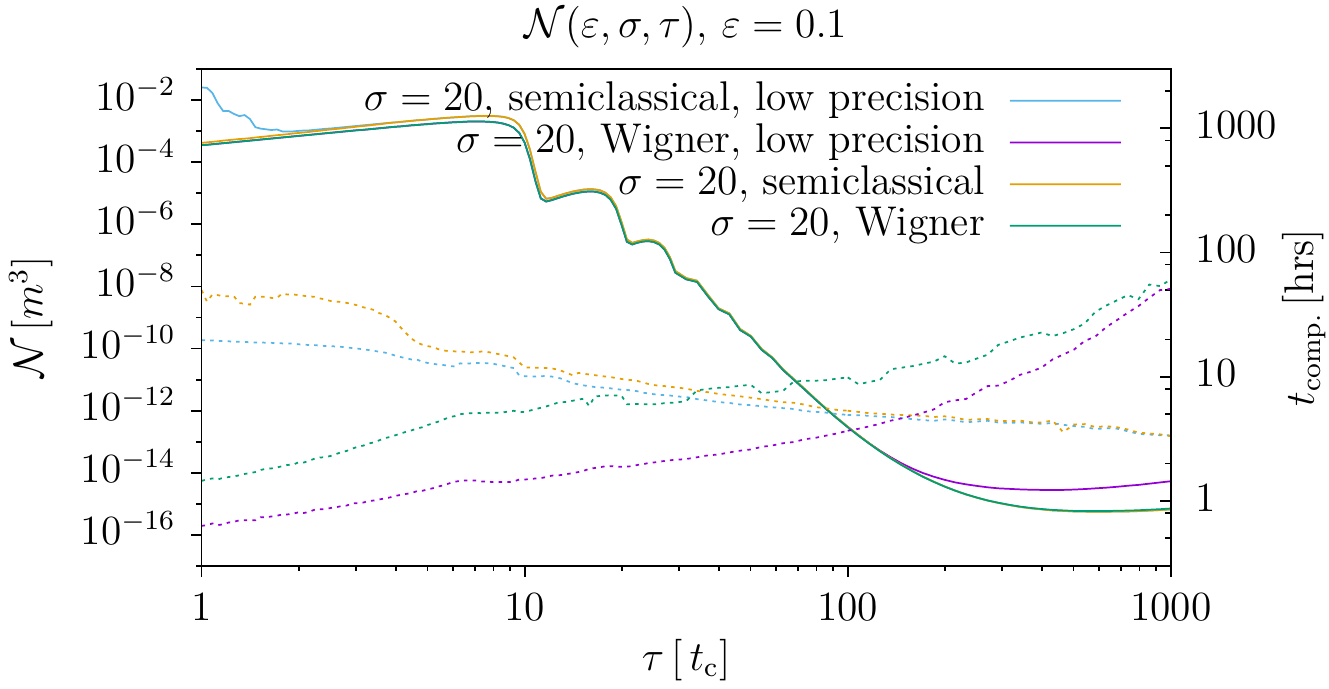}
 \caption[Comparison of the total particle number per Compton volume of rotating Sauter pulses as a function of the pulse length $\tau$ for different methods.]
 {
   Comparison of the total particle number per Compton volume of the rotating Sauter pulse for \(\epsilon=0.1,\,\sigma=20\) as a function of the pulse length \(\tau\) for different settings regarding precision.
   Solid lines show particle yield, dashed lines show processor time per spectrum.
   The Wigner method quickly increases in computational time with increasing pulse length, until it becomes numerically unfeasible.
   The semiclassical method, on the other hand, becomes numerically cheaper for longer pulses.
 }
 \label{fig:total_yield_20}
\end{figure}
In order to compare the momentum spectra calculated by all three methods, let us choose an illustrative example.
Let $\tau=10/m$ and $\sigma=6$.
\Fig\ref{fig:spectra_methods} shows the spectrum as computed by all the available methods.
In the semiclassical result and the result of the LCRFA the sum over both solutions according to \Eqref{eq:MomentumSpectrumTotal} is taken.
It turns out that the semiclassical method overestimates the pair production probability by roughly 12 percent as compared to the result of the Wigner method, which should be considered as exact within the chosen numerical precision.
The result in LCRFA also has the same order of magnitude as the other results but underestimates certain features of the momentum spectrum.
This is due to the small number of field cycles as the approximation gets better for bigger \(\sigma\), see \Fig\ref{fig:total_yield}.

We can compute the total particle yield per volume from the momentum spectrum by integrating over momentum space
\begin{align*}
  \mathcal{N}_\mathrm{sc} \definedby \int  \frac{\dd[3]\cm}{(2\pi  )^3}\, W_s(\vv{\cm})\,.
\end{align*}
Comparing the results we find that the methods agree for an intermediate range of pulse length \(\tau\) (see \Fig\ref{fig:total_yield}).

\pagebreak
We also find that the semiclassical method is not stable for short pulses.
This can be explained by looking at the non-rotating Sauter pulse which is studied in more detail in Appendix~\ref{app:Sauter}.
Taking the turning points given by \Eqref{eq:TPSauter} into consideration, we find that, for decreasing \(eE_0\tau/m\), the turning points get closer in the complex plane (see \Fig\ref{fig:TPSauter} for a plot of \(|\omega(t)|\) around the turning points).
The approximation performed in Sec.~\ref{sec:semiclass} assumes that \Eqref{eq:APPROX} holds for every turning point.
This is not the case if the different turning points get too close to each other in the complex plane.

For longer pulses the numerics of the Wigner method become challenging.
This is due to the fact that the integration from \(t=-10\tau\) to \(t=10\tau\), which is performed analytically in the semiclassical method, needs more steps the longer the pulse becomes.
For pulses that are too long the precision of the result is limited by computational errors which leads to an overestimation of the total particle yield due to summing up numerical noise.

We find that for \(\sigma=20\) both numerical methods have a comparable computation time for pulse durations \(\tau\in (40\,\tc,100\,\tc)\), see \Fig\ref{fig:total_yield_20}.
For shorter pulses the Wigner method is computationally faster, while for longer ones the semiclassical method should be preferred.

The computation time when using the LCRFA is in comparison to the numerical semiclassical method negligible.
We find that the approximation gets better for a longer pulse length \(\tau\) and a higher \(\sigma\) (see \Fig\ref{fig:total_yield}, observe, that within the LCRFA calculations, the number of considered turning points has been fixed to nine, in contrast to the adaptive method used for the numerical method described in Appendix~\ref{sec:numsemi}).
This can be explained by the fact that the approximation of the pulse being locally constant becomes better for longer pulses, and that a larger \(\sigma\) means more rotations per pulse length and hence larger rotation effects compared to pulse shape effects.

\subsubsection{Interpretation of the Independent Solutions}
\begin{figure}
 \centering
 \includegraphics[width=\linewidth]{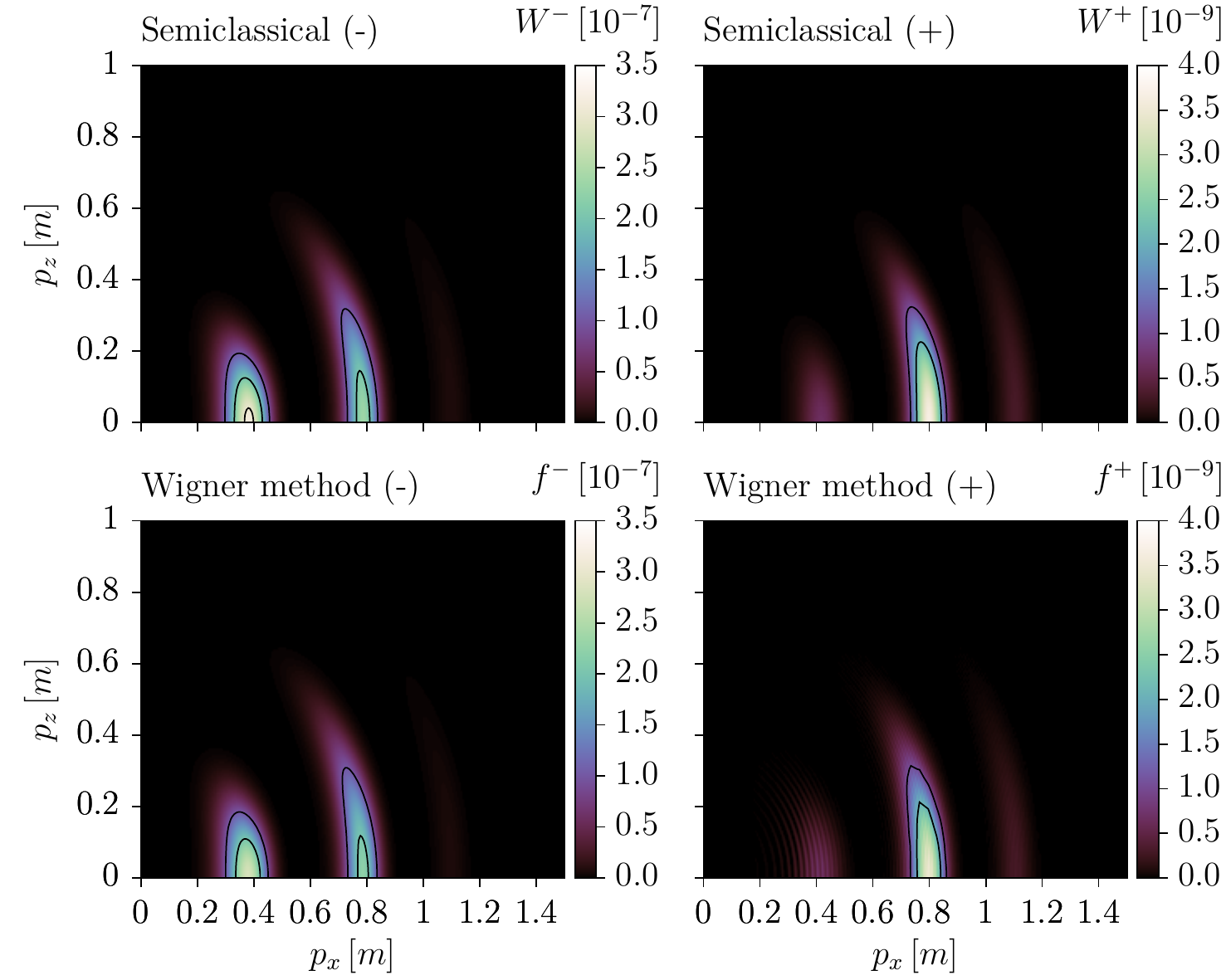}
 \caption[Spectra of the particle yield for the independent solutions of the semiclassical method compared to the corresponding projections of the Wigner function.]
 {
   Comparison of spectra for the semiclassical $+$ and $-$ solutions (see \Eqref{eq:MomentumSpectrum}) as well as corresponding spectra from the Wigner method (see \Eqref{eqn:wigner_semiclassical_projected_f}).
   The levels of the contour lines are indicated by the marks of the color box.
   The pulse parameters are $\tau=46.42\,\tc$, $\sigma=20$.
   Except for an interference pattern in the bottom right plot around \(p_x\sim0.4\), which is not found in the corresponding semiclassical result (top right plot), the spectra of the two methods agree with each other.
 }
 \label{fig:spectra_spin_cmp}
\end{figure}
\begin{figure}
 \centering
 \includegraphics{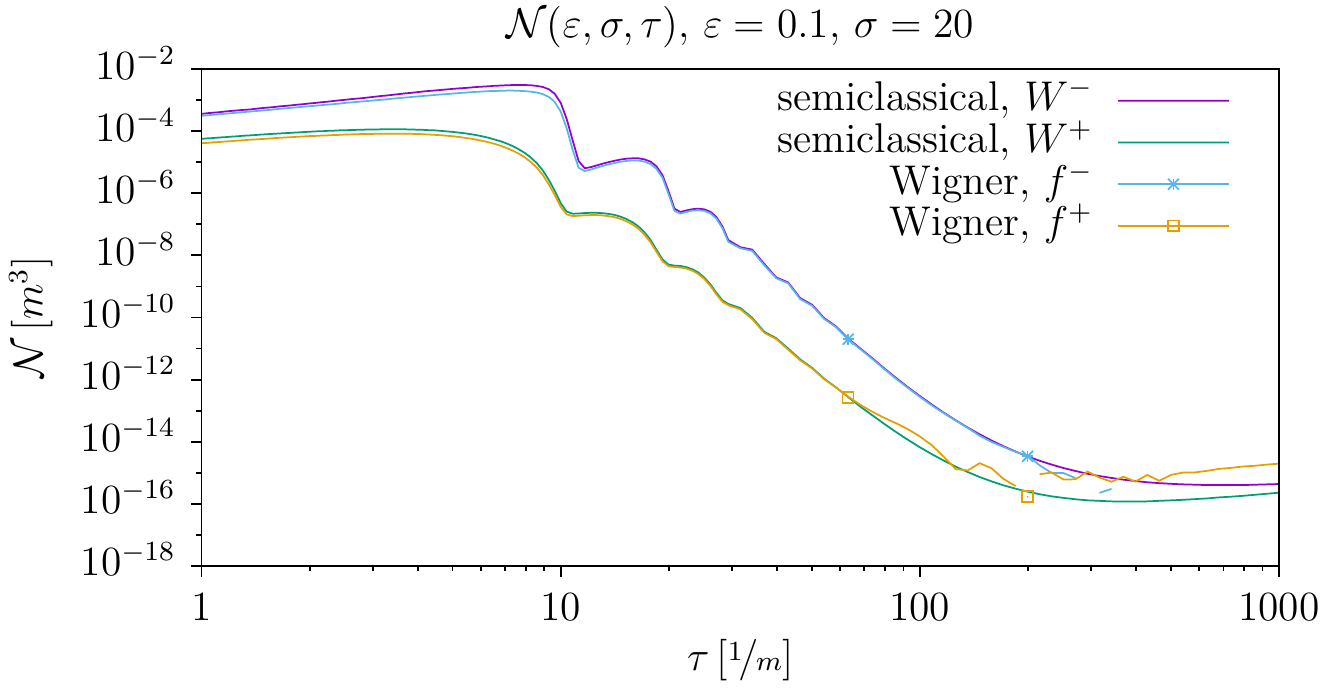}
 \caption[Comparison of the particle yield for the independent solutions of the semiclassical method with the corresponding projections of the Wigner function.]
 {
   Comparison of the particle yield for the semiclassical $W^+$ and $W^-$ solutions (see \Eqref{eq:MomentumSpectrum}) with the corresponding Wigner function projections $f^+$ and $f^-$ (see \Eqref{eqn:wigner_semiclassical_projected_f}).
   The Wigner method data suffer from a lack of precision for \(\tau>100\), which results in some artifacts in the blue and yellow lines.
   The data points on top of the lines have been calculated using higher precision.
   We find that the results agree with each other.
 }
 \label{fig:spin_plot_20}
\end{figure}
In Sec.~\ref{sec:semiclass} we found two independent solutions of the Dirac equation which were interpreted as different spin components in \Ref\cite{Strobel:2014tha}.
To compare these with the Wigner method we can construct a projector $P_s$, by requiring
\begin{align*}
  P_s\cdot \psi_{\vv{\cm},s}&=\psi_{\vv{\cm},s}&&\text{and}&
 P_s\cdot \psi_{\vv{\cm},-s}&=0
\end{align*}
for the two independent solutions of the Dirac equation defined in \Eqref{eq:firstAnsatz}.

The resulting projectors read
\begin{align}
 P_s \definedby \frac{1}{2}\mathbbm{1}-s\frac{1}{2}\frac{\cm_z\gamma_5+m\,\gamma_5\gamma^3}{\epsilon_\perp} \nonumber
 \\
 \label{eq:projector_interpret}
 &= \frac{1}{2}\mathbbm{1} -s\frac{1}{2}\left( \frac{\cm_z}{\epsilon_\perp} (P_r-P_l) + \frac{m}{\epsilon_\perp}\left( P_{\mu_z^+} - P_{\mu_z^-} \right)\right)\,,
\end{align}
they are idempotent and orthogonal
\begin{align*}
 P_s\cdot P_s&=P_s\,,
 &
 P_s\cdot P_{-s}&=0.
\end{align*}
They are also complete, i.\,e.,
\begin{align*}
 P_s+ P_{-s}&=\mathbbm{1}\,.
\end{align*}

Accordingly they project onto the two parts of the spectrum which correspond to these solutions.
In \Eqref{eq:projector_interpret} it is evident that the two solutions from the scattering method correspond to a linear combination of chirality and magnetic momentum.

While in the context of the Wigner function, which contains the full spinor information, \(\delta f_\mathrm{c}\) and \(\delta f_{\mu_z}\), given in \Eqs\eqref{eqn:asymm_chiral_homog} and \eqref{eqn:asymm_magnetic_homog} respectively, are the physically meaningful observables, we will construct \(f^s\) to show the connection to the solutions of the semiclassical method.
The projected one-particle function, as defined in \Eqref{eqn:general_f_projected}, using this projection is given by
\begin{align}
 \nonumber
 f^s&= f\left[P_s\left( \mathcal{W} - \mathcal{W}_\mathrm{vac.} \right)\right]\\
 \label{eqn:wigner_semiclassical_projected_f}
&= \frac{1}{2} \left( f-s \, \delta f_\mathrm{sc} \right)
\end{align}
with the corresponding asymmetry $\delta f_\mathrm{sc}$.
The latter can be related to the chiral and magnetic momentum asymmetries $\delta f_\mathrm{c}$ and $\delta f_{\mu_z}$ respectively as
\begin{align*}
 \delta f_\mathrm{sc} &= \frac{q_z}{\epsilon_\perp}\delta f_\mathrm{c}+\frac{m}{\epsilon_\perp}\delta f_{\mu_z}\,.
\end{align*}
Using this we find that the data for the semiclassical and Wigner method agree (see \Figs\ref{fig:spectra_spin_cmp} and \ref{fig:spin_plot_20}).
This shows that the two independent solutions of the semiclassical method represent spinor eigenstates of the linear combination of the chirality and magnetic moment projection specified in \Eqref{eq:projector_interpret}.

\subsection{Total Particle Yield}\label{sec:simprotTY} %
In order to get a general picture of what is happening, the first interesting observable is the total particle yield, disregarding the complexities of the shape of the spectra.
For a fixed amplitude of $\varepsilon=0.1$ the two-dimensional parameter space of $\sigma\in[0,50]$ and $\tau\in[1\,\tc,1000\,\tc]$ was scanned.
Both the Wigner method and the semiclassical method were used where appropriate.

\begin{figure}
 \centering
 \includegraphics{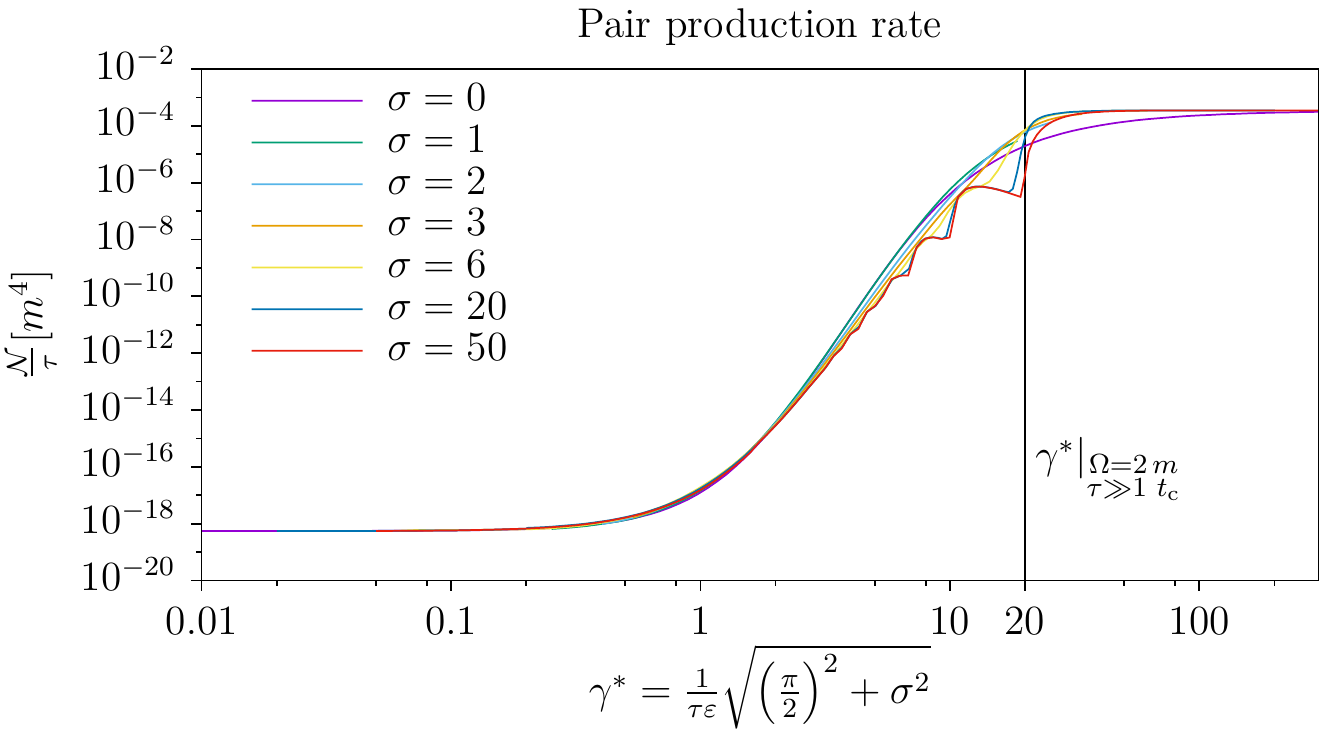}
 \caption[Total particle yield of the rotating Sauter pulse.]
 {
   Total particle yield of the rotating Sauter pulse.
   The data of the two-dimensional parameter scan are closely aligned by normalizing the particle yield to the pulse duration and using the combined Keldysh parameter.
 }
 \label{fig:total_yield_unified}
\end{figure}
\enlargethispage{-\baselineskip}
It has been shown previously \autocite{Blinne:2013via} that if the rotational frequency $\Omega=\frac{\sigma}{\tau}$ is combined with the inverse time scale of the pulse duration to find a combined Keldysh parameter $\widetilde{\gamma^*}=\frac{1}{\varepsilon m}\sqrt{\left( \frac{1}{\tau}  \right)^2 + \Omega^2}$, a lot of the particle yield data fall on a common line.
Further parameter scans demonstrated, that the alignment of the data can be improved if the particle yield is normalized by the pulse duration $\tau$ and if the combined Keldysh parameter is modified to the definition in \Eqref{eq:keldysh_combined}.

\Fig\ref{fig:total_yield_unified} shows that the particle yield is mainly a function of the combined Keldysh parameter modified with effects of multiphoton resonances which are visible in the interval $\gamma^*\in[4,30]$ and are more pronounced for larger $\sigma$.
This is because, for large $\sigma$,  $\gamma^*=\gamma_\Omega$ (see \Eqref{eqn:keldysh_combined_sigmatoinf}) and $\gamma_\tau\to0$ (see \Eqref{eqn:keldysh_combined_intermsof}), which indicates clearly separated regimes and clear domination of multiphoton pair production.

\FloatBarrier
\subsection{Typical Spectra}\label{sec:typical_spectra} %
\begin{figure}
 \centering
 \input{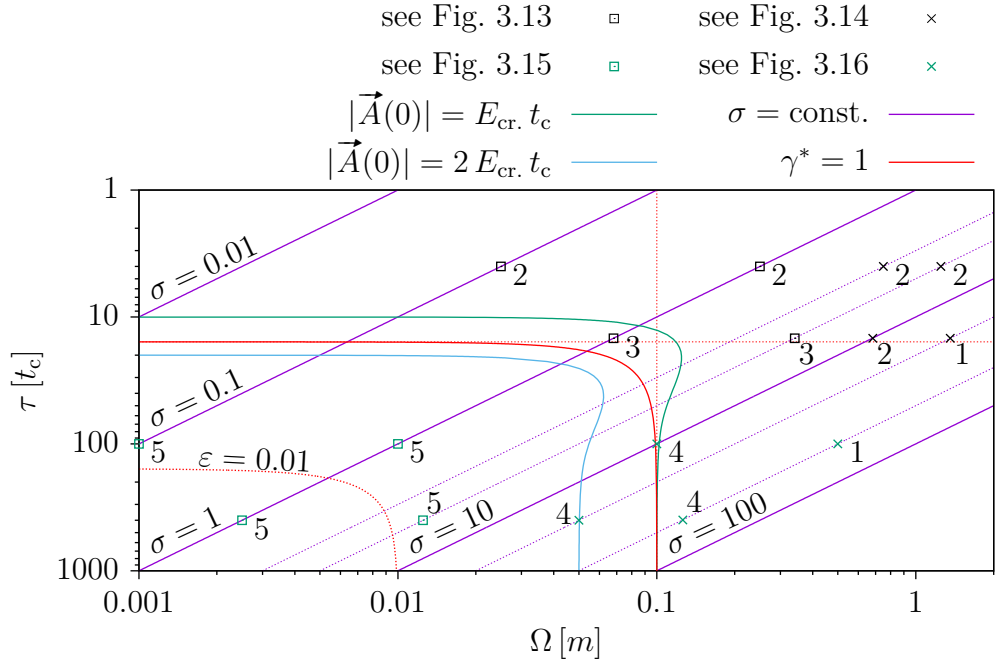}
 \caption[Overview over parameter range of rotating Sauter pulse]{
   Overview over parameter range of rotating Sauter pulse.
   Curves of constant parameters $|\vv{A}|$, $\gamma^*$ and $\sigma$ are given in different colors as indicated above the plot.
   The dashed red lines give $\gamma_\Omega=1$ (vertical red dashed line) and $\gamma_\tau=\frac{2}{\pi}$ (horizontal red dashed line).
   Where relevant an amplitude of $\varepsilon = 0.1$ is assumed, with the only exception of the red dotted line which is marked $\varepsilon=0.01$.
   The latter gives $\gamma^*=1$ for a lower amplitude, which will be of relevance in \Sec\ref{sec:Bichromatic}.
   Markings indicate example spectra in \Figs\ref{fig:typical_spectra_1a} to \ref{fig:typical_spectra_2b}, as indicated above the plot.
   The arrangement of the black and green markings corresponds to the corresponding example spectra on pages~\pageref{fig:typical_spectra_1a}f and \pageref{fig:typical_spectra_2a}f, respectively.
   Numbered labels on markings indicate interpretation of example spectra according to the classification given in \Sec\ref{sec:typical_spectra}.
 }
 \label{fig:omega_tau_1}
\end{figure}
\begin{figure}[p]
\begin{leftfullpage}
 \centering
 \includegraphics{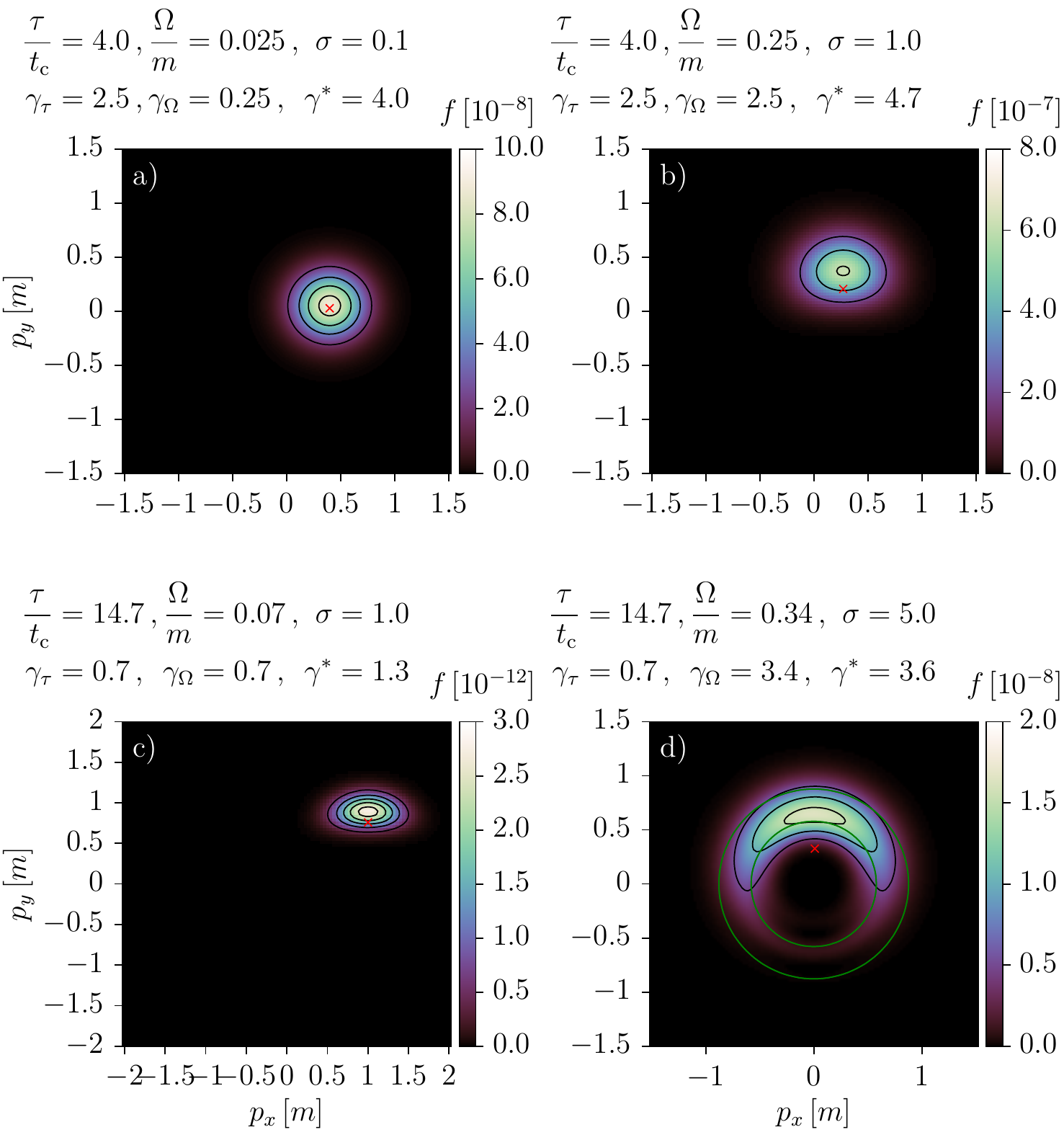}
 \caption[Collection of pair production spectra for the rotating Sauter pulse (1/4)]{
   A collection of pair production spectra for the rotating Sauter pulse as marked in \Fig\ref{fig:omega_tau_1} as black boxes (\includegraphics[height=1ex,valign=B]{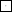}).
   The levels of the contour lines are indicated by the marks of the color box.
   The Parameters for the rotating Sauter pulse are given on top of the individual plots.
   The red cross marks the vector potential $\vv{A}$ at $t=0$.
   The green circles, if present, show the locations of the expected multiphoton rings for appropriate values of $n$.
 }
 \label{fig:typical_spectra_1a}
\end{leftfullpage}
\end{figure}
\begin{figure}[p]
\begin{fullpage}
 \centering
 \includegraphics{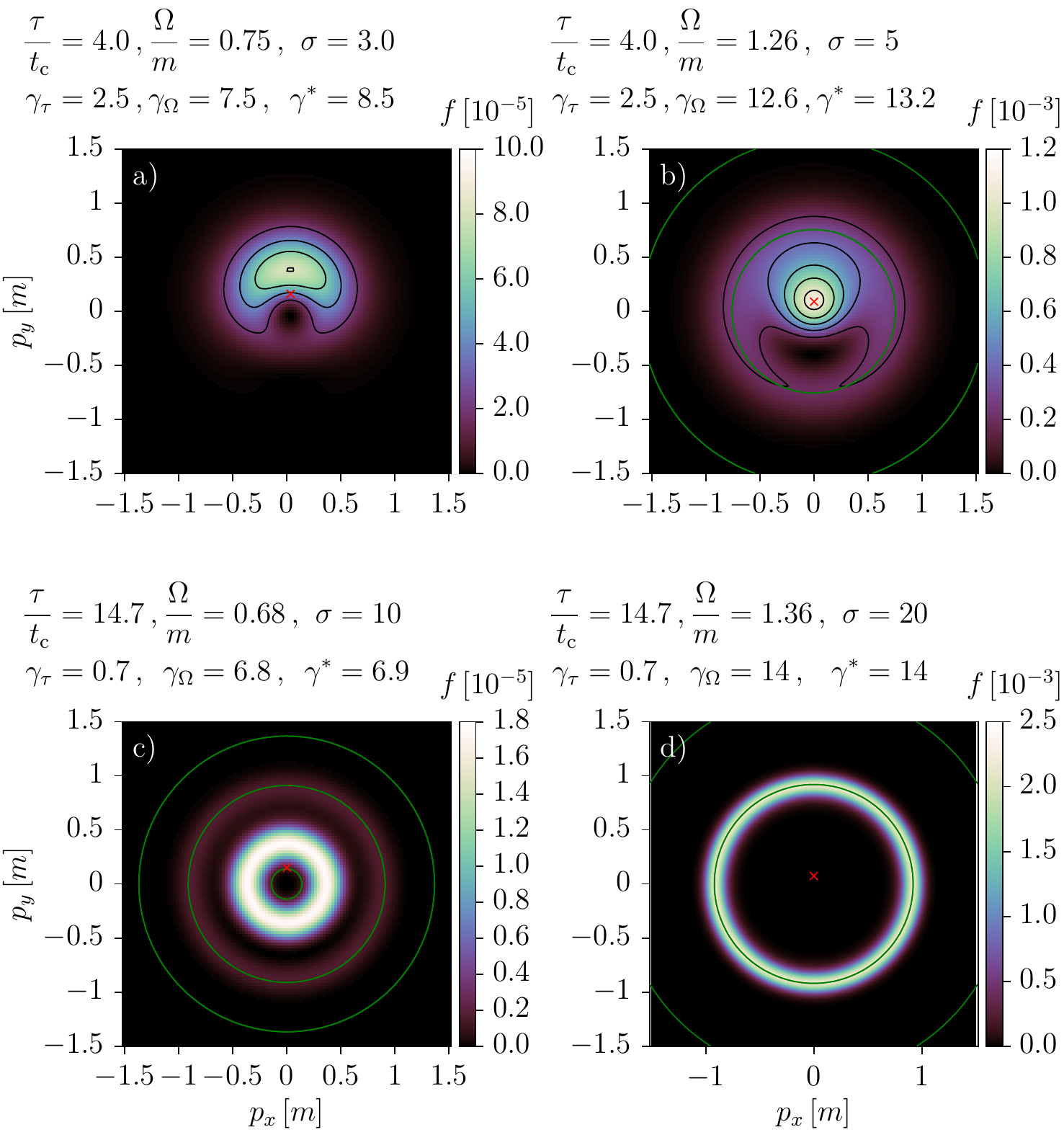}
 \caption[Collection of pair production spectra for the rotating Sauter pulse (2/4)]{
   A collection of pair production spectra for the rotating Sauter pulse as marked in \Fig\ref{fig:omega_tau_1} as black crosses (\includegraphics[height=1ex,valign=B]{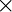}).
   The levels of the contour lines are indicated by the marks of the color box.
   The Parameters for the rotating Sauter pulse are given on top of the individual plots.
   The red cross marks the vector potential $\vv{A}$ at $t=0$.
   The green circles, if present, show the locations of the expected multiphoton rings for appropriate values of $n$.
 }
 \label{fig:typical_spectra_1b}
\end{fullpage}
\end{figure}
\begin{figure}[p]
\begin{leftfullpage}
 \centering
 \includegraphics{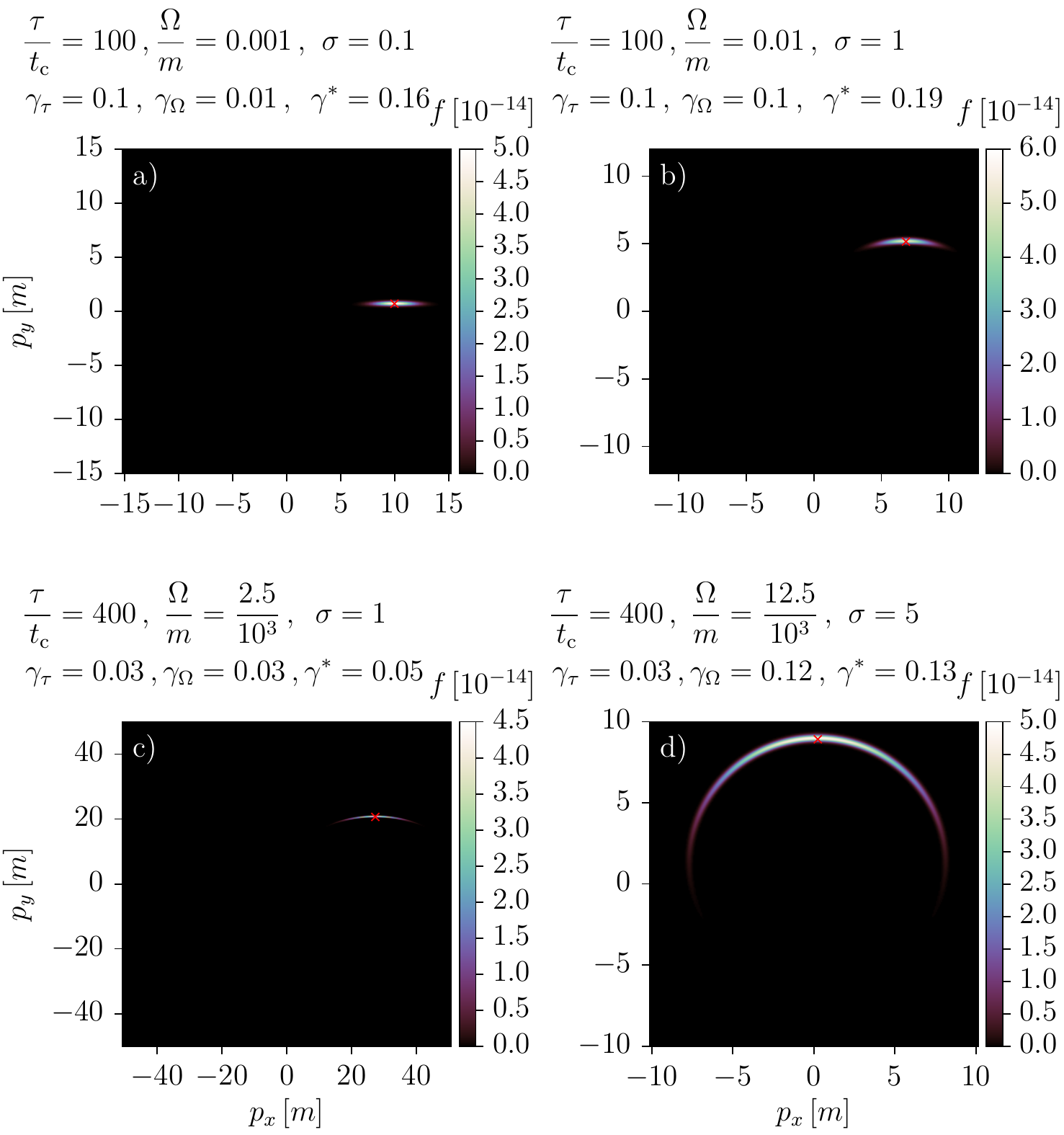}
 \caption[Collection of pair production spectra for the rotating Sauter pulse (3/4)]{
   A collection of pair production spectra for the rotating Sauter pulse as marked in \Fig\ref{fig:omega_tau_1} as green boxes (\includegraphics[height=1ex,valign=B]{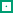}).
   The levels of the contour lines are indicated by the marks of the color box.
   The Parameters for the rotating Sauter pulse are given on top of the individual plots.
   The red cross marks the vector potential $\vv{A}$ at $t=0$.
 }
 \label{fig:typical_spectra_2a}
\end{leftfullpage}
\end{figure}
\begin{figure}[p]
\begin{fullpage}
 \centering
 \includegraphics{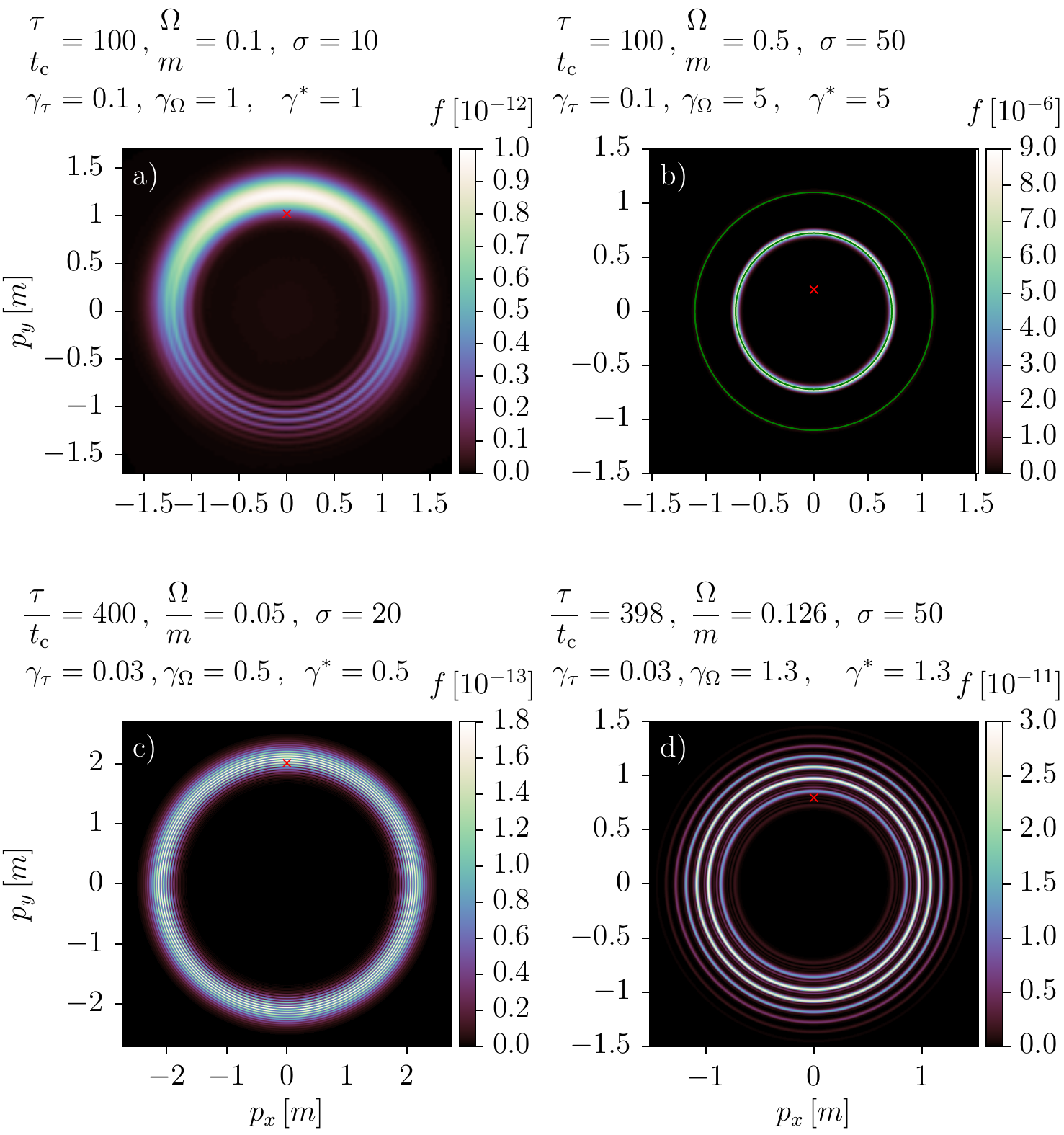}
 \caption[Collection of pair production spectra for the rotating Sauter pulse (4/4)]{
   A collection of pair production spectra for the rotating Sauter pulse as marked in \Fig\ref{fig:omega_tau_1} as green crosses (\includegraphics[height=1ex,valign=B]{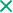}).
   The levels of the contour lines are indicated by the marks of the color box.
   The Parameters for the rotating Sauter pulse are given on top of the individual plots.
   The red cross marks the vector potential $\vv{A}$ at $t=0$.
   The green circles, if present, show the locations of the expected multiphoton rings for appropriate values of $n$.
 }
 \label{fig:typical_spectra_2b}
\end{fullpage}
\end{figure}
In \Sec\ref{sec:expectations} the different parameter regimes regarding multiphoton and Schwinger pair production have been discussed from a phenomenological viewpoint.
In \Fig\ref{fig:omega_tau_1} the parameter space spanned by the pulse duration $\tau$ and the pulse frequency $\Omega$ is displayed for fixed amplitude $\varepsilon=0.1$.
The theoretical boundaries $\gamma^*=1$, $\abs{\vv{A}}(0)=\varepsilon\,E_\crit\,\tau$ and $\abs{\vv{A}}(0)=2\,\varepsilon\,E_\crit\,\tau$ are plotted and additionally a collection of lines with constant $\sigma=\Omega\tau$.
The latter are important, because for a fixed $\sigma$ the locus of the vector potential always has the same shape and is only scaled by $\varepsilon\,\tau$.
In this section a number of example spectra out of this parameter space will be discussed according to the parameters and the shape of the spectrum.
These example spectra are organized into 4 groups (\Figs\ref{fig:typical_spectra_1a} to \ref{fig:typical_spectra_2b}) of 4 spectra (``a)'' to ``d)'') each, as can be told by the different kinds and colors of markings in \Fig\ref{fig:omega_tau_1}, referring to these figures.
The first double page, showing \Figs\ref{fig:typical_spectra_1a} and \ref{fig:typical_spectra_1b}, features short pulse durations, whereas the second double page with \Figs\ref{fig:typical_spectra_2a} and \ref{fig:typical_spectra_2b} features long pulse durations, both with increasing frequencies from left to right.

It is not possible to classify every spectrum as either multiphoton pair production or Schwinger pair production.
A large portion of the parameter space does not clearly fall into either category and the spectrum can only be explained by a mixture of both processes.
There are a number of different possible interpretations which we assign to each of the spectra.
\begin{enumerate}
 \item Clear multiphoton pair production
 \item Multiphoton pair production, modified by influences of the vector potential
 \item A complete mixture of both processes with no recognizable tendency
 \item Schwinger pair production, modified by influences of multiphoton pair production
 \item Clear Schwinger pair production
\end{enumerate}
The assignments of the example spectra regarding this classification are already noted in \Fig\ref{fig:omega_tau_1} by the numbers close to the markings.

As stated earlier, for large enough $\sigma$, it is expected that the spectra show a rotational symmetry.
This is the case for all the plots with $\sigma\geq20$ (\Fig\ref{fig:typical_spectra_1b}d), \Figs\ref{fig:typical_spectra_2b}b), c) and d)), regardless of the pulse duration being quite small as in \Fig\ref{fig:typical_spectra_1b}d) or large as in \Fig\ref{fig:typical_spectra_2b}c).

\pagebreak
\Fig\ref{fig:typical_spectra_1b}c) and \Fig\ref{fig:typical_spectra_2b}a) both have the same value of $\sigma$, but only the first has rotational symmetry, so $\sigma=10$ is not large enough to guarantee rotational symmetry on its own.
In \Fig\ref{fig:typical_spectra_1b}c) $\gamma_\Omega$ is already large enough to allow multiphoton pair production, which improves the rotational symmetry, but only the outer ring lines up with the expected momentum, so the multiphoton interpretation is not complete.

\begin{figure}
  \centering
  \includegraphics{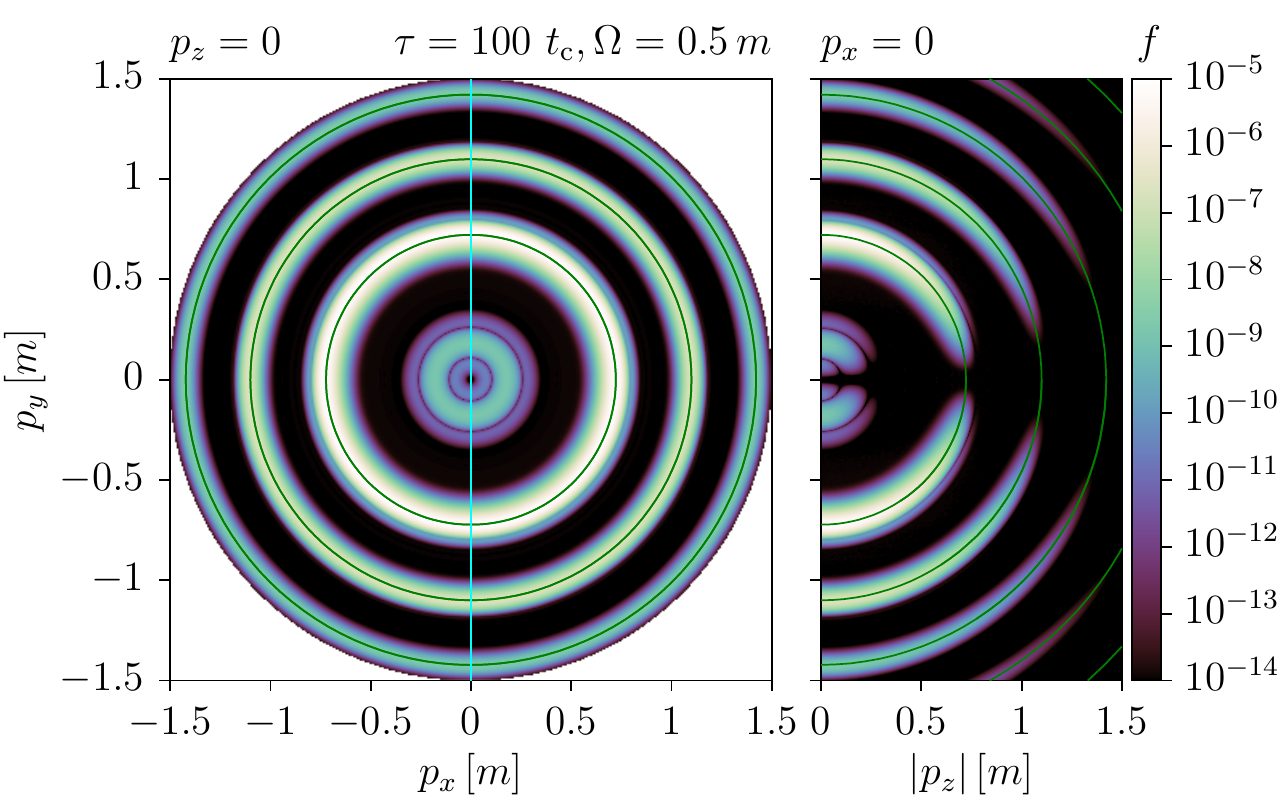}
  \caption[Slices of a three dimensional spectrum (1/3)]{
    This plot shows two different slices of a three dimensional spectrum of pairs produced by a $\sigma=50$ pulse.
    The pulse parameters are identical to \Fig\ref{fig:typical_spectra_2b}b), the spectrum only seems different due to the logarithmic color scale.
    The vertical cyan line in the large plot shows the position of the slice that is shown in the smaller plot.
    The green circles show the expected position of pairs due to multiphoton pair production.
    The innermost green circle corresponds to the energy\hyp{}momentum\hyp{}relation with the energy of 5 photons.
  }
  \label{fig:slices_2}
\end{figure}
We can clearly see multiphoton pair production in \Fig\ref{fig:typical_spectra_2b}b) and \Fig\ref{fig:typical_spectra_1b}d).
The produced pairs line up exactly with the prediction of \Eqref{eqn:multiphoton_circle_effmass}, and the Keldysh parameters are large enough to support multiphoton pair production.
This interpretation is backed up by the full three dimensional data from which a selection of slices is displayed in \Fig\ref{fig:slices_2}.
The multiphoton rings that are visible in the $p_z=0$ slice are in fact part of multiphoton spheres with suppressed pair production around their pole regions at maximum and minimum $p_z$.
The 4\hyp{}photon\hyp{}process is suppressed, because 4 photons do not have enough total energy ($4\Omega=2\,m$) to overcome the pair production threshold $2\,m^*>2\,m$.
However, due to the pulse length the Fourier spectrum of the electric field pulse is broadened and some photons with energies larger than $\Omega=0.5\,m$ exist and enable the 4\hyp{}photon\hyp{}process, which leads to pairs produced in the inner region of \Fig\ref{fig:slices_2}.
The spectra \Fig\ref{fig:typical_spectra_1b}d) and \Fig\ref{fig:typical_spectra_2b}b) are assigned to interpretation $\#1$.

\begin{figure}
  \centering
  \includegraphics{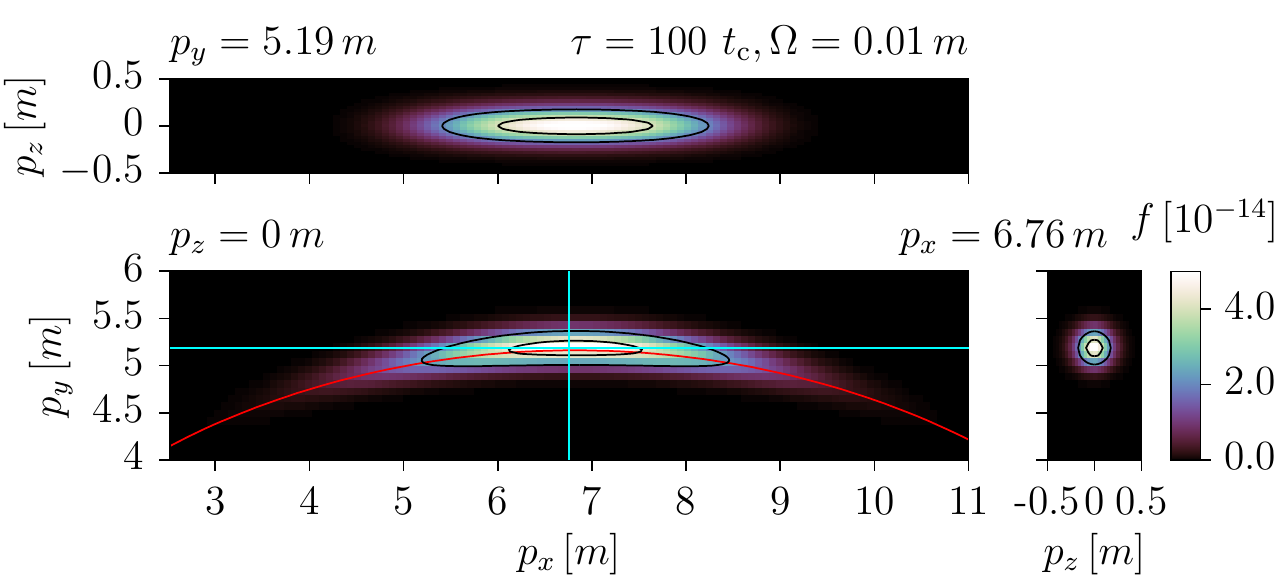}
  \caption[Slices of a three dimensional spectrum (2/3)]{
    This plot shows three different slices of a three dimensional spectrum of pairs produced by a $\sigma=1$ pulse.
    The $p_z=0$ slice was already shown in \Fig\ref{fig:typical_spectra_2a}b).
    The levels of the contour lines are indicated by the marks of the color box.
    The cyan lines in the large plot show the position of the slices that are shown in the smaller plots.
    The red curve shows the locus of the vector potential as a function of time.
  }
  \label{fig:slices_3}
\end{figure}
\begin{figure}
  \centering
  \includegraphics{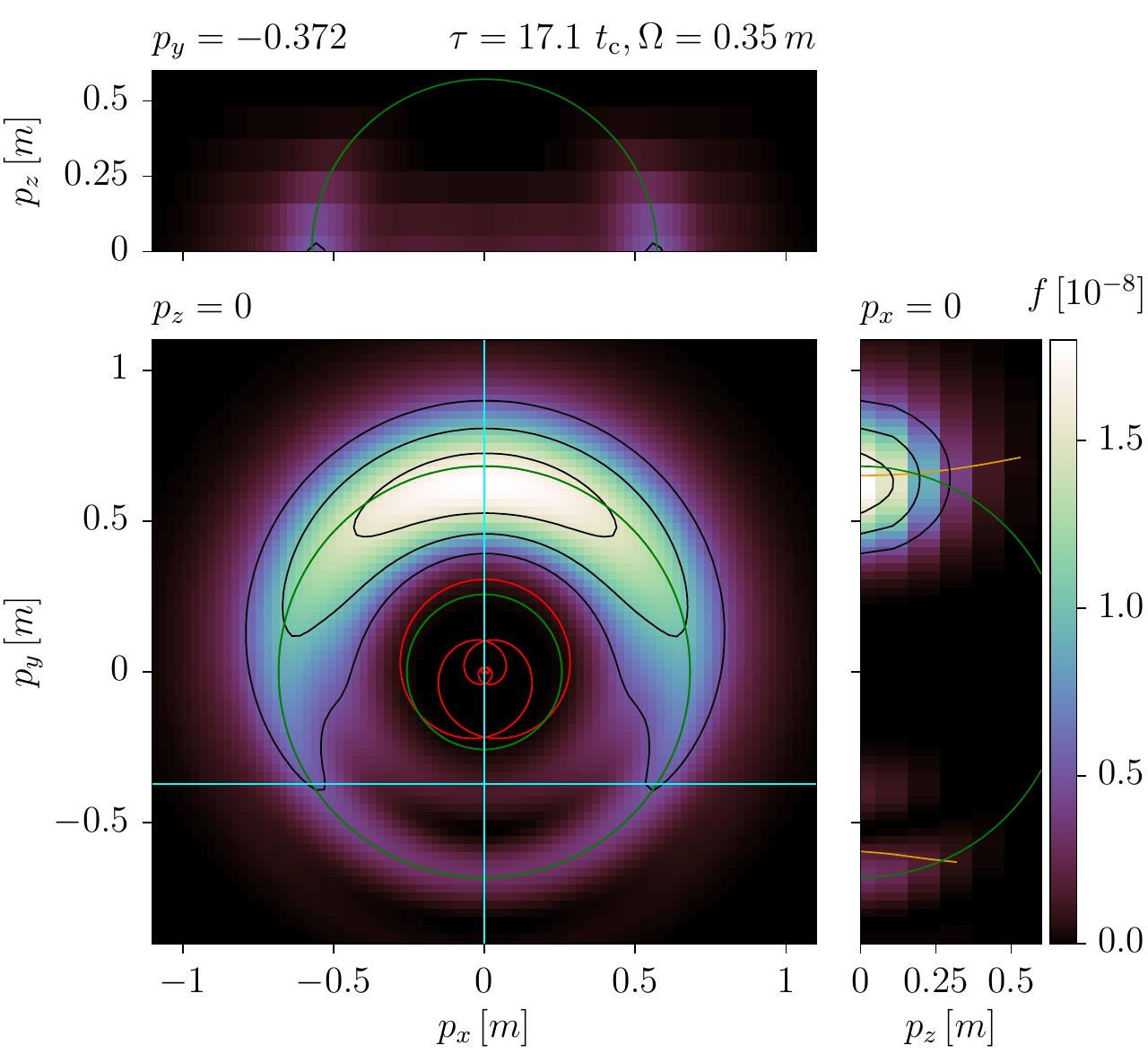}
  \caption[Slices of a three dimensional spectrum (3/3)]{
    This plot shows three different slices of a three dimensional spectrum of pairs produced by a $\sigma=6$ pulse.
    The pulse parameters are not identical but similar to \Fig\ref{fig:typical_spectra_1a}d).
    The levels of the contour lines are indicated by the marks of the color box.
    The cyan lines in the large plot show the position of the slices that are shown in the smaller plots.
    The green circles show the expected position of pairs due to multiphoton pair production.
    The red curve shows the locus of the vector potential.
    The yellow curve shows the value of $p_y$ averaged over pairs with either $p_y<0$ or $p_y>0$ and $p_x=0$ for all $p_z$.
  }
  \label{fig:slices_1}
\end{figure}
On the other hand we can clearly see Schwinger pair production in all of \Fig\ref{fig:typical_spectra_2a}.
All the produced pairs have momenta along the locus of the vector potential and the spectrum is peaked at $\vv{A}(0)$.
In \Fig\ref{fig:slices_3} the three dimensional data for \Fig\ref{fig:typical_spectra_2a}b) are displayed.
It is clearly visible that the number of produced pairs decays with increasing $p_z$ but does not follow a spheroidal shape as in the multiphoton case.
The loci of the vector potential for different values of $\sigma$ are shown in \Fig\ref{fig:vector_potential_loci} for comparison.
These spectra are assigned to interpretation $\#5$.

All the remaining spectra display features of both pair production processes and will be assigned to the intermediate interpretations $\#2$ to $\#4$.
Please note that these assignments are due to interpreting the pair production spectra w.\,r.\,t. the phenomenology explained in \Sec\ref{sec:phenomenology} and are not stringent in a mathematical way.
We propose them to develop a better understanding of the pair production regimes and their boundaries.

For short pulses we see pair production peaked near $\vv{A}(0)$, which would suggest Schwinger pair production.
The Keldysh parameters rather indicate a different interpretation, not allowing Schwinger pair production in a significant quantity.
This apparent contradiction is resolved by noting that the broad bandwidth of the short rotating Sauter pulse enables multiphoton pair production at $\vv{p}=0$ at $t=0$.
Those pairs are subsequently accelerated by the electric field which explains the shape of the spectra.
Conclusively, the spectra \Figs\ref{fig:typical_spectra_1a}a), b), \Figs\ref{fig:typical_spectra_1b}a), b) and c) will be assigned to interpretation $\#2$.

Another kind of explanation can be applied to \Fig\ref{fig:typical_spectra_2b}a), c) and d).
The Keldysh parameters do not strongly support multiphoton pair production.
Pairs are produced close to the vector potential at $t=0$, but as $\sigma$ is large enough in the latter two cases we expect and observe cylindrical symmetry.
One could argue that the vector potential takes a number of cycles, while pairs are produced by the Schwinger effect and interferences between those pairs account for the filigree rings in these spectra.
However it must be noted, that the distance between the rings is exactly what would be expected in multiphoton pair production, while the inner rings with lower photon numbers are simply missing.
The Keldysh parameters and the absence of the inner rings discount multiphoton pair production as the dominating process and these spectra will be assigned to interpretation $\#4$.

The only two remaining spectra of all displayed examples, \Figs\ref{fig:typical_spectra_1a}c) and d), show features of multiphoton pair production and Schwinger pair production and no clear tendency seems to be evident.
They are assigned to interpretation $\#3$.

\Fig\ref{fig:slices_1} shows the three dimensional spectrum for a pulse with parameters close to those assigned to $\#3$.
The pair production peak in the upper part of the spectrum does not seem to line up with the multiphoton sphere (green circle), as indicated by the average $p_y$ of those pairs with $p_y>0$ displayed as the upper yellow line.
In the lower part of the $p_x=0$ plot it should be noted that a portion of the produced pairs indeed lines up with the calculated multiphoton sphere with $|\vv{p}|\approx0.66\,m$.
A somewhat weaker partial pair production sphere with a radius of $|\vv{p}|\approx0.37\,m$ is also visible, regardless of the radius fits any multi-photon order.
If an average of $p_y$ for pairs with $p_y<0$ is calculated, which averages over both of these partial spheres, the result also does not line up with the expectation (lower yellow line).
This suggests that the misalignment of the upper yellow line is also just a result of averaging over two overlapping partial spheres.

When the assignments are taken into account and mapped over the parameter space, see \Fig\ref{fig:omega_tau_1}, it is noted that the assignment seems to be continuous w.\,r.\,t. the parameters and that the Keldysh parameters and the lines of constant absolute value of the vector potential at $t=0$ are sufficient guides to distinguish some of the interpretations.
Clear multiphoton pair production seems to require a large enough $\sigma\gtrsim20$ and $\Omega\gtrsim0.3\,m$.

\subsection{Magnetic Moment}\label{sec:spin_states}
\begin{figure}
\begin{leftfullpage}
 \centering
 \includegraphics{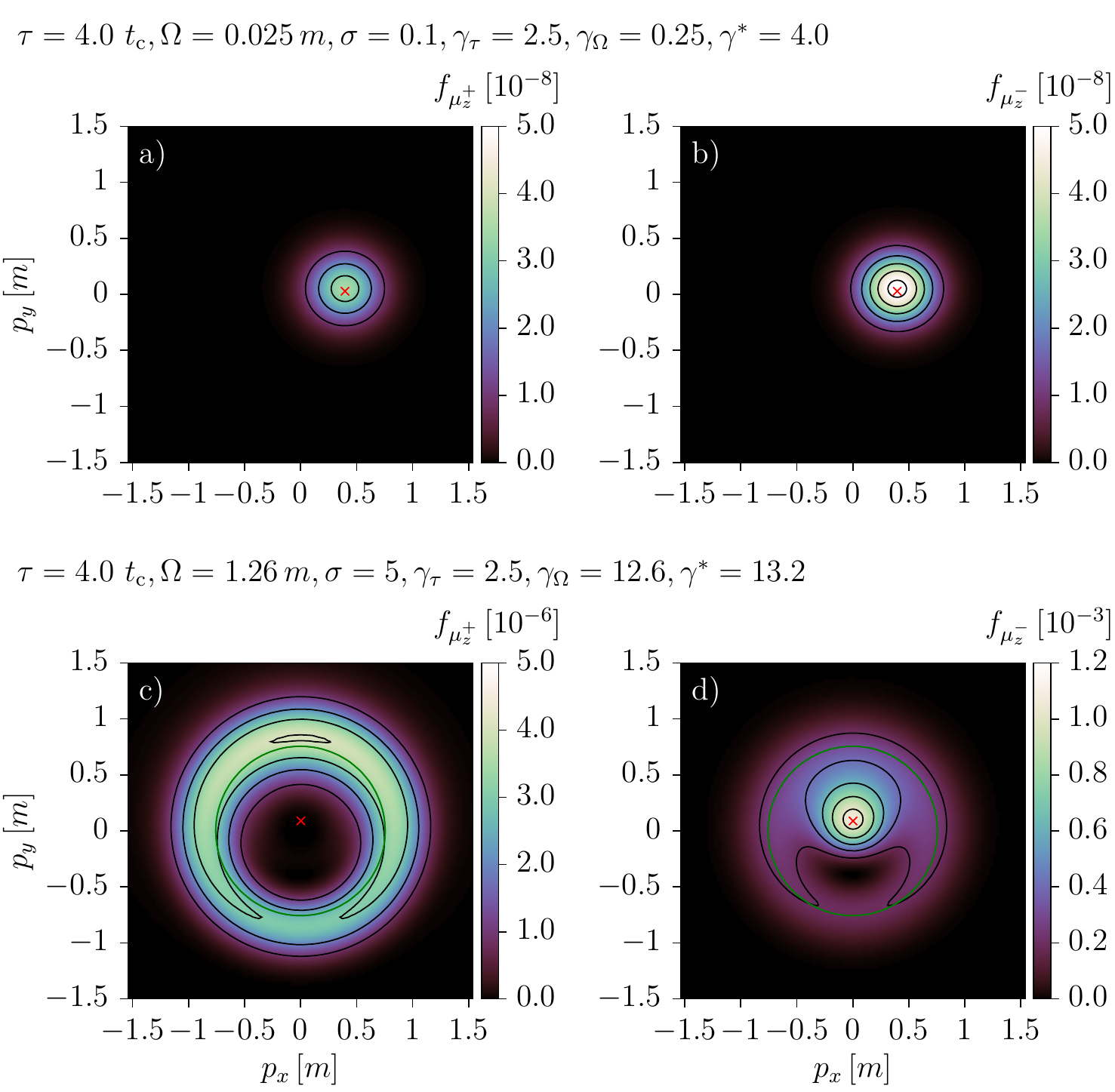}
 \caption[Spectra of pairs with specific magnetic moment (1/2)]{
   Panels a) and b): Pair production spectrum \Fig\ref{fig:typical_spectra_1a}a) separated by magnetic moment.
   Panels c) and d): Pair production spectrum \Fig\ref{fig:typical_spectra_1b}b) separated by magnetic moment.
   The levels of the contour lines are indicated by the marks of the color box.
   The red crosses indicate the vector potential at $=0$.
   The green circles, if present, show the expected position of pairs due to multiphoton pair production.
 }
 \label{fig:spin_states_1}
\end{leftfullpage}
\end{figure}
\begin{figure}
\begin{fullpage}
 \centering
 \includegraphics{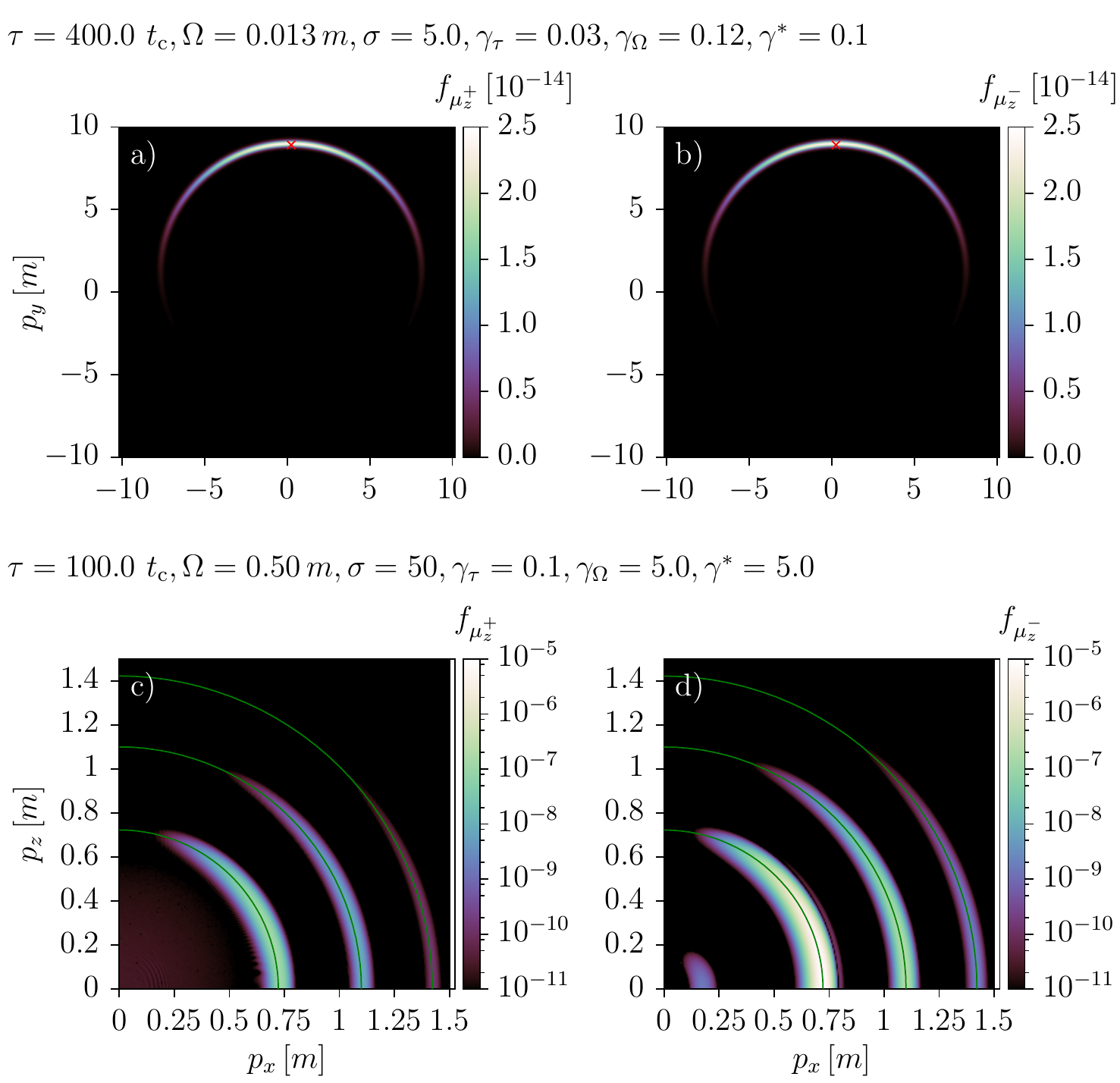}
 \caption[Spectra of pairs with specific magnetic moment (2/2)]{
   Panels a) and b): Pair production spectrum \Fig\ref{fig:typical_spectra_2a}d) separated by magnetic moment.
   Panels c) and d): Pair production spectrum \Fig\ref{fig:typical_spectra_2b}b) separated by magnetic moment.
   The red crosses indicate the vector potential at $=0$.
   The green circles, if present, show the expected position of pairs due to multiphoton pair production.
 }
 \label{fig:spin_states_2}
\end{fullpage}
\end{figure}
As discussed in \Sec\ref{sec:quantumstates} it is possible to gain access to information about various quantum states of the produced pairs.
It turned out that the magnetic moment, a combination of spin and charge, is a candidate for meaningful information.

\pagebreak
The spectra in \Figs\ref{fig:spin_states_1} and \ref{fig:spin_states_2} show the pair production spectra projected onto specific states of magnetic moment, according to \Eqref{eqn:asymm_magnetic_homog} and
\begin{align*}
  f_{\mu_z^\pm} = \frac12\left( f \pm \delta f_{\mu_z}  \right)\,.
\end{align*}
Interestingly, in some cases there is nearly no asymmetry, as in \Figs\ref{fig:spin_states_1}a) and b) and \Figs\ref{fig:spin_states_2}a) and b), while in other cases the asymmetry is drastic.
In \Fig\ref{fig:spin_states_1}c) pair production of pairs in the state $\mu_z^+$ is suppressed by three orders of magnitude.
The pair production spectra for both states also have a completely different shape in this case.
The suppression in \Fig\ref{fig:spin_states_2}c) is weaker, but nevertheless apparent on a logarithmic scale.
The relevance of spin effects to multiphoton pair production has been demonstrated in \Ref\cite{arXiv:1605.03476}.

We conjecture that the susceptibility to spin effects is linked with the characteristic frequency of the pulse.
In pulses of long duration as in \Fig\ref{fig:spin_states_2}, the Schwinger-dominated spectra are symmetric w.\,r.\,t. magnetic moment, while the multiphoton dominated spectra show an asymmetry.

When the pulse duration is very short as in \Fig\ref{fig:spin_states_1}, the picture is not so clear.
The broad spectrum of short pulses enables resonant pair production which peaks at $\vv{p}=\vv{A}(0)$, this is the case in \Figs\ref{fig:spin_states_1}a), b) and d).
In the first two cases the spectra are nearly symmetric w.\,r.\,t. magnetic moment, while in \Figs\ref{fig:spin_states_1}c) the resonant pair production process due to the broadened spectrum seems to be completely suppressed and a multiphoton ring remains which is slightly modified by the vector potential.

\section{Generalized Polarization} %
\label{sec:general_polarization}
\begin{figure}[p]%
\begin{leftfullpage}
 \centering
 \includegraphics{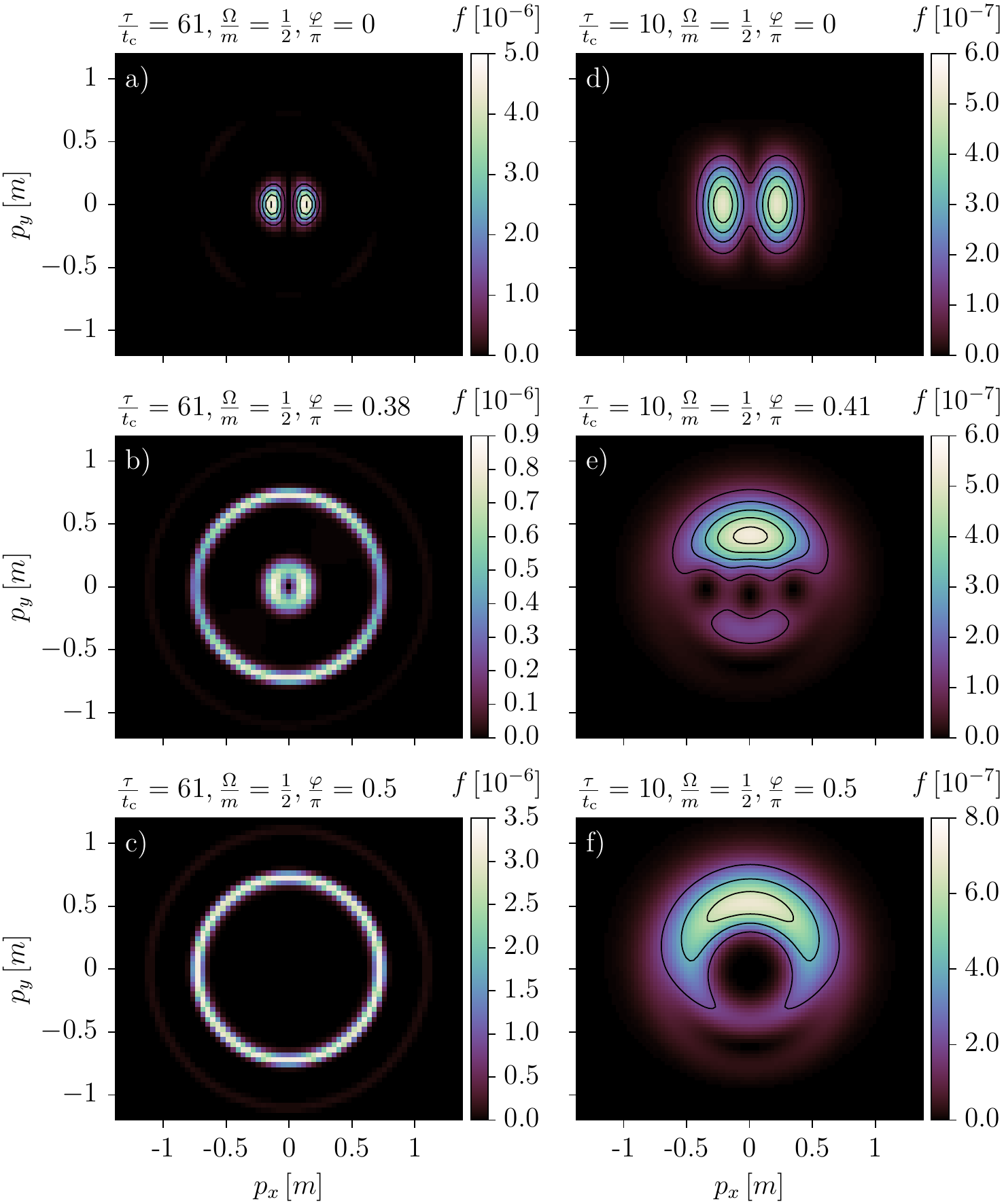}
 \caption[Spectra of pair production in general polarization (1/2)]{
   Example spectra of pair production in linear, elliptical and circular polarization for different parameters.
   The levels of the contour lines are indicated by the marks of the color box.
   The amplitude is $\varepsilon=0.1$ in all cases.
   The left hand set of spectra has $\sigma=30.5$ and thus cylindrical symmetry is expected and present in the circular case.
   Already in the elliptical regime the symmetry can be observed.
   The right hand set of spectra has $\sigma=5$ and is far away from cylindrical symmetry in the elliptical case.
 }
 \label{fig:elli_multi_1}
\end{leftfullpage}
\end{figure}
\begin{figure}[p]%
\begin{fullpage}
 \centering
 \includegraphics{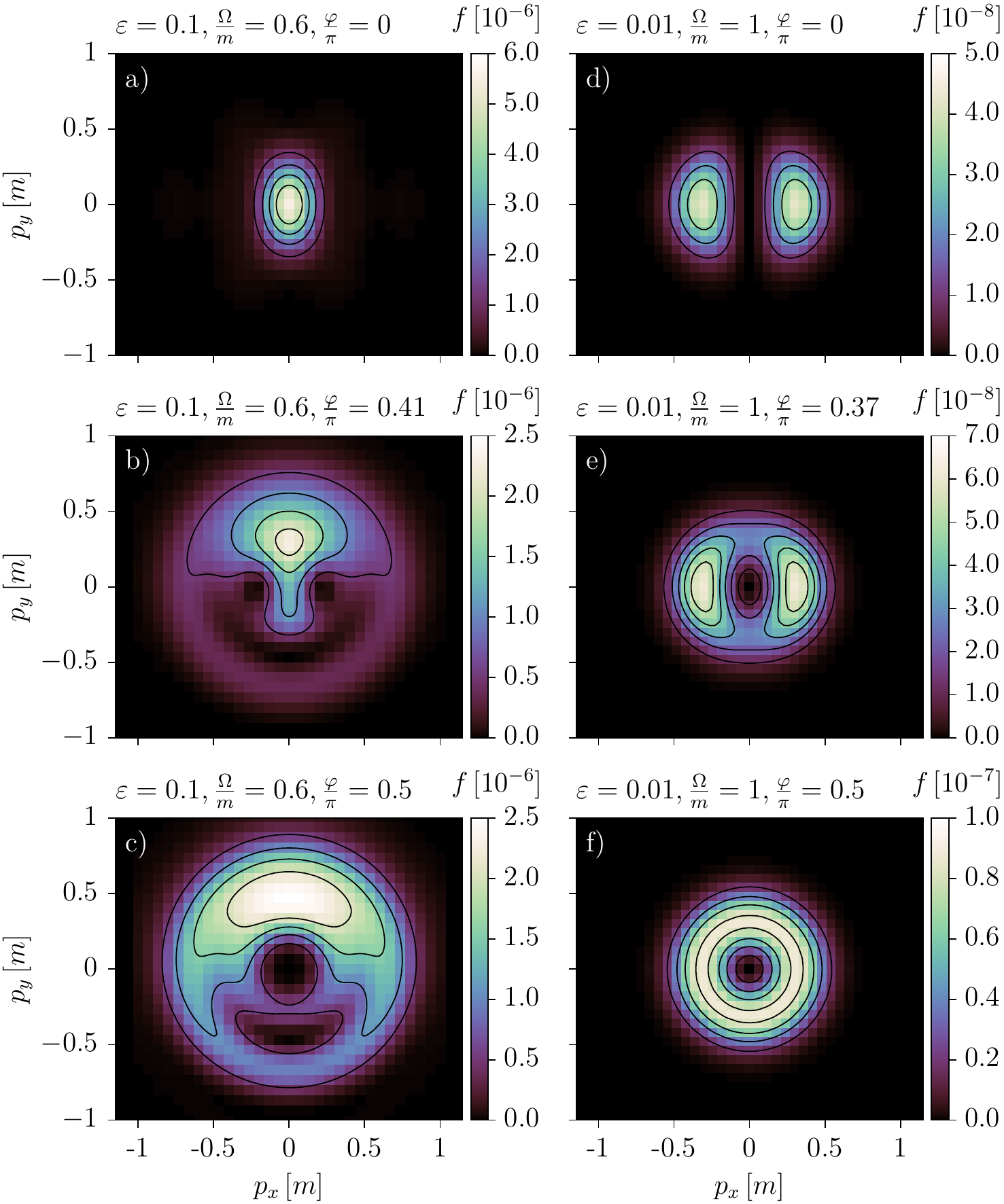}
 \caption[Spectra of pair production in general polarization (2/2)]{
   Example spectra of pair production in linear, elliptical and circular polarization for short pulses with $\tau=10\,\tc$.
   The levels of the contour lines are indicated by the marks of the color box.
   In these cases the amplitude $\varepsilon$ and the frequency $\Omega$ are varied.
   The left hand set of spectra has $\sigma=6$ and the circular case is also featured in \Fig\ref{fig:spectra_methods}a).
   The spectra on the right hand side use a weaker field with a higher frequency of $\sigma=10$.
 }
 \label{fig:elli_multi_2}
\end{fullpage}
\end{figure}
In an experiment real laser light would never have perfect circular polarization, or even perfect linear polarization.
It is as such necessary to look into the intermediate regime of elliptic polarization.
While some studies have already been conducted \autocite{Li:2014nua,arXiv:1504.06051,Li:2015cea,Wollert:2015kra}, there is still a large parameter region unexplored.
The parameter that interpolates between the linear and circular regimes is the phase shift between the orthogonal components of the electric field.
If it is introduced to both, the $x$ and $y$ component in a symmetric way, we can rotate the resulting field such that it coincides with the linear oscillating pulsed field in \Ref\cite{Kohlfurst:2013ura} for $\varphi=0$, resulting in
\begin{align*}
   \vv{E}(t)\definedby\frac{\varepsilon\,E_\mathrm{crit.}}{\sqrt{2}\cos(\frac{\varphi}{2})\cosh^2\left(\nicefrac{t}{\tau}\right)} \mathcal{R}\left(-\frac{\pi}{4}\right) \begin{pmatrix}
                    \cos(\Omega t+\frac{\varphi}{2}) \\
                    \cos(\Omega t-\frac{\varphi}{2}) \\
                    0
                  \end{pmatrix}
        \\
    &=\frac{\varepsilon\,E_\mathrm{crit.}}{\cosh^2\left(\nicefrac{t}{\tau}\right)}
         \begin{pmatrix}
            \cos(\Omega t) \\
            \tan(\frac{\varphi}{2})\sin(\Omega t) \\
            0
         \end{pmatrix}\,.
\end{align*}
Due to the normalization factor $\nicefrac{1}{\left(\sqrt{2}\cos(\frac{\varphi}{2})\right)}$ this also coincides with the rotating Sauter pulse from \Eqref{eqn:puls-sauter-rot} for $\varphi=\nicefrac{\pi}2$.
Note that this definition is slightly different from that used in Refs.~\autocite{arXiv:1504.06051,Li:2015cea}, but equivalent.
Also the dimensionless parameter $\sigma=\Omega\tau$ is the same as in the rotating Sauter pulse.

\begin{figure}
 \centering
 \includegraphics{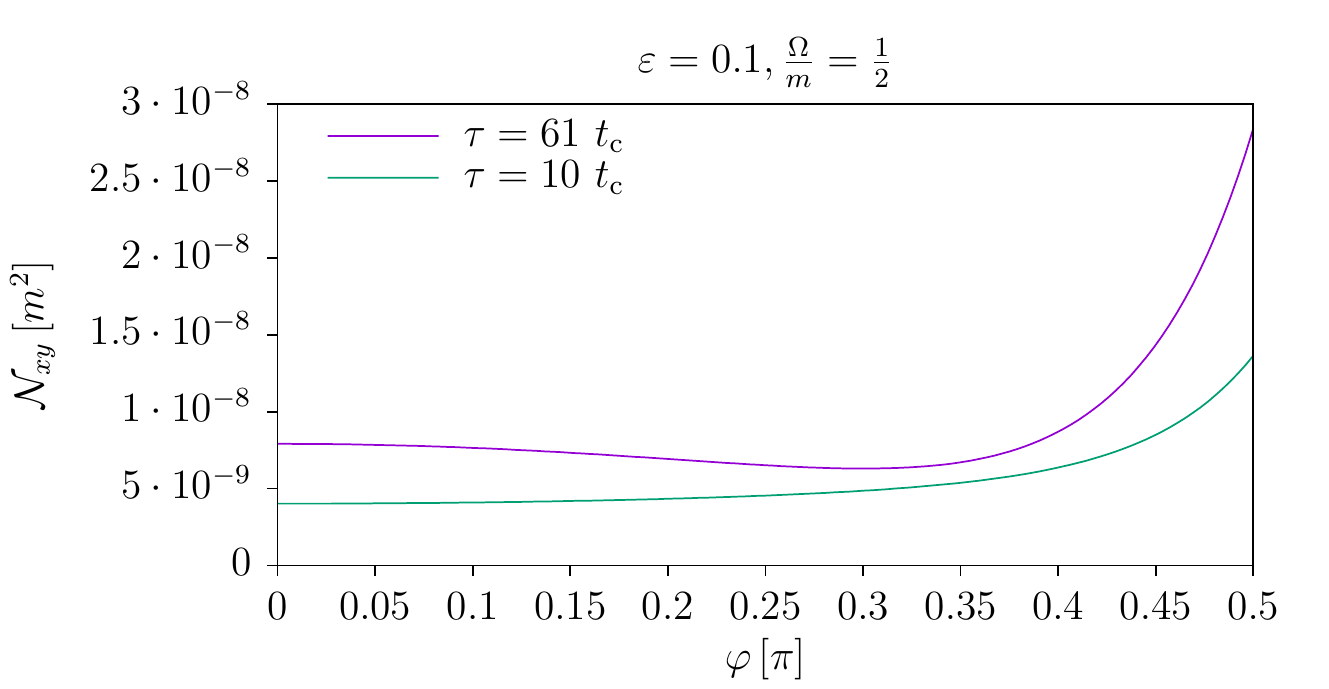}
 \caption[Particle yield in general polarization (1/2)]{
   Two dimensional particle yield in general polarization for spectra shown in \Fig\ref{fig:elli_multi_1}.
 }
 \label{fig:elli_total_1}
\end{figure}
\begin{figure}
 \centering
 \includegraphics{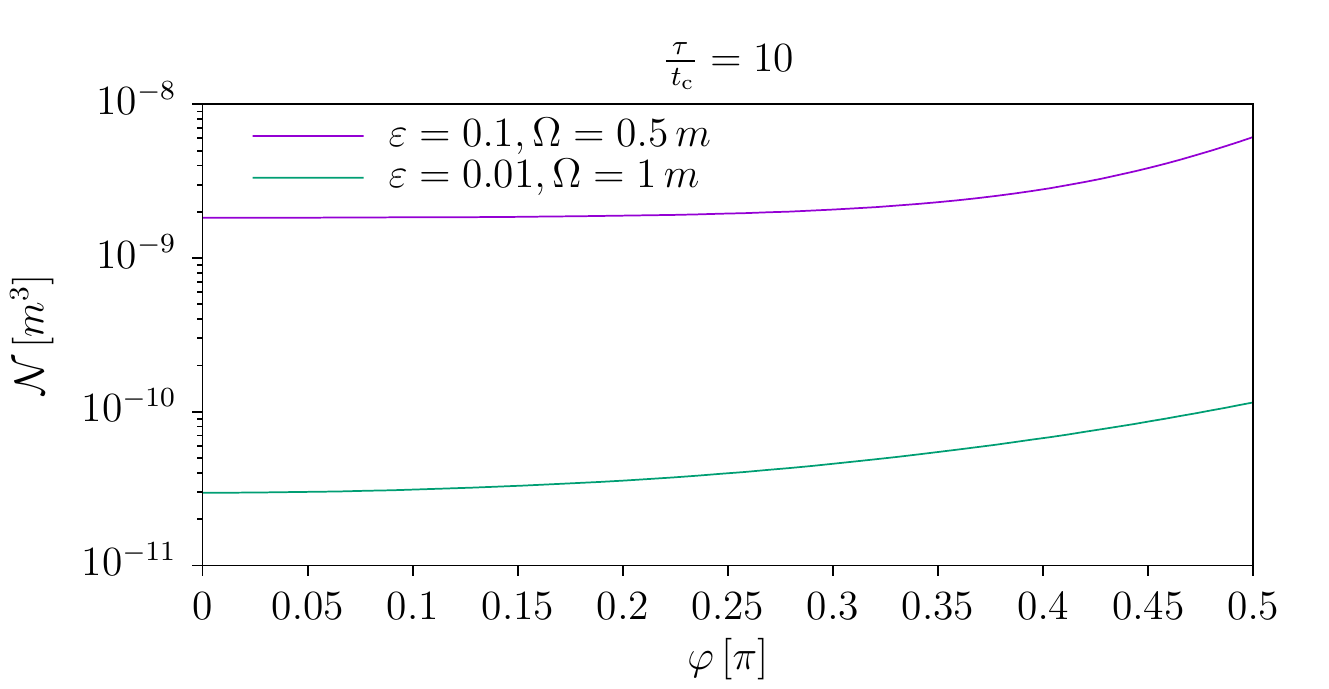}
 \caption[Particle yield in general polarization (2/2)]{
   Total particle yield in general polarization for spectra shown in \Fig\ref{fig:elli_multi_2}.
 }
 \label{fig:elli_total_2}
\end{figure}
A set of example spectra are depicted in \Figs\ref{fig:elli_multi_1} and \ref{fig:elli_multi_2}.
The spectra for the elliptical fields interpolate smoothly between the spectra for the corresponding linear or rotating fields.
Depending on the variety of features of the spectra in the extreme cases, the intermediate cases can also show a lot of variety.

If the number of rotations $\sigma$ is large and the spectrum in the rotating case is cylindrically symmetric (see \Fig\ref{fig:elli_multi_1}c)), the intermediate spectra (e.\,g. \Fig\ref{fig:elli_multi_1}b)) develop this symmetry already for moderate $\varphi$.
In other cases, for example \Fig\ref{fig:elli_multi_1}e), no cylindrical symmetry appears but the pair production spectrum resembles a collimated electron bunch.

The pair production spectra in \Figs\ref{fig:elli_multi_2}a) to c) have parameters quite similar to those in \Figs\ref{fig:elli_multi_1}d) to f).
Only the frequency is slightly higher, which in the linear case (\Figs\ref{fig:elli_multi_1}d) and \ref{fig:elli_multi_2}a)) which is enough of a difference to have two symmetric peaks in one case and one centered peak in the other.
Two symmetric peaks for the lower frequency can also be interpreted as a minimum at $\vv{p}=0$ which is present also in the elliptic case.
The collimated bunch still exists in the higher frequency case.

\pagebreak
Additional spectra corresponding to \Figs\ref{fig:elli_multi_2}a) to c) with $\varphi$ covering in the interval $[0,\nicefrac{\pi}{2}]$ are printed on the top right of the odd pages in this dissertation and can be viewed like a flip-book.
\enlargethispage{-\baselineskip}

In \Figs\ref{fig:elli_multi_2}d) to f) the amplitude of $\varepsilon=0.01$ is lower, completely suppressing Schwinger pair production.
In combination with the sufficiently large $\Omega=1\,m$ those spectra clearly demonstrate multiphoton pair production.

For the spectra in \Fig\ref{fig:elli_multi_1} only two dimensional data with $p_z=0$ are available and \Fig\ref{fig:elli_total_1} displays the two dimensional particle yield $\mathcal{N}_{xy\,}$, as defined in \Eqref{eqn:numberofparticles2d}, for varying ellipticity.
The total yield in the spectra for $\tau=61\,\tc$ has a minimum at $\varphi\approx0.3\,\pi$ as can be seen in \Fig\ref{fig:elli_total_1}, while the yield with a pulse duration of  $\tau=10\,\tc$ is monotonic.
For the spectra in \Fig\ref{fig:elli_multi_2} the complete three dimensional dataset was calculated and the total yield is displayed in \Fig\ref{fig:elli_total_2}.
The question if the minimum in the $\tau=61\,\tc$ case exists only in the two dimensional particle yield or also in the total particle yield can not be answered with the available data.

\section{Chirped Pulses}
\begin{figure}[p]%
\begin{leftfullpage}
 \centering
 \includegraphics{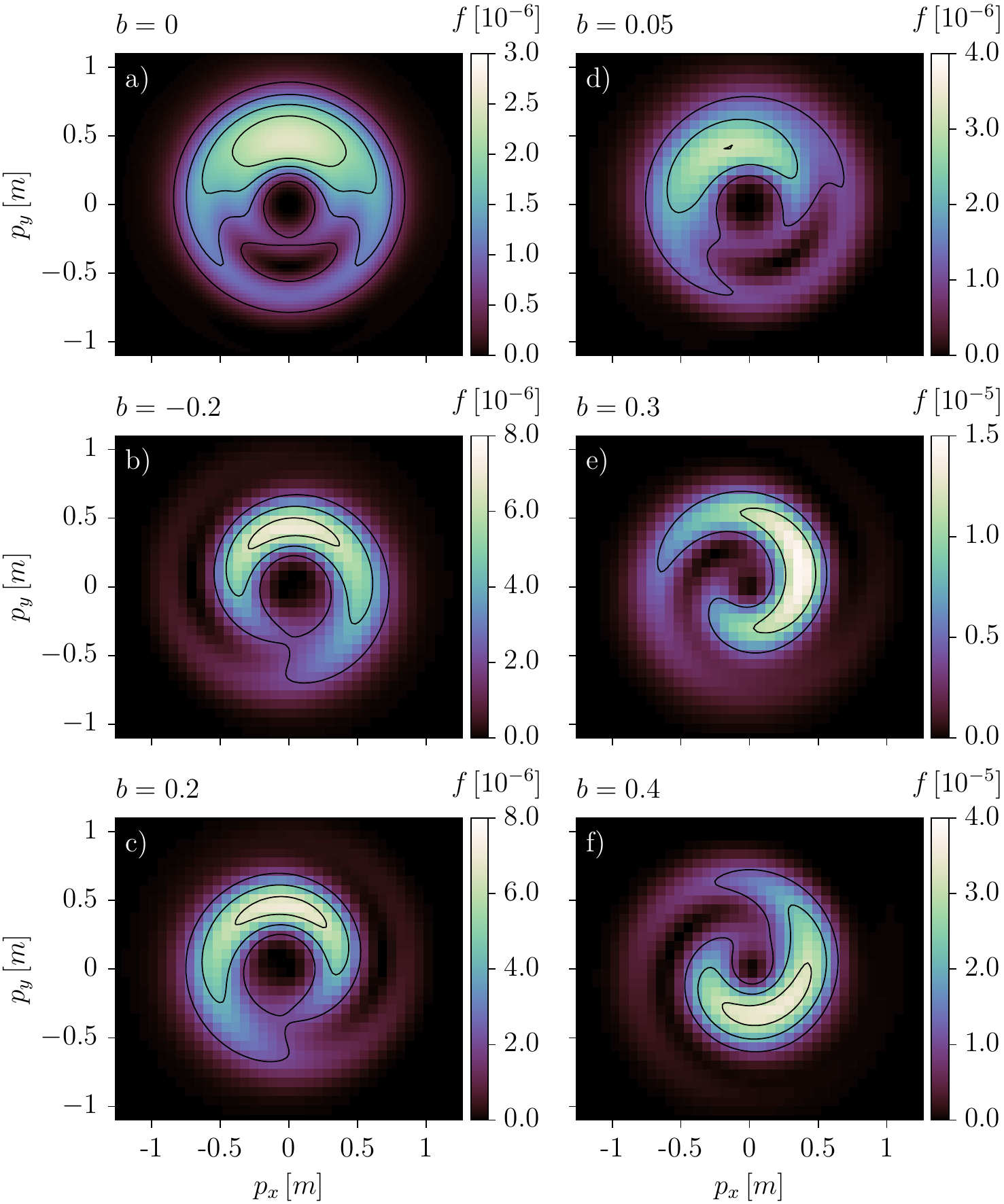}
 \caption[Spectra of pair production in chirped pulses (1/2)]{
   A set of example spectra of pair production in chirped pulses using the model in \Eqref{eqn:pulse_chirped}.
   The levels of the contour lines are indicated by the marks of the color box.
   The Pulse duration is $\tau=10\,\tc$ with a frequency of $\omega=0.6\,m$\,, and consequently $\sigma=6$.
 }
 \label{fig:chirp_multi_1}
\end{leftfullpage}
\end{figure}
\begin{figure}[p]%
\begin{fullpage}
 \centering
 \includegraphics{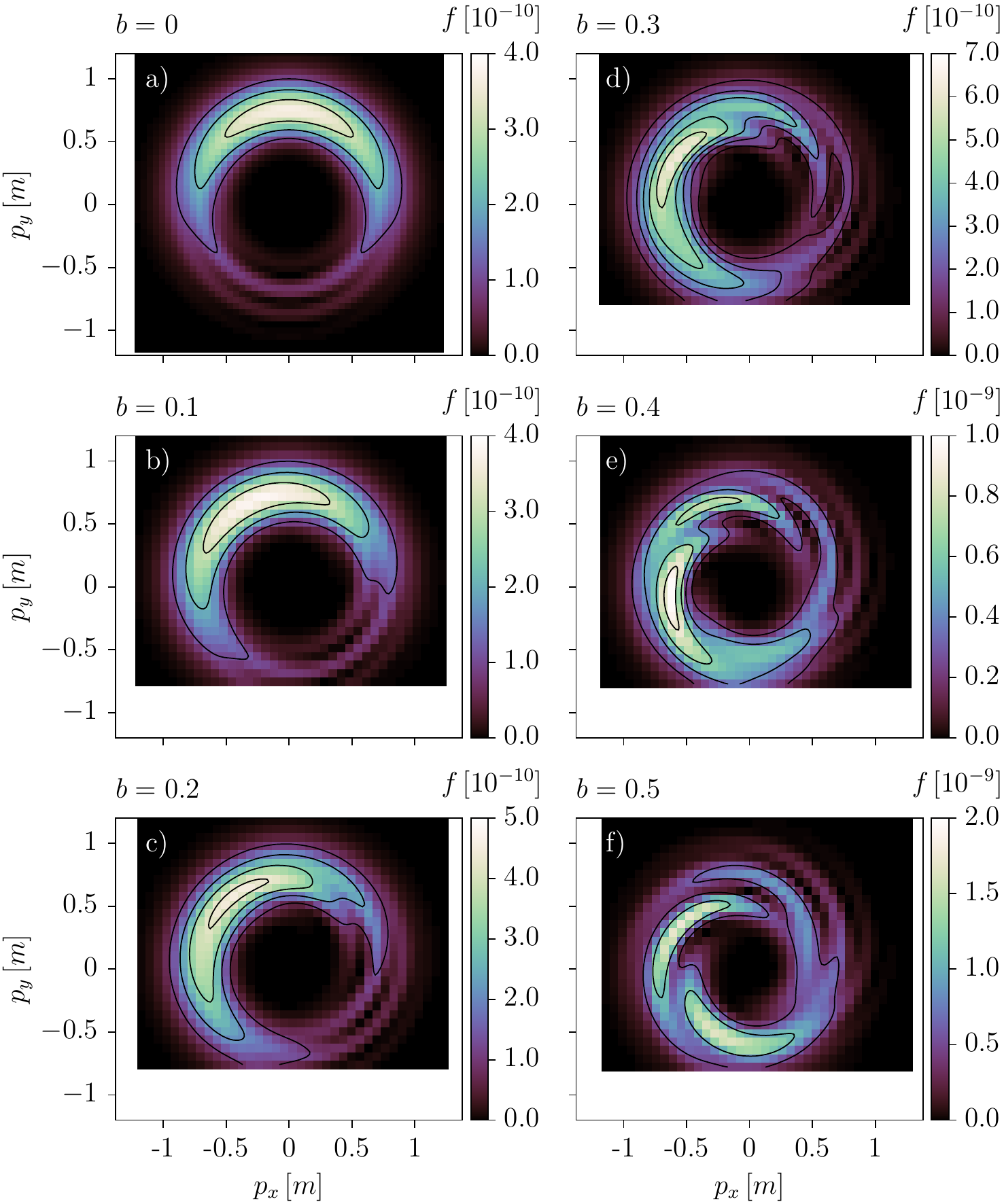}
 \caption[Spectra of pair production in chirped pulses (2/2)]{
   A set of example spectra of pair production in chirped pulses using the model in \Eqref{eqn:pulse_chirped}.
   The levels of the contour lines are indicated by the marks of the color box.
   The Pulse duration is $\tau=30\,\tc$ with a frequency of $\omega=\frac{7}{30}m\approx0.23\,m$\,, and consequently $\sigma=7$.
 }
 \label{fig:chirp_multi_2}
\end{fullpage}
\end{figure}
Laser pulses that were amplified using a technique called chirped pulse amplification (CPA) may have a residual chirp after compression.
A chirped pulse offers a broader bandwidth of available photon energies for pair production and might introduce interesting effects.
Using the Wigner method it is possible to calculate pair production spectra for those kinds of pulses.
A model for a chirped Gaussian pulse is
\begin{align}
  \label{eqn:pulse_chirped}
  \vv{E}(t) &=\varepsilon\,E_\mathrm{crit.}\,e^{-\left(\frac{t}{\tau}\right)^2}
    \begin{pmatrix}
      \cos\left[\sigma\left(b\left(\frac{t}{\tau}\right)^2 + \frac{t}{\tau}\right)\right]
      \\[5pt]
      \sin\left[\sigma\left(b\left(\frac{t}{\tau}\right)^2 + \frac{t}{\tau}\right)\right]
      \\
      0
    \end{pmatrix}\,.
\end{align}

Using this model a few examples have been calculated and show quite complex behavior of pair production in chirped pulses, see \Figs\ref{fig:chirp_multi_1} and \ref{fig:chirp_multi_2}.
The chirp distorts the spectrum a lot and also increases the particle yield as higher photon modes enter the pulse spectrum.
When the sign of $b$ is flipped, the spectrum is flipped $p_x\to-p_x$ as can be seen in \Fig\ref{fig:chirp_multi_1}b) and c).

\section{Bichromatic Fields} %
\label{sec:Bichromatic}
A popular idea in the field of pair production in strong fields is dynamically assisted pair production.
It is based on superposing a low amplitude, high frequency pulse and a high amplitude, low frequency pulse.
The result would be a deliberate combination of multiphoton and Schwinger pair production.
The tunneling process is helped by absorbing some photons, resulting in a lower tunneling barrier.
If this is the case, then Schwinger pair production should still have an influence on where the particles end up in the spectrum and pair production should be strongest at those $t$ where the field strength peaks.

\subsection{Bichromatic Rotating Fields}
Assume a superposition of two rotating Sauter pulses with amplitudes $\varepsilon_{0,1}$, frequency ratio $\abs{n}>1$ and phase shift $\varphi$
\begin{equation}
   \label{eqn:bichromatic_rotating_sauter}
   \vv{E}(t)=\frac{E_\crit}{\cosh\left( \frac{t}{\tau} \right)^2} \begin{pmatrix}
                    \varepsilon_0 \cos(\Omega t) + \varepsilon_1 \cos(n\Omega t+\varphi) \\
                    \varepsilon_0 \sin(\Omega t) + \varepsilon_1 \sin(n\Omega t+\varphi)\\
                    0
                  \end{pmatrix}\,.
\end{equation}
We will call the $\varepsilon_0$ component the ``fundamental'' pulse and the $\varepsilon_1$ component the ``harmonic'' pulse, having integer $n$ in mind.
The frequency ratio does not necessarily have to be integer, but for simplicity we will restrict ourselves to integer $n$ most of the time.
If $n$ is negative, the two components are counter\hyp{}rotating.
Atomic ionization with counter- and co\hyp{}rotating circularly polarized lasers has been experimentally studied in \Ref\cite{PhysRevA.91.031402,PhysRevA.93.023415}.
Both components on their own could be classified according to the scheme explained in \Sec\ref{sec:typical_spectra}.
Depending on the relation of those two classifications, the outcome in the superposed case could be quite different.

To be close to the initial idea, superposing a low amplitude, high frequency pulse and a high amplitude, low frequency pulse, we should choose parameters such that pair production by the fundamental pulse would be dominated by the Schwinger effect.
The harmonic pulse should at least fall into an intermediate regime, but does not have to allow for a lot of multiphoton pair production on its own.
The dotted line in \Fig\ref{fig:omega_tau_1} shows, that for a small amplitude of $\varepsilon_1=0.01$ the relation $\gamma^*\gg1$ of the combined Keldysh parameter  holds for the majority of parameter choices.

According to the Schwinger effect, pair production should occur where the field strength has a maximum.
In the superposition of pulses given in \Eqref{eqn:bichromatic_rotating_sauter}, the absolute field strength has not only one global maximum, but a set of local maxima as well.
If the envelope is disregarded and
\begin{equation*}
  \vv{E}_0(t)=E_\crit \begin{pmatrix}
                    \varepsilon_0 \cos(\Omega t) + \varepsilon_1 \cos(n\Omega t+\varphi) \\
                    \varepsilon_0 \sin(\Omega t) + \varepsilon_1 \sin(n\Omega t+\varphi)\\
                    0
                  \end{pmatrix}
\end{equation*}
is considered, it follows that the electric field strength
\begin{equation*}
  \lvert\vv{E}_0(t)\rvert = E_\crit \sqrt{\varepsilon_0^2+\varepsilon_1^2+2\varepsilon_0\varepsilon_1\cos\big[ \varphi+(n-1)\Omega\, t \big]}
\end{equation*}
has an infinite number of local maxima at
\begin{equation}
  \label{eqn:maxima}
  t_{0k} = \frac{2\pi k-\varphi}{(n-1)\Omega}\,,\, k\in\mathbb{Z}\,.
\end{equation}
This corresponds to
\begin{equation}
  \label{eqn:bichromatic_maxima_count}
  N=\abs{n-1}
\end{equation}
maxima for each cycle of the fundamental wave.
In the pulsed case this is still true where $\abs{t_{0k}}\lesssim\tau$.
In that case the location of the maxima can be calculated in a good approximation from \Eqref{eqn:maxima} via series expansion of $\lvert\vv{E}\rvert$ up to linear order at $t_{0k}$ and making use of $\tanh(x)\approx x$ which results in
\begin{equation*}
  t_k = t_{0k}\left( 1-\frac{1}{1-3\left(\frac{t_{0k}}{\tau} \right)^2+\frac{(n-1)^2\varepsilon_0\varepsilon_1\sigma^2}{2\left( \varepsilon_0+\varepsilon_1 \right)^2}} \right)\,.
\end{equation*}
When pairs are produced at those times and gain momentum only by classical acceleration, their final momenta are given by the vector potential as in
\begin{equation}
  \label{eqn:final-peaks}
  \vv{p}_{\hspace{-2pt}k} =\int_{t_k}^\infty e\vv{E}(t)\,\dd t = \vv{A}(t_k) - \vv{A}(\infty) = \vv{A}(t_k)\,.
\end{equation}
We expect pairs to be produced in the vicinity of those points as is the case in atomic ionization \autocite{PhysRevA.91.031402,PhysRevA.93.023415}.

\subsubsection{Dynamically Assisted Schwinger Pair Production in Rotating Sauter Pulses}
\begin{figure}
 \centering
 \includegraphics{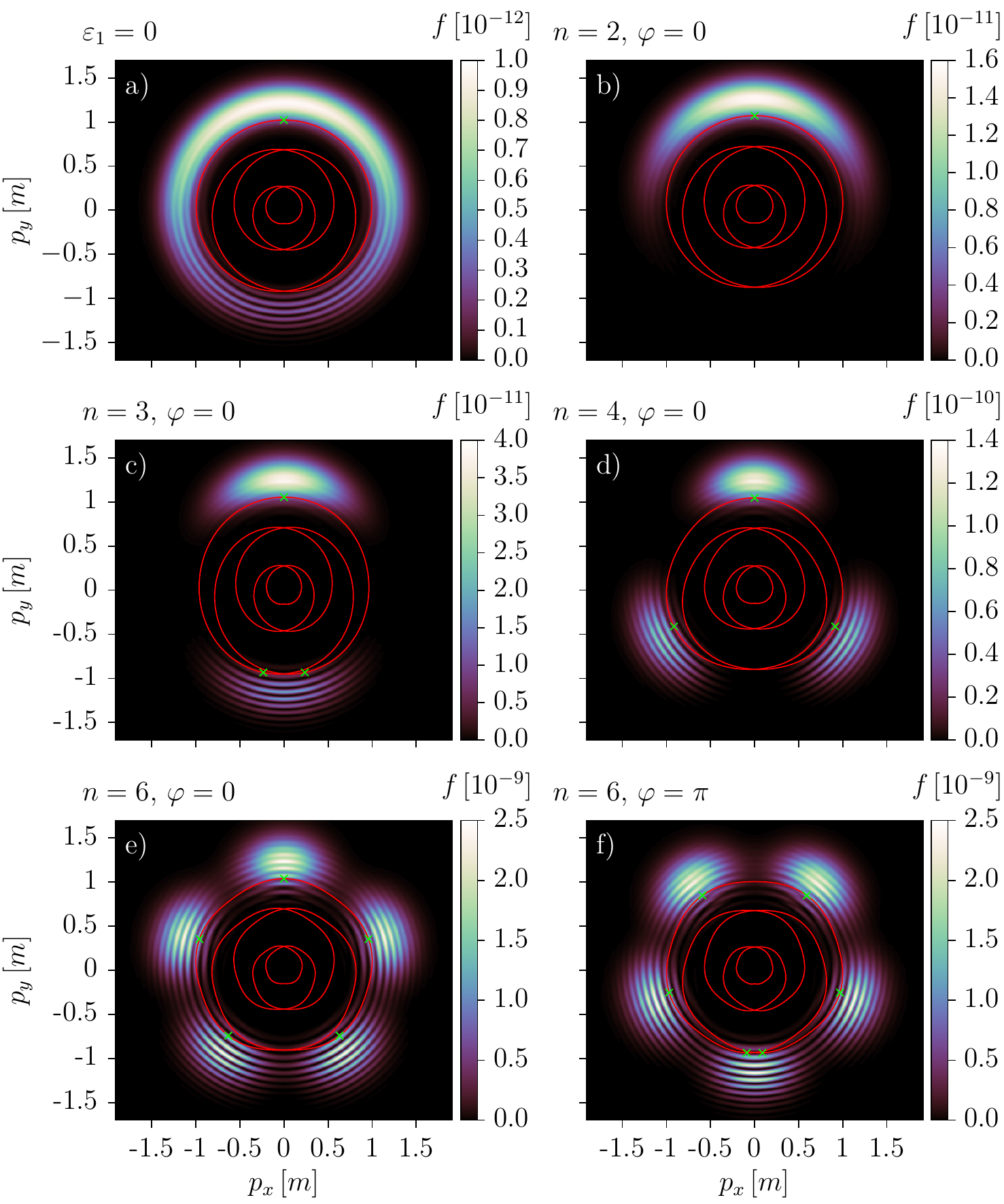}
 \caption[Example spectra for dynamically assisted Schwinger pair production in co\hyp{}rotating Sauter pulses.]{
   Example spectra for dynamically assisted Schwinger pair production in co\hyp{}rotating Sauter pulses with harmonic order $n$.
   The red lines display the locus of the vector potential.
   The green crosses give the predicted pair production peaks as given by \Eqref{eqn:final-peaks}.
   Amplitudes are $\varepsilon_0=0.1$ and $\varepsilon_1=0.01$ at a pulse duration of $\tau=100\,\tc$ and fundamental frequency $\Omega=0.1\,m$.
 }
 \label{fig:2dtau100-0}
\end{figure}
Choosing a fundamental that is dominated by Schwinger pair production and a weaker harmonic that is dominated by multiphoton pair production, the numerical results  confirm the predictions made in the previous section.

The fundamental pulse has the amplitude $\varepsilon_0=0.1$, the duration $\tau=100\,\tc$ and the frequency $\Omega=0.1\,m$.
Its resulting spectrum was already shown in \Fig\ref{fig:typical_spectra_2b}a) and it was assigned interpretation \#4, as it exhibits mostly Schwinger pair production.
The spectrum, together with the locus of the vector potential is again displayed in \Fig\ref{fig:2dtau100-0}a).

The harmonic order is varied from 2 to 6, resulting in harmonic frequencies of $n\Omega=0.2m,\dotsc,0.6m$.
Combined with the lower amplitude of $\varepsilon_1=0.01$ this puts pair production by the harmonic pulses into the multiphoton regime.
On its own the harmonic pulses with $n\leq5$ have their distribution function $f$ below the numeric accuracy of $10^{-14}$, meaning no ascertainable pair production.
In the $n=6$ case the harmonic pulse with frequency $\Omega_\mathrm{harm.}=6\Omega=0.6\,m$ produces a multiphoton ring at $|\vv{p}|=\frac12\sqrt{\left( 4\cdot0.6\,m \right)^2-\left( 2m^* \right)^2}=0.663\,m$ with $m^*=1.00014\,m$.

\begin{figure}
 \centering
 \includegraphics{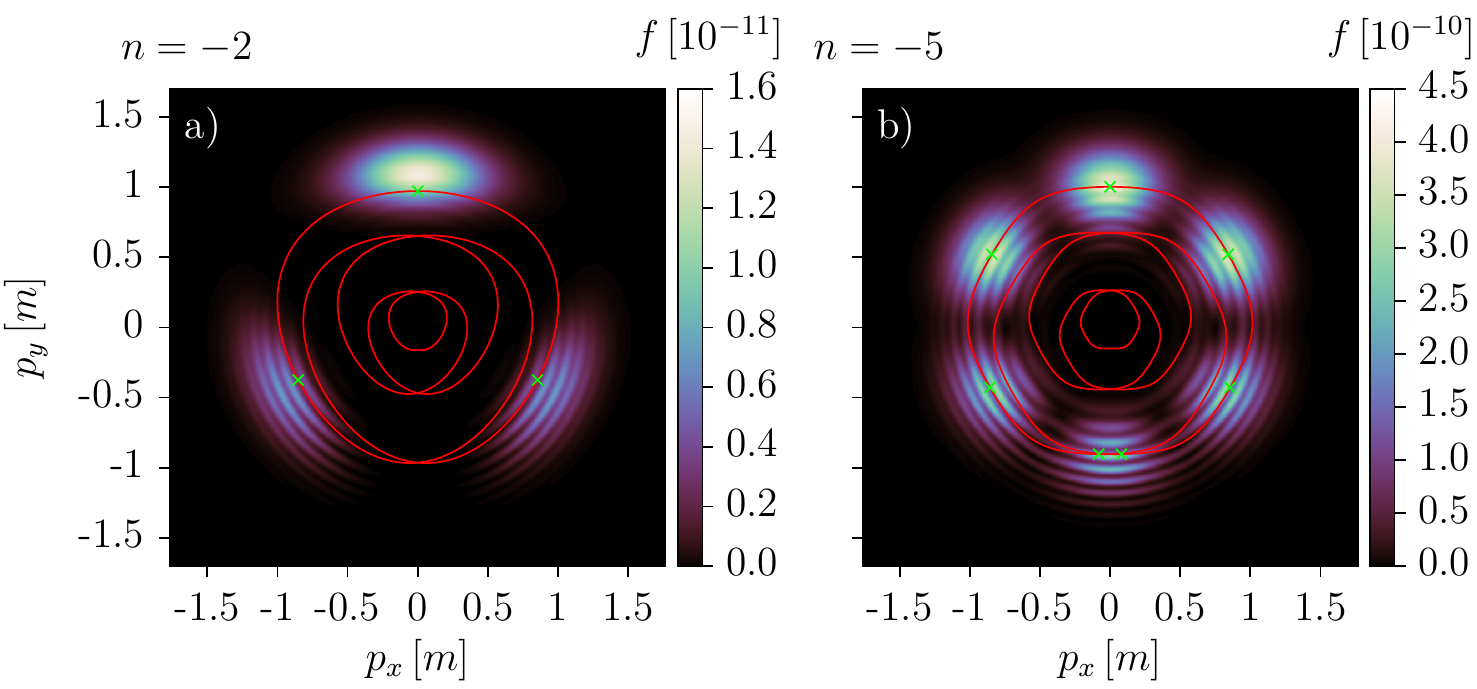}
 \caption[Example spectra for dynamically assisted Schwinger pair production in counter\hyp{}rotating Sauter pulses.]{
   Example spectra for dynamically assisted Schwinger pair production in counter\hyp{}rotating Sauter pulses with harmonic order $n$.
   The red lines display the locus of the vector potential.
   The green crosses give the predicted pair production peaks as given by \Eqref{eqn:final-peaks}.
   Amplitudes are $\varepsilon_0=0.1$ and $\varepsilon_1=0.01$ at a pulse duration of $\tau=100\,\tc$ and fundamental frequency $\Omega=0.1\,m$.
   Subplot a) shows the pair production spectrum by the fundamental pulse only.
 }
 \label{fig:2dtau100-counter}
\end{figure}

As can be observed in \Figs\ref{fig:2dtau100-0}b) to f), the produced pairs are close to the predicted momenta, while the deviations behave much like it was the case with only the fundamental pulse.
The number of pair production peaks in the spectrum is in accordance with \Eqref{eqn:bichromatic_maxima_count}.
Changing the relative phase $\varphi$ results in rotating the positions of the pair production peaks in the spectrum as can be seen in \Figs\ref{fig:2dtau100-0}e) and f).
\Fig\ref{fig:2dtau100-counter} shows that this is also true for counter\hyp{}rotating fields $n<0$.
\begin{figure}
 \centering
 \includegraphics{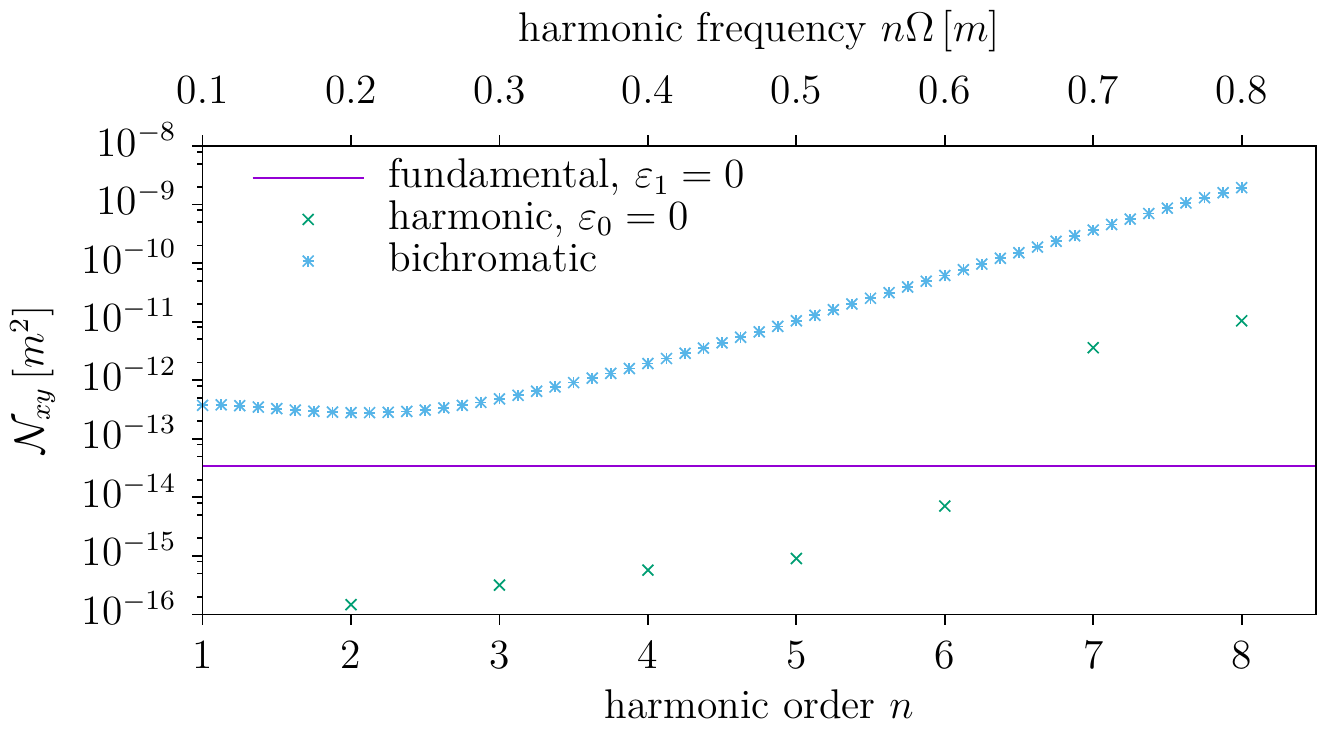}
 \caption[Pair production yield of dynamically assisted Schwinger pair production in rotating Sauter pulses.]{
   Pair production yield of dynamically assisted Schwinger pair production in rotating Sauter pulses.
   Amplitudes are $\varepsilon_0=0.1$ and $\varepsilon_1=0.01$ at a pulse duration of $\tau=100\,\tc$ and fundamental frequency $\Omega=0.1\,m$.
   Two dimensional integral over particle yield.
   Example spectra given in \Fig\ref{fig:2dtau100-0}.
 }
 \label{fig:bichromatic_rotating_long_yield}
\end{figure}
Spectra similar to \Figs\ref{fig:2dtau100-0}b) and \ref{fig:2dtau100-counter}a) have already been observed in atomic ionization using a superposition of a circularly polarized laser beam and its second harmonic in a co- and counter\hyp{}rotating fashion \autocite{PhysRevA.91.031402}.
The strong enhancement of pair production in this dynamically assisted Schwinger effect is displayed in \Fig\ref{fig:bichromatic_rotating_long_yield}.
The data of the particle yield of the bichromatic field extend to $n=1$.
At $n=1$ the fundamental and harmonic pulse are, apart from their amplitude, the same and their superposition is nothing but the fundamental pulse with a slightly higher amplitude.
According to the Schwinger formula from \Eqref{eqn:schwinger_formula} this slightly higher amplitude alone results in more than one magnitude of enhancement in pair production yield, which is in accordance with the numerical data.
This is one more indicator that Schwinger pair production is indeed the dominant process in the fundamental pulse.

\pagebreak
\subsubsection{Phase Dependency in Short Bichromatic Rotating Sauter Pulses}
\begin{figure}
 \centering
 \includegraphics{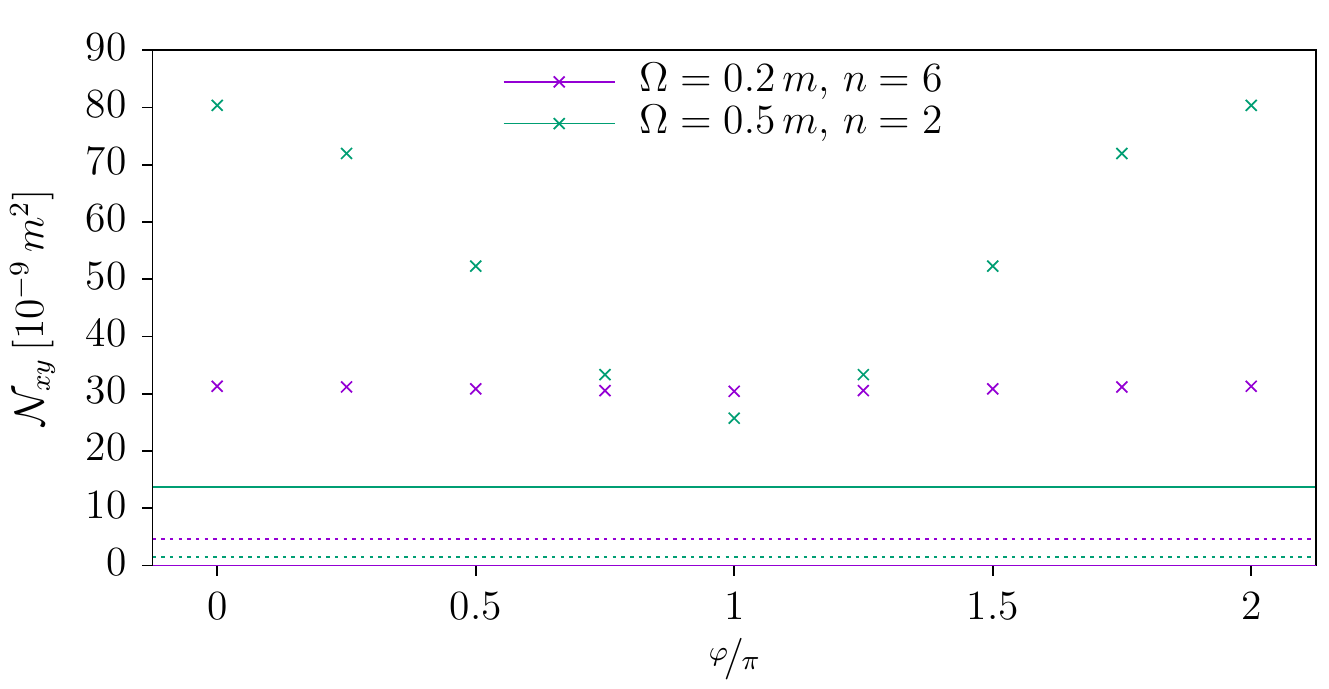}
 \caption[Particle yield in bichromatic rotating Sauter pulses]{
   Particle yield in bichromatic rotating Sauter pulse, depending on relative phase $\varphi$.
   Solid lines: particle yield by fundamental pulse.
   Dashed lines: particle yield by harmonic pulse.
   Crosses: particle yield by bichromatic pulse.
 }
 \label{fig:bichromatic_rotating_short_yield}
\end{figure}
\begin{figure}
 \centering
 \includegraphics{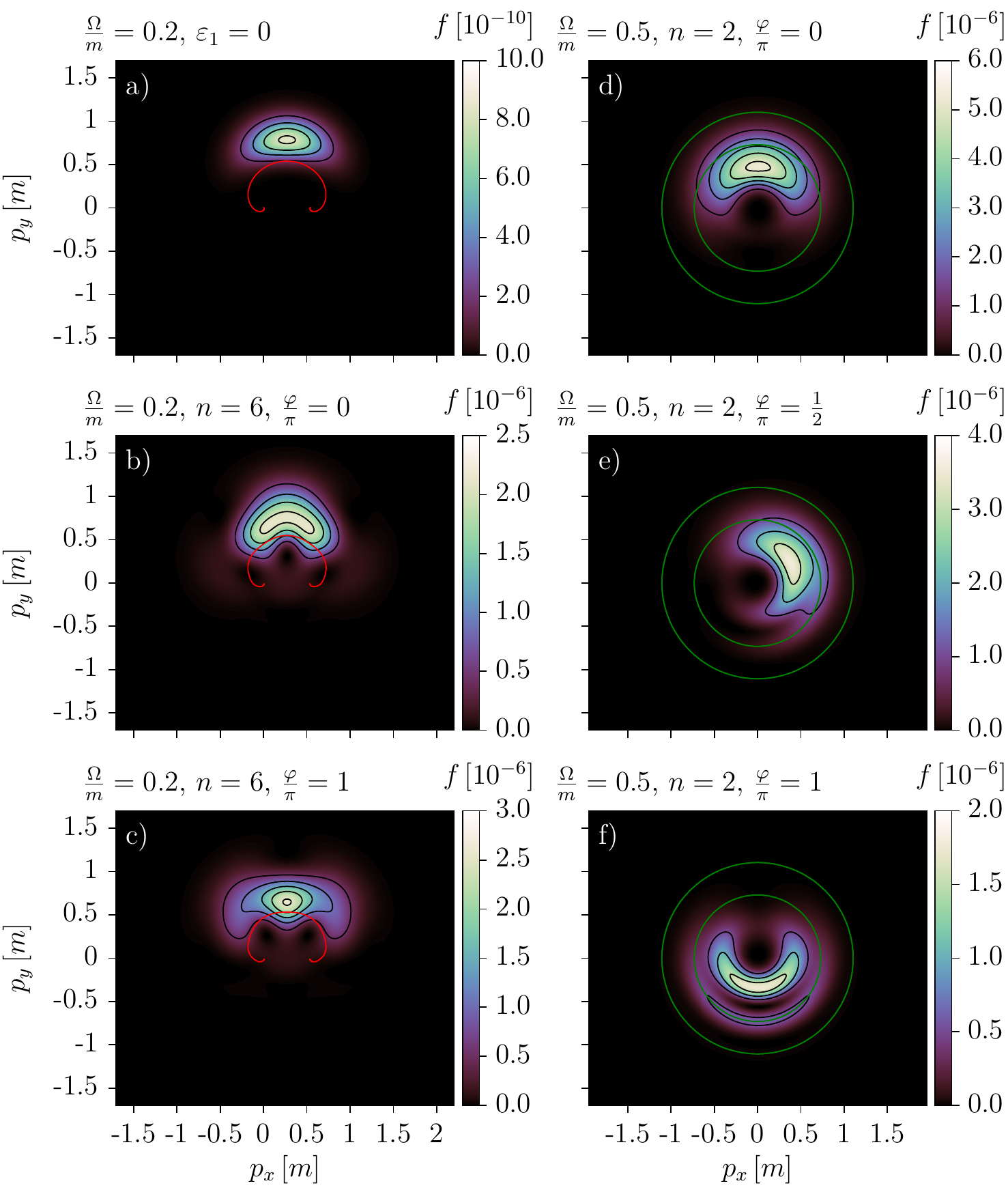}
 \caption[Spectra for a selection of bichromatic rotating Sauter pulses with varying relative phase $\varphi$]{
   Spectra for a selection of bichromatic rotating Sauter pulses with varying relative phase $\varphi$.
   The levels of the contour lines are indicated by the marks of the color box.
   The red lines display the locus of the vector potential.
   The green circles show the expected position of pairs due to multiphoton pair production.
   Amplitudes are $\varepsilon_0=0.1$ and $\varepsilon_1=0.01$ at a pulse duration of $\tau=10\,\tc$.
   Subplots a) to c) have fundamental frequency $\Omega=0.2\,m$ and harmonic order $n=6$ while
   subplots d) to f) have $\Omega=0.5\,m$ and $n=2$.
 }
 \label{fig:bichromatic_rotating_short_multi}
\end{figure}
In short rotating Sauter pulses, the variation introduced by changing the relative phase $\varphi$ is much larger.
Due to the short pulse duration already the fundamental pulse is influenced by multiphoton pair production.
In the first example, when the relative phase is changed from $\varphi=0$ in \Fig\ref{fig:bichromatic_rotating_short_multi}b) to $\varphi=\pi$ in \Fig\ref{fig:bichromatic_rotating_short_multi}c) the pair production peak gets much narrower, resulting in a more collimated electron bunch.

In the second example, by changing the relative phase, a multiphoton signature, that is not visible at $\varphi=0$ in \Fig\ref{fig:bichromatic_rotating_short_multi}d), can be made directly visible at $\varphi=\pi$ in \Fig\ref{fig:bichromatic_rotating_short_multi}f).
In the latter spectrum  a partial multiphoton ring is clearly present.
As can be seen in \Fig\ref{fig:bichromatic_rotating_short_yield} the particle yield is nearly constant in the first case, while in the second case it is drastically lowered when $\varphi\approx\pi$.

\FloatBarrier
\subsection{Bichromatic Linear Fields}
Bichromatic linear fields have already been studied in a number of cases \autocite{Schutzhold:2008pz,Dunne:2009gi,Fey:2011if,Akal:2014eua}.
While previous studies focused on infinitely oscillating fields or pulsed fields with a uniform envelope, we will analyze examples of bichromatic, linear oscillating Sauter pulses.
The electric field for these studies is given by
\begin{equation*}
  E_x(t) = \frac{E_\crit}{\cosh\left( \nicefrac{t}{\tau} \right)^2}\bigl( \varepsilon_0\cos(\Omega t+\varphi_0) + \varepsilon_1\cos(n\Omega t+\varphi) \bigr)\,,
\end{equation*}
where not only the relative phase $\delta\varphi=\varphi-\varphi_0$, but also an overall carrier envelope phase $\varphi_0$ is included.
However it is expected that this does only play a role for few cycle pulses, so we will set $\varphi=0$.

As the pair production spectra have the same symmetries as the oscillating Sauter pulses studied in \Ref\cite{Kohlfurst:2013ura}, we can calculate the particle yield according to \Eqref{eqn:numberofparticles_isotropic} using
\begin{align}
  \label{eqn:numberofparticles_linear}
  \mathcal{N}&=\frac{1}{\left( 2\pi \right)^2}\int_{-\infty}^{\infty} \dd{p_x} \int_0^{\infty} \dd{p_y}  p_y\,f(p_y,p_x)\,.
\end{align}

\subsubsection{Dynamically Assisted Schwinger Pair Production in Linear Sauter Pulses}
\begin{figure}
 \centering
 \includegraphics{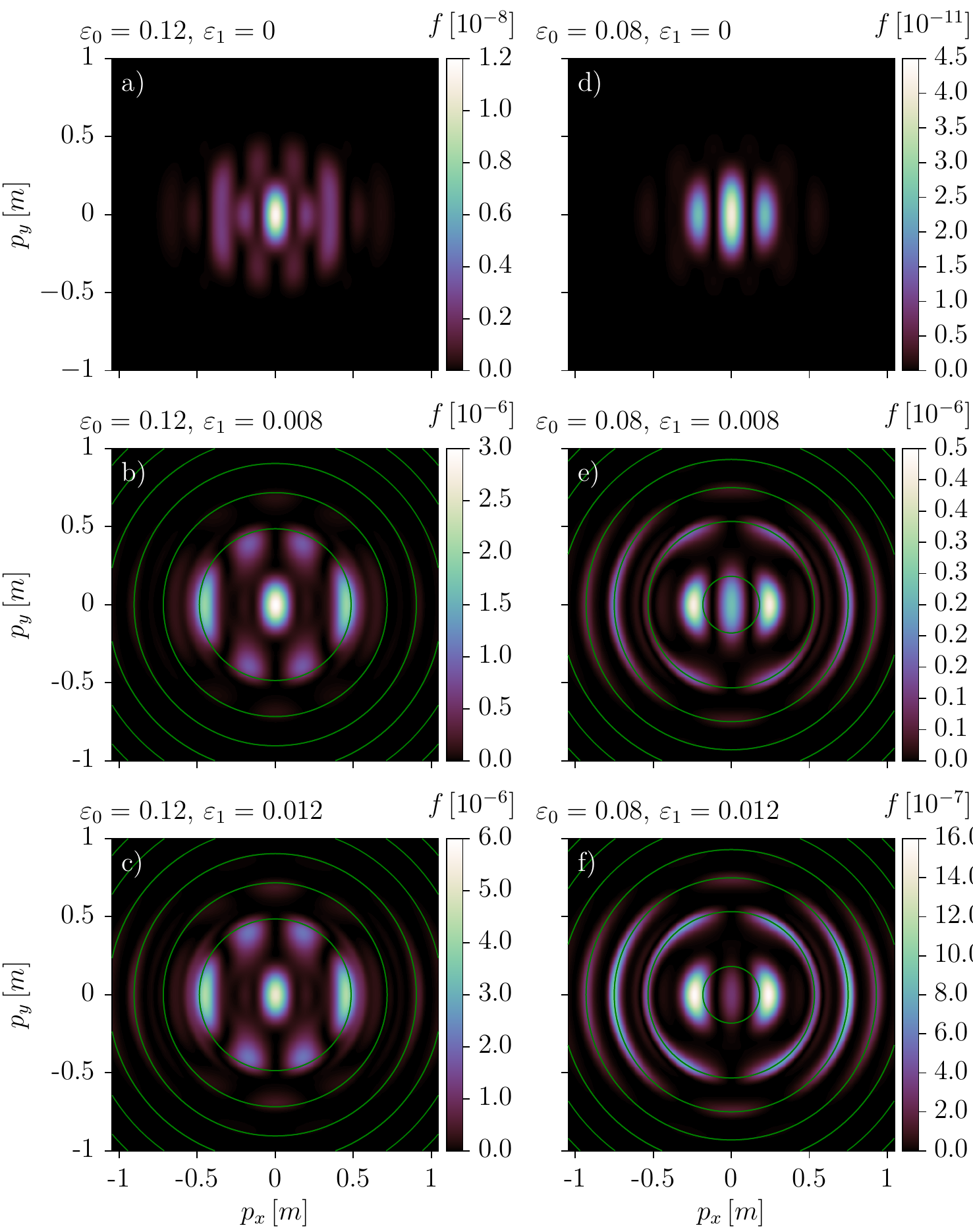}
 \caption[Spectra of pair production in bichromatic linear Sauter pulses (1/2)]{
   Bichromatic linear Sauter pulses with $\tau=50\,\tc$, $\Omega=0.23\,m$ and $n=5$ and varying field strengths.
   Panels a) and d): pairs produced by fundamental pulse only.
   Remaining panels: pairs produced by bichromatic pulses.
   The green circles, if present, show the locations of the expected multiphoton rings for appropriate values of $n$.
 }
 \label{fig:bichromatic_linear_long_multi}
\end{figure}
\begin{figure}
 \centering
 \includegraphics{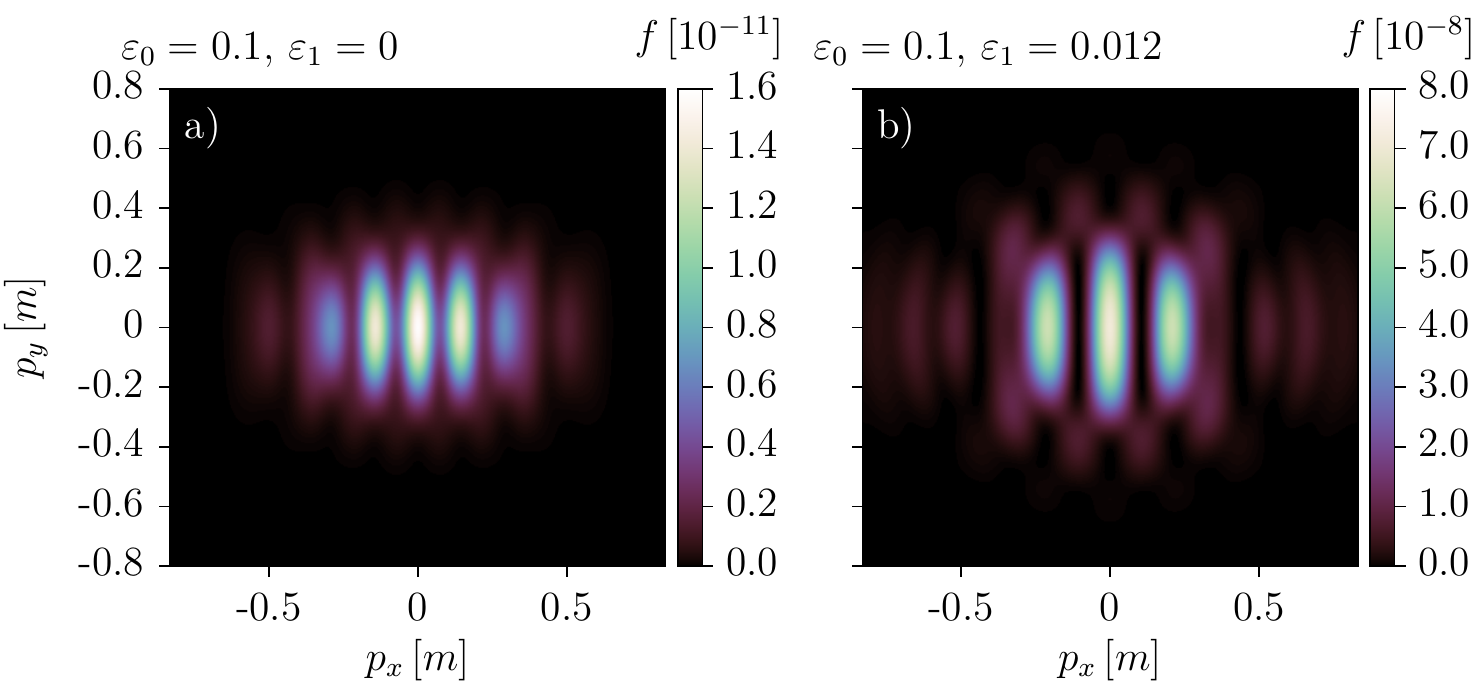}
 \caption[Spectra of pair production in bichromatic linear Sauter pulses (2/2)]{
   Bichromatic linear Sauter pulse with $\tau=50\,\tc$, $\Omega=0.15\,m$ and $n=5$.
   Left panel: pairs produced by fundamental pulse only.
   Right panel: pairs produced by bichromatic pulse.
 }
 \label{fig:bichromatic_linear_long_multi_2}
\end{figure}
\begin{figure}
 \centering
 \includegraphics{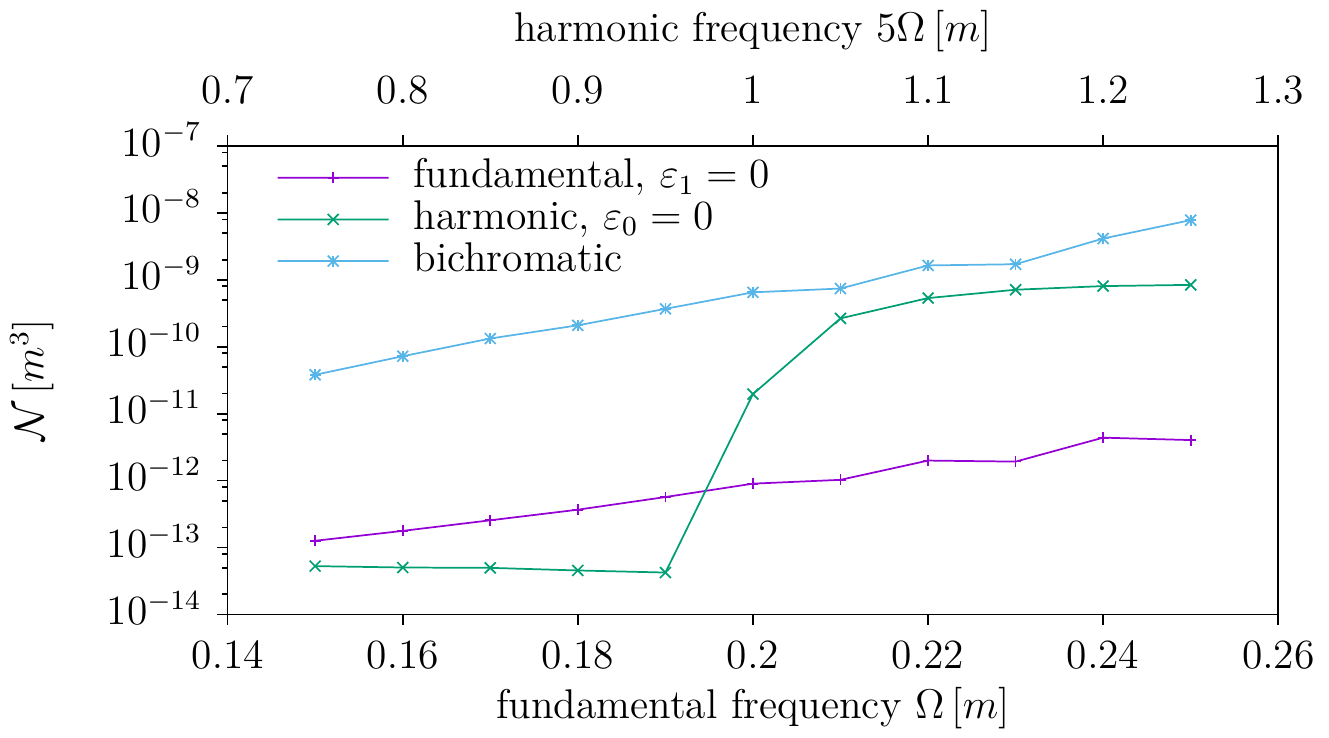}
 \caption[Dynamically assisted Schwinger pair production in linear Sauter pulses]
 {
   Dynamically assisted Schwinger pair production in linear Sauter pulses with frequency ratio $n=5$ and amplitudes $\varepsilon_0=0.12$ and $\varepsilon_1=0.01$.
   Note the two\hyp{}photon resonance of the harmonic wave at $5\Omega=1\,m$.
 }
 \label{fig:total-tau50-1d}
\end{figure}
Two different frequencies of the fundamental pulse are considered, $\Omega=0.23\,m$ in \Fig\ref{fig:bichromatic_linear_long_multi} and $\Omega=0.15\,m$ in \Fig\ref{fig:bichromatic_linear_long_multi_2}.
In both cases the pulses are paired with their 5th harmonic at a pulse duration of $\tau=50\,\tc$.
The lowest Keldysh parameter for the fundamental pulse in these examples is $\gamma_\Omega=1.5$ in \Fig\ref{fig:bichromatic_linear_long_multi_2}a), the highest is $\gamma_\Omega=2.88$ in \Fig\ref{fig:bichromatic_linear_long_multi}d).
We can thus expect beginning multiphoton pair production under the influence of the vector potential, which explains the absence of rings in \Fig\ref{fig:bichromatic_linear_long_multi}a) and the qualitative change of the spectrum when the amplitude is changed in \Fig\ref{fig:bichromatic_linear_long_multi}d).

When the harmonic wave is added, the changes to the shape of the pair production spectrum are not fundamental, however additional features might occur.
In all cases, as expected, the addition of the higher harmonic shifts the spectrum toward multiphoton behavior.
On the one hand this effect is stronger in those cases with a lower amplitude for the fundamental pulse, which accounts for a bigger Keldysh parameter, which can be observed by comparing \Fig\ref{fig:bichromatic_linear_long_multi}b) to \Fig\ref{fig:bichromatic_linear_long_multi}e) or \Fig\ref{fig:bichromatic_linear_long_multi}c) to \Fig\ref{fig:bichromatic_linear_long_multi}f).
On the other hand this effect is also stronger when the amplitude of the harmonic pulse is increased, which can be seen by comparing \Fig\ref{fig:bichromatic_linear_long_multi}b) to \Fig\ref{fig:bichromatic_linear_long_multi}c) or \Fig\ref{fig:bichromatic_linear_long_multi}e) to \Fig\ref{fig:bichromatic_linear_long_multi}f).

The total particle yield may be strongly enhanced by the added harmonic wave as shown in \Fig\ref{fig:total-tau50-1d}.
The multiphoton resonance is visible in the particle yield of the harmonic pulse, but not in the respective data for the bichromatic pulse.
This can be understood in the context of dynamically enhanced Schwinger pair production, because the necessary number of absorbed photons does not suddenly decrease.
Instead a smaller but constant number of photons is absorbed and the increasing frequency continuously lowers the tunneling barrier.

\subsubsection{Off-Axis Particle Acceleration}
\begin{figure}
 \centering
 \includegraphics{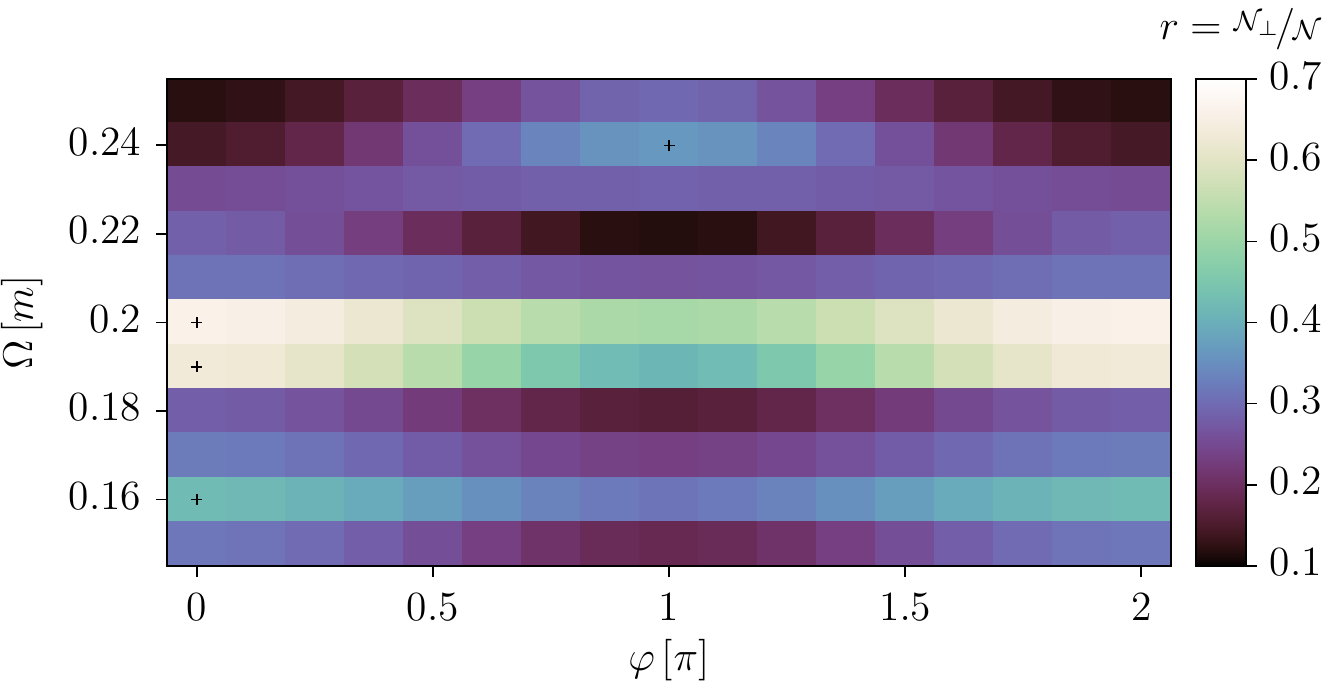}
 \caption[Transverse portion of the particle yield in bichromatic linear Sauter pulses]{
   Portion of particles accelerated in transverse direction as defined by \Eqref{eqn:transverse_part}.
   Pluses mark the parameters for the examples in \Fig\ref{fig:bichromatic_linear_long_multi_3}.
   Amplitudes are $\varepsilon_0=0.08$ and $\varepsilon_1=0.01$, harmonic order is $n=5$.
   The low resolution of the plot and apparent discontinuity w.\,r.\,t. the frequency $\omega$ is due to the limited set of available data.
 }
 \label{fig:bichromatic_linear_transverse_part}
\end{figure}
\begin{figure}
 \centering
 \includegraphics{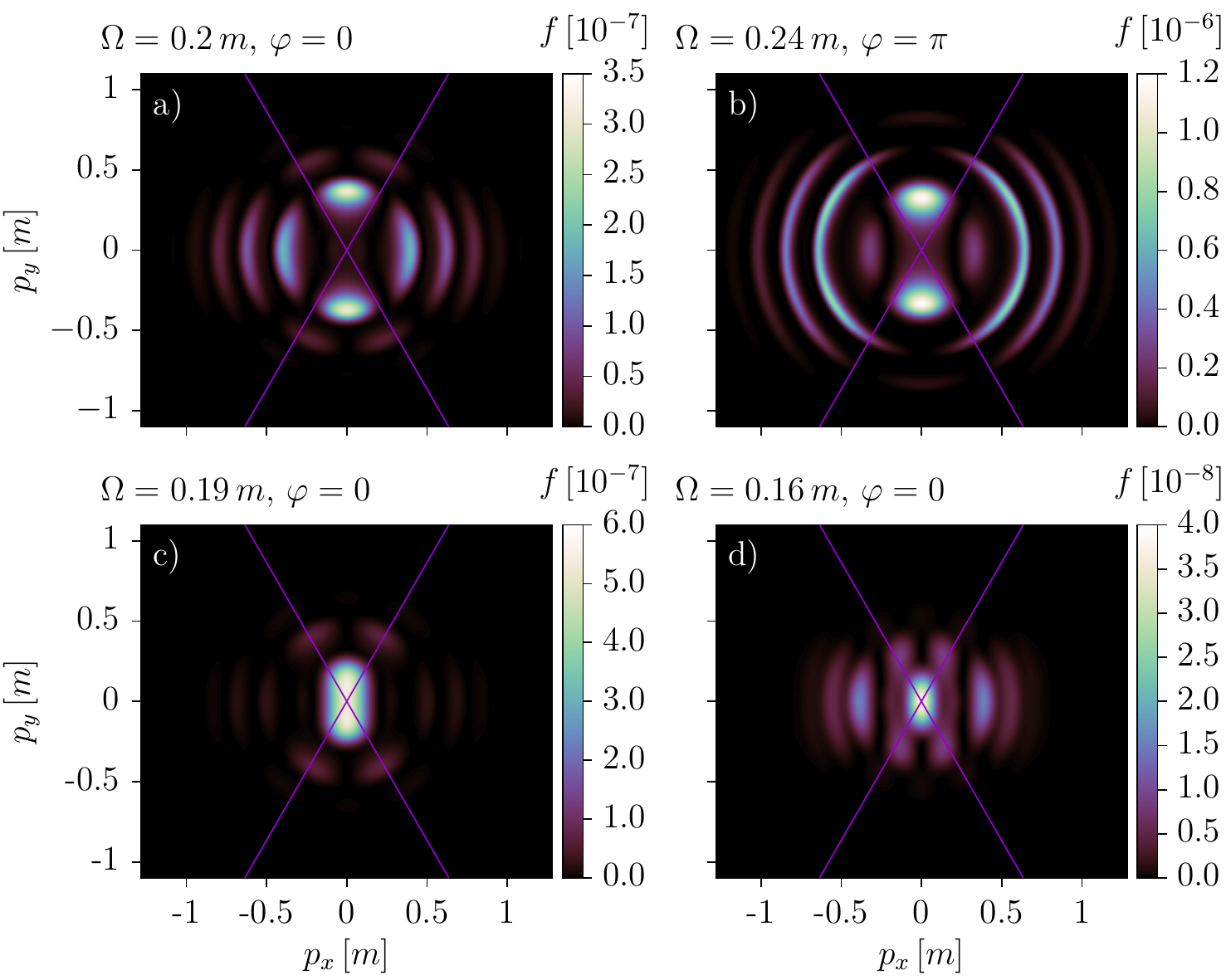}
 \caption[Example spectra from bichromatic linear Sauter pulses with transverse acceleration]{
   Example spectra from bichromatic linear Sauter pulses with transverse acceleration as marked in \Fig\ref{fig:bichromatic_linear_transverse_part}.
   Pulse duration is $\tau=50\,\tc$, amplitudes are $\varepsilon_0=0.08$ and $\varepsilon_1=0.01$, harmonic order is $n=5$.
   Purple lines display the boundary for transverse acceleration as given by $\arctan\left( \frac{|p_y|}{|p_x|} \right) = 60^\circ$.
 }
 \label{fig:bichromatic_linear_long_multi_3}
\end{figure}
An additional phenomenon in bichromatic linear Sauter pulses is that there exist a number of cases, where a large portion of the created particles are accelerated perpendicularly to the electric field.
This offers a possibility to maximize the portion of produced pairs that hit the detector in an experimental setup.
In order to quantify this behavior, we will consider every created particle to be transversely accelerated, if its momentum encloses an angle of more than $60^\circ$ with the longitudinal $p_x$ axis.

\pagebreak
Thus, by multiplying the integrand of \Eqref{eqn:numberofparticles_linear} with a Heaviside function, we can define the transverse particle yield
\begin{align*}
  \mathcal{N}_\perp \definedby
  \frac{1}{\left( 2\pi \right)^2}\int_{-\infty}^\infty \dd{p_x} \int_0^\infty \dd{p_y} p_y f(p_x,p_y) \Theta\left( \arctan\left( \frac{|p_y|}{|p_x|} \right) - 60^\circ \right) \\
  &= \frac{1}{\left( 2\pi \right)^2}\int_{-\infty}^\infty \dd{p_x} \int_0^\infty \dd{p_y} p_y f(p_x,p_y) \Theta(|p_y|-\sqrt{3}|p_x|) \\
  &= \frac{1}{\left( 2\pi \right)^2}\int_{-\infty}^\infty \dd{p_x} \int_{\frac{1}{\sqrt{3}}\abs{p_x}}^\infty \dd{p_y} p_y f(p_x,p_y)
\end{align*}
and the relative transverse particle yield
\begin{align}
  \label{eqn:transverse_part}
r \definedby \frac{\mathcal{N}_\perp}{\mathcal{N}}\,.
\end{align}

As can be seen in \Fig\ref{fig:bichromatic_linear_transverse_part}, the transverse particles peak at a number of different parameter sets.
Some example spectra are given in \Fig\ref{fig:bichromatic_linear_long_multi_3} and their parameters are marked with plus symbols in \Fig\ref{fig:bichromatic_linear_transverse_part}.
There are at least two quite different types of spectra with transverse acceleration of particles.
In some cases, collimated particle bunches with high velocity are accelerated in a transverse direction.
Two examples of this are displayed in \Fig\ref{fig:bichromatic_linear_long_multi_3}a) and b).

In other cases, particles are accelerated in all directions with a slight tendency towards the $p_\perp$\hyp{}direction.
Due to the cylinder symmetry this accounts for a large relative transverse particle yield.

  \chapter{Inclusion of Magnetic Fields}
\label{chap:magnetic}
Previous attempts of including magnetic fields in the DHW formalism have considered inhomogeneous fields \autocite{Kohlfurst:2015niu,Kohlfurst:2015zxi}.
In this case not only the special case $\vv{B}=0$ is abandoned, but also spatial homogeneity is discarded.
The latter is making calculations much more difficult, because the simplification from \Eq\eqref{eqn:diff-ops-simp} can no longer be used.
Instead the full pseudo differential operators have to be taken into account or a derivative expansion has to be executed.

Pair production in unidirectional electric fields in the presence of a collinear constant magnetic field has been studied in \Ref\cite{Tarakanov:2002mn}.
There a kinetic equation similar to QKT has been derived using a different set of basis states that accommodate for the constant magnetic field.
It is not straightforward to similarly include a temporally constant and spatially homogeneous magnetic field in the Wigner formalism, because this would introduce Landau levels and one would have to reconsider the initial condition.
On the other hand it is also technically not possible to have temporally changing magnetic fields and spatially homogeneous electric fields, because of Faraday's law of induction $\nabla\times\vv{E}=-\dot{\vv{B}}$.

When pair production in homogeneous electric fields is discussed, the homogeneous fields are often a model for fields that change over length scales that are much larger than the Compton wavelength.
If the scale of the spatial inhomogeneities of the electric field is large enough and the approximation by a locally constant field is correct, we should be able to locally calculate the pair production rate and integrate over the region where the field is localized to calculate the particle yield.
When the length scale of the inhomogeneities is decreased the total particle yield should at some point start to differ from the integrated local particle yield.
In \Ref\cite{Kohlfurst:2015niu} a spatially localized electric field with a length scale $\lambda$ was studied using the Wigner formalism.
There also a magnetic field was introduced and $\nabla\times\vv{E}=-\dot{\vv{B}}$ was kept valid.
Those results were compared to a calculation that artificially fixed $\vv{B}=0$ and it was shown that the magnetic field is important to achieve correct results for spatial inhomogeneities at scales smaller than 50 Compton wavelengths at an electric field strength of $\varepsilon=0.707$.
At the same time it was shown that the inhomogeneous result for the particle yield is the same as the result from integrating over the particle yield for locally homogeneous fields at a variation scale of 50 Compton wavelengths or larger.

When the magnetic field is changing, Faraday's law requires us to also introduce a spatial inhomogeneity of the electric field of the same order.
If we make sure that this spatial inhomogeneity of the electric field has a large enough scale we should be allowed to neglect it in the calculation.
In other words we technically include the correct spatially inhomogeneous electric field to ensure the validity of Faraday's law while at the same time approximating the inhomogeneous electric field by a homogeneous field for the calculations.

We could thus start the time evolution at a point in time where there is no magnetic field and the initial conditions \Eqref{eqn:wigner_vac} are valid and slowly switch the magnetic field on afterwards.
Before ending the time evolution, the magnetic field must be switched off in a similar manner.
Switching the magnetic field on and off is implemented by a function
\begin{align*}
  \tilde{\Theta}(x) \definedby \begin{cases}
                                              0 & x<0 \\
                                              1 & x>1 \\
                                              6x^5-15x^4+10x^3 & \mathrm{otherwise,}
                                            \end{cases}
\end{align*}
which is also known as the smootherstep function \autocite{Ebert:2002:TMP:572337}.
It has zero first and second order derivatives at $x=0$ and $x=1$.
Its first derivative has a maximum at $x=\nicefrac12$ with $\tilde{\Theta}'(\nicefrac12)=\nicefrac{15}{8}<2$.
The maximum rate of change for the magnetic field when
\begin{align}
  \label{eqn:magnetic_field_switched}
B(t)\definedby B_0\cdot\tilde{\Theta}\left( \frac{t}{\Delta t} \right)
\end{align}
is $g_B=\frac{2B_0}{\Delta t}$ which should not exceed $g_\mx$.
Requiring $g_B<g_\mx$ yields $\Delta t>2000\,\tc\cdot \frac{B}{B_\crit}$.
Thus for $B=0.01B_\crit$, with the critical field strength $B_\crit=\nicefrac{m^2c^2}{e\hbar}\approx 4.4\mathrm{GT}$, the time scale for switching the magnetic field should be $\Delta t>20\,\tc$.

\section{Equation of Motion}
In the case of homogeneous electric and magnetic fields the pseudo differential operators from \Eq\eqref{eqn:diff-ops} simplify to
\begin{align}
  \label{eqn:diff-ops-hom}
  \begin{split}
   \Dt &=\partial_t + e\vv{E}(t)\cdot\nablap\,,
\\
  \vDx &= e\vv{B}(t)\times\nablap\,,
\\
  \vv{P} &=\vv{p}\,.
  \end{split}
\end{align}

In this case the equation of motion does not decouple and all the 16 components may be nonzero, thus we cannot use \Eq\eqref{eqn:Wigner-EoM-Homogen} but have to start over from \Eq\eqref{eqn:Wigner-EoM-Fierz}.
Because now different differential operators apply to the various functions, it is not possible to turn the resulting equation into an ordinary differential equation by the method of characteristics as has been done in the purely electric case.
Solving a time evolving partial differential equation can be done in two steps.
The first step is to use an appropriate form of basis decomposition to represent the spatial dependence of the functions at each point in time with a finite set of numbers.
Those numbers could be the function values at the vertices of a grid or the coefficients for a set of basis functions.
This decomposition should be able to provide the spatial derivatives of the discretized functions.
Using those it is possible to calculate the temporal derivatives of the function values at the grid points or of the coefficients, if the PDE is put into the form
\begin{align}
  \label{eqn:PDE_ODE_form}
  \partial_t \mathcal{W} &= F(t, \mathcal{W}, \vv{\nabla}\mathcal{W})\,.
\end{align}
The second step is then solving the resulting set of ODEs with an appropriate solver.

\pagebreak
The equations of motion  \Eq\eqref{eqn:Wigner-EoM-Fierz} can be easily brought into the form \Eqref{eqn:PDE_ODE_form} by inserting \Eq\eqref{eqn:diff-ops-hom} and sorting the terms accordingly.
This results in
\begin{equation}
  \label{eqn:Wigner-EoM-Fierz-Hom}
  \begin{rarray}{1.5pt}[c]{rllll}
      \qquad\partial_t \bbs
      &=-e\left( \vv{E}\cdot\nablap \right) \bbs
      &
      &+2\vv{p}\cdot\vv{\bbt}^{1}
    \\
      \partial_t \bbp
      &=-e\left( \vv{E}\cdot\nablap \right) \bbp
      &
      &-2\vv{p}\cdot \vv{\bbt}^{2}
      &-{2m \, \bba^0 }
    \\
      \partial_t \bbv^0
      &=-e\left( \vv{E}\cdot\nablap \right) \bbv^0
      &-{e\left( \vv{B}(t)\times\nablap \right)\cdot\vv{\bbv}}
    \\
      \partial_t \bba^0
      &=-e\left( \vv{E}\cdot\nablap \right) \bba^0
      &-{e\left( \vv{B}(t)\times\nablap \right)\cdot\vv{\bba}}
      &
      &+{2m\,\bbp}
    \\
      \partial_t \vv{\bbv}
      &=-e\left( \vv{E}\cdot\nablap \right) \vv{\bbv}
      &-{e\left( \vv{B}(t)\times\nablap \right)\cdot \bbv^0}
      &-{2\vv{p}\times \vv{\bba} }
      &-{2 m\, \vv{\bbt}^{1}}
    \\
      \partial_t \vv{\bba}
      &=-e\left( \vv{E}\cdot\nablap \right) \vv{\bba}
      &-{e\left( \vv{B}(t)\times\nablap \right)\cdot \bba^0}
      &-{2\vv{p}\times \vv{\bbv} }
    \\
      \partial_t \vv{\bbt}^{1}
      &=-e\left( \vv{E}\cdot\nablap \right) \vv{\bbt}^{1}
      &-{  e\left( \vv{B}(t)\times\nablap \right)\times  \vv{\bbt}^{2} }
      &-{2\vv{p}\, \bbs}
      &+{2m\, \vv{\bbv}}
    \\
      \partial_t \vv{\bbt}^{2}
      &=-e\left( \vv{E}\cdot\nablap \right) \vv{\bbt}^2
      &+{ e\left( \vv{B}(t)\times\nablap \right)\times \vv{\bbt}^{1}}
      &+{2\vv{p}\, \bbp}\,.
  \end{rarray}
\end{equation}

If this system of equations would be used, the numerical difficulties in calculating the one-particle distribution function $f$ as described in the homogeneous case in \Sec\ref{sec:wigner_method} would reappear.
It is possible to do a similar substitution also in this case, but due to the more complex structure of \Eqref{eqn:Wigner-EoM-Fierz-Hom}, a lot more additional terms arise.
The task of carrying out this substitution and turning the resulting equations into C++ code was much too tedious to do by hand.

In order to automate the process, a \textit{Mathematica} \autocite{mathematica} notebook has been written, which created the C++ code automatically.
For this to work, the system of equations was brought into the form
\begin{align}
  \nonumber\partial_t \bbW &= C_i \left( \partial_{p_i} \bbW \right) + M\bbW   \\
  \label{eqn:Wigner-EoM-Fierz-Hom-GeneralForm} &= C_i \bbW_{,p_i} + M\bbW   \\
  \nonumber\text{with } \bbW &= \begin{pmatrix}
    \bbs & \bbp & \bbv^0 & \vv{\bbv} & \bba^0 & \vv{\bba} & \vv{\bbt}^1 & \vv{\bbt}^2
  \end{pmatrix}^\intercal\,.
\end{align}
Here the short-hand notation $\bbW_{,p_i}$ is introduced for $\partial_{p_i} \bbW$.
Determining the Matrices $C_i$ is a matter of carefully evaluating the dot products, double cross products and triple products in \Eq\eqref{eqn:Wigner-EoM-Fierz-Hom}.
The details can be found in \App\ref{app:c-matrices}, where $C_i$ and $M$ are given in \Eqs\eqref{eqn:definition_c_matrices} and \eqref{eqn:definition_m_matrix}, respectively.
The substitution from \Eqref{eqn:wigner_subst} can be written in terms of the vacuum solution \Eqref{eqn:wigner_vac}
\begin{align*}
      \bbs(\vv{p}, t) &= (1-f(\vv{p}, t))\, \bbs_\vac(\vv{p}) - \vv{p} \cdot \vv{v}(\vv{p}, t) \,, \\
 \vv{\bbv}(\vv{p}, t) &= (1-f(\vv{p}, t))\, \vv{\bbv}_\vac(\vv{p}) + \vv{v}(\vv{p}, t)\,.
\end{align*}
In terms of
\begin{align}
  \nonumber W(\vv{p},t) &= \begin{pmatrix}
    f & \bbp & \bbv^0 & \vv{v} & \bba^0 & \vv{\bba} & \vv{\bbt}^1 & \vv{\bbt}^2
  \end{pmatrix}^\intercal
\intertext{and}
  \nonumber \bbW_\vac(\vv{p}) &= \begin{pmatrix}
    \bbs_\vac & 0 & 0 & \vv{\bbv}_\vac & 0 & \vv{0} & \vv{0} & \vv{0}
  \end{pmatrix}^\intercal
  \intertext{it takes the form}
  \label{eqn:wigner_subst_mag}
  \bbW(\vv{p},t) &= A(\vv{p})\, W(\vv{p},t)+ \bbW_\vac(\vv{p}) \\
\intertext{with}
  \nonumber A(\vv{p}) &=
  \begin{pmatrix}
-\bbs_\vac & 0 & 0 & -\vv{p}^\intercal & 0 & 0 & 0 & 0\\
0 & 1 & 0 & 0 & 0 & 0 & 0 & 0\\
0 & 0 & 1 & 0 & 0 & 0 & 0 & 0\\
-\vv{\bbv}_\vac & 0 & 0 & \id & 0 & 0 & 0 & 0\\
0 & 0 & 0 & 0 & 1 & 0 & 0 & 0\\
0 & 0 & 0 & 0 & 0 & \id & 0 & 0\\
0 & 0 & 0 & 0 & 0 & 0 & \id & 0\\
0 & 0 & 0 & 0 & 0 & 0 & 0 & \id
  \end{pmatrix}\,,
\end{align}
where block matrix notation has been used.
When \Eq\eqref{eqn:wigner_subst_mag} is inserted into \Eq\eqref{eqn:Wigner-EoM-Fierz-Hom-GeneralForm} and solved for $\partial_t W$ the resulting equation reads
\begin{equation}
  \label{eqn:Wigner-EoM-Mag}
  \partial_t W = \left( A^{-1}C_iA \right)W_{,p_i}+ A^{-1}\left( C_iA_{,p_i} + MA \right) W + A^{-1}C_i \partial_{p_i} \bbW_{\vac}\,,
\end{equation}
realizing that $M\bbW_\vac=0$.
Using this result and the known matrices it is possible to write automatically the 16 equations of motion into C++ code.
The $W_{,p_i}$ need to be calculated before this code can be executed, this is done by a Fourier transformation.

All the given equations up to this point were supposed to describe pair production in a three dimensional momentum space $\vv{p}$ that evolves with time $t$.
In the Wigner method, due to the method of characteristics, the trajectories in momentum space decouple and do not exchange information during their evolution.
This means, that every point $\vv{p}$ of the final momentum spectrum can be calculated independently.
In the context of the partial differential equation \Eqref{eqn:Wigner-EoM-Mag} this is not true, a complete grid of points in momentum space has to be stored and evolved together.
At this point the number of dimensions should be reduced in order to have a manageable memory usage.
Setting $E_z=B_x=B_y=0$ has the effect that all derivatives with respect to $p_z$ are dropped from the equations, because the matrix $C_z$, defined in \Eqref{eqn:definition_c_matrices}, is identically zero.
This enables two-dimensional calculations with a $p_x$-$p_y$ grid, for any given value of $p_z$.

Setting $\vv{B}\to\vv{0}$ in the resulting equation of motion results in \Eqref{eqn:Wigner-EoM-NoB},
which is equivalent to \Eq\eqref{eqn:Wigner-EoM-ModQKT} plus another set of equations for the remaining functions, that completely decoupled from the one-particle distribution function.

\section{Numerical Implementation}
Following the previous section, there are now various possibilities to combine the available forms of the equation of motion with the numerical methods.
They are listed together with their development name in \Tab\ref{tab:codenames} in chronological order.
\begin{table}[ht]
\centering
\begin{threeparttable}
  \caption{Table listing the various numerical codes.}
  \label{tab:codenames}
  \begin{tabular}{lcll}
   \toprule
   differential eq.                   & subst.?\tnote{1} & $\vv{B}$ & name \\\midrule
   ODE, \Eqref{eqn:Wigner-EoM-Homogen}\tnote{2} & no & $=\vv{0}$ & no name, first attempts with Mathematica \\
   ODE, \Eqref{eqn:Wigner-EoM-ModQKT}  & yes & $=\vv{0}$ & Wigner method (ppsolve, charwigner) \\
   PDE, \Eqref{eqn:Wigner-EoM-NoB}\tnote{3} & yes & $=\vv{0}$ & fftwignerh \\
   PDE, \Eqref{eqn:Wigner-EoM-Mag} & yes & $\neq\vv{0}$ & fftwignerb \\
   PDE, \Eqref{eqn:Wigner-EoM-Fierz-Hom}\tnote{4} & no & $\neq\vv{0}$ & fftwignerb-nosubst \\\bottomrule
  \end{tabular}
  \begin{tablenotes}
    \item [1] Indicates whether the variables have been substituted to solve for $f$ directly. Otherwise still a substitution is applied to remove the vacuum solution.\!\tnote{2,4}
    \item [2] The system of equation for this case results from \Eqref{eqn:Wigner-EoM-Homogen} by substituting $\bbw\to\bbw'+\bbw_\vac$ and solving for $\dot{\bbw}'$.
    \item [3] \Eqref{eqn:Wigner-EoM-NoB} follows from \Eqref{eqn:Wigner-EoM-Mag} by inserting $\vv{B}=\vv{0}$.
    \item [4] The system of equation for this case results from \Eqref{eqn:Wigner-EoM-Fierz-Hom} by substituting $(\bbs,\vv{\bbv})\to(\bbs',\vv{\bbv}')+(\bbs_\vac,\vv{\bbv}_\vac)$ and solving for $\partial_t(\bbs',\vv{\bbv}')$.
  \end{tablenotes}
\end{threeparttable}
\end{table}

The solution scheme is based upon setting the PDE \Eqref{eqn:Wigner-EoM-Mag} onto a grid in $\vv{p}$-space and calculating $W_{,p_i}$ using a Fourier transformation.
This removes the derivatives w.\,r.\,t. $\vv{p}$ and turns the system of 16 continuous PDEs into a system of $16\cdot N_x\cdot N_y\cdot N_z$ ODEs which can be solved using, e.\,g., a Runge\hyp{}Kutta method.
The momentum grid covers in each direction a range $p_i\in(-P_i,P_i)$.
Usually two-dimensional calculations are done using $N_x=N_y=N$ and $P_x=P_y=P$.

One of the fastest available implementations of the fast Fourier transformation is the \textit{FFTW} \autocite{FFTW05}, which is a pure C library.
As the library exists in multiple variants in order to support multiple floating point data types, most of the function names depend on the used floating point type.
A C++ wrapper to the pure C library was built which has the task of selecting the correct function names automatically at compilation time.
This wrapper is basically a C++ struct, which takes the data type as a template parameter and contains references to the correct functions.
Such wrappers do already exist, but they also try to provide an abstract interface that hides the details of the FFT.
Doing so they only implement the basic \textit{FFTW} functionality, which does not allow for carrying out multiple independent transform within one array.
Using the full \textit{FFTW} interface it is possible to store the complete state in a single array, using \textit{FFTW}'s built-in parallelism.
This memory layout was also preferred in order to minimize memory access when evaluating the right hand side of the equation of motion using parallel threads, accessing the individual functions using \textit{Blitz++} \autocite{veldhuizen1998arrays}.

\begin{figure}
 \centering
 \includegraphics{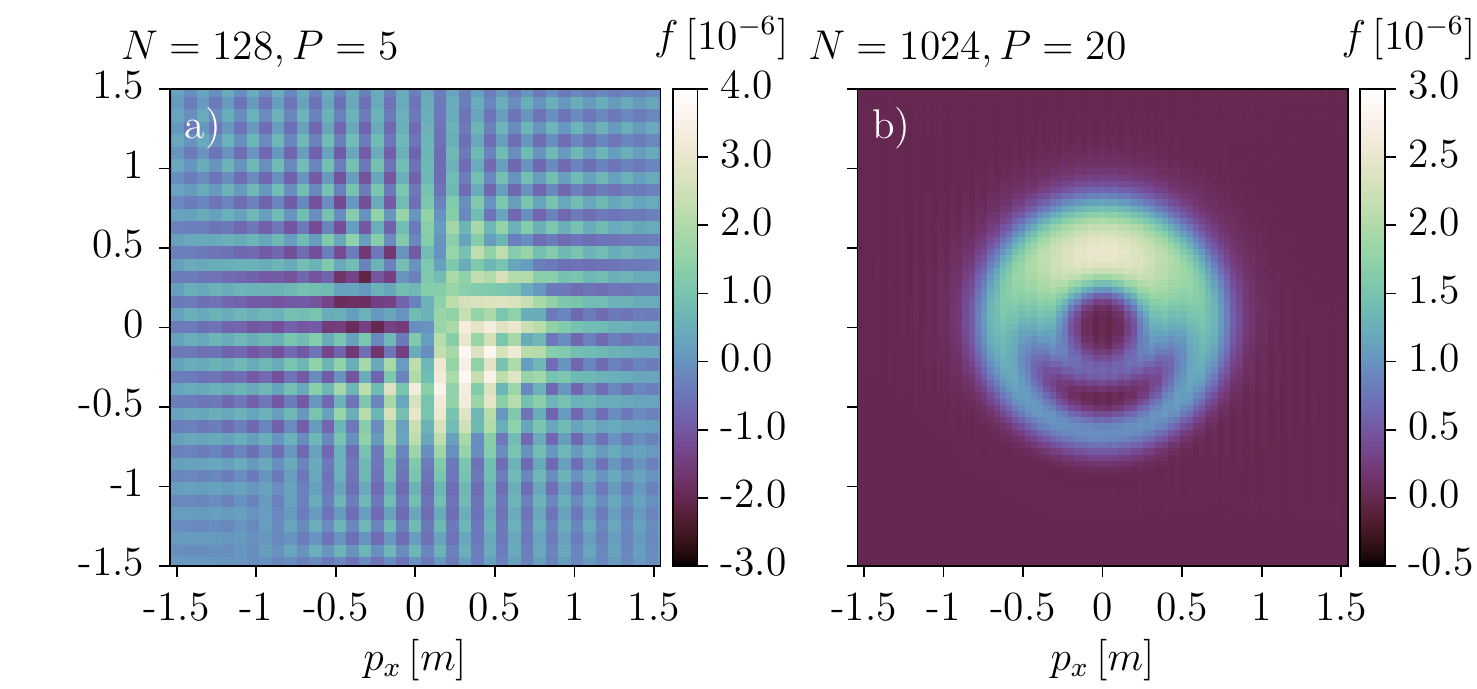}
 \caption[Example spectra of \textit{fftwignerh} without damping]{
   Two example spectra calculated by \textit{fftwignerh} without damping.
   The electric field is given by the rotating Sauter pulse \Eqref{eqn:puls-sauter-rot} with parameters $\tau=10\,\tc$, $\Omega=0.6\,m$.
   A reference spectrum for this pulse as calculated by the other available methods can be found in \Fig\ref{fig:spectra_methods}.
   Panel a): Grid size $N=128$, momentum extent $P=5$.
   The calculation breaks down after high Fourier modes are populated exponentially.
   Panel b): Grid size $N=1024$, momentum extent $P=20$.
   The spectrum seems to agree with the reference spectrum, but artifacts of high Fourier modes are visible.
 }
 \label{fig:fftwignerb_multi_1}
\end{figure}
\begin{figure}
 \centering
 \includegraphics{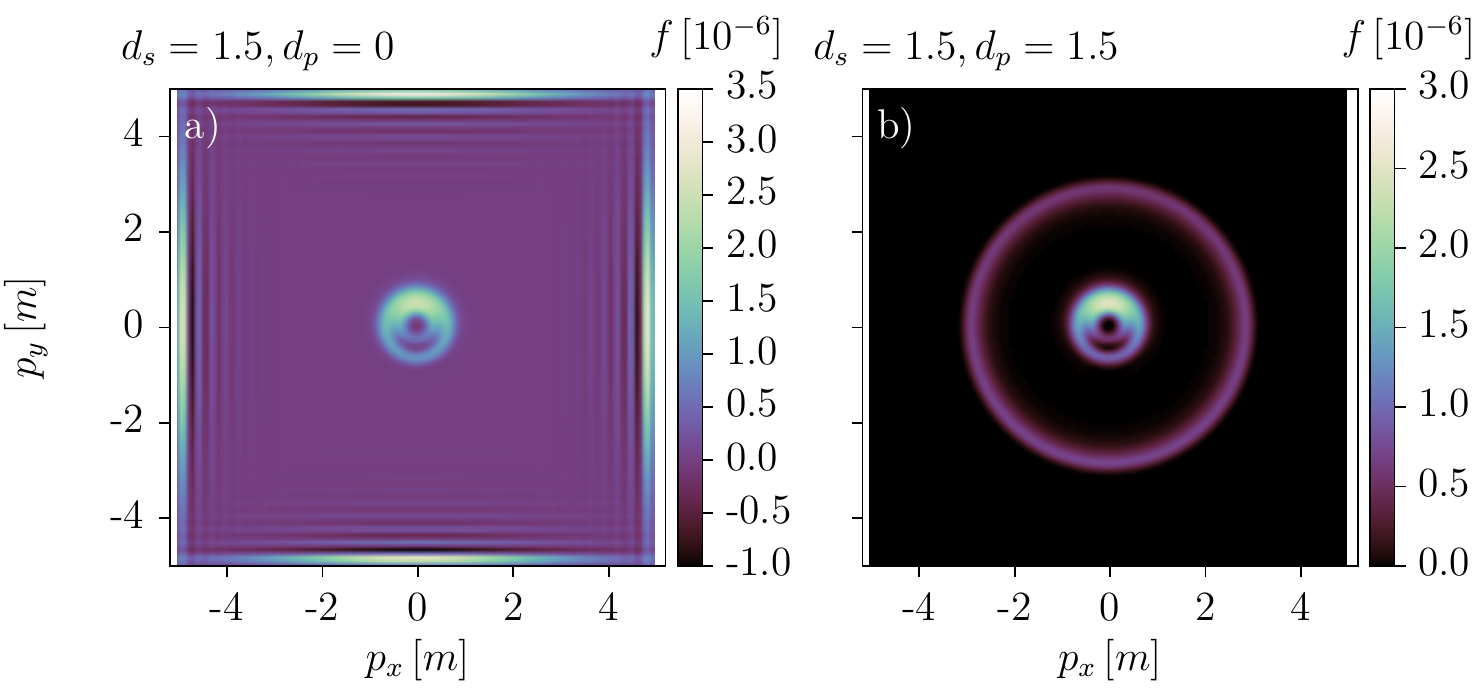}
 \caption[Example spectra of \textit{fftwignerh} with damping]{
   Two example spectra calculated by \textit{fftwignerh} with damping.
   Grid size in both cases $N=128$, momentum extent $P=5$.
   The electric field is given by the rotating Sauter pulse \Eqref{eqn:puls-sauter-rot} with parameters $\tau=10\,\tc$, $\Omega=0.6\,m$.
   Common damping parameters are $x_p=x_s=0.7$ and $h_p=h_s=20$.
   Panel a): Dampening only in Fourier space.
   The central region of the spectrum agrees with the reference spectrum \Fig\ref{fig:spectra_methods}, but some artifacts remain at the boundaries.
   Panel b): Both damping terms enabled.
   The central region of the spectrum agrees with the reference spectrum, the only artifact is a smooth ring at the border of the damping term.
 }
 \label{fig:fftwignerb_multi_2}
\end{figure}
As a first step towards the full problem of \Eq\eqref{eqn:Wigner-EoM-Mag}, a code has been written that solves \Eq\eqref{eqn:Wigner-EoM-NoB} using a Fourier basis decomposition.
This code was called \textit{fftwignerh} and was used to test and improve the stability of the numerical algorithm.
Once this code worked well enough the equation of motion was changed to include the additional functions and terms and the code renamed \textit{fftwignerb}.
\enlargethispage{\baselineskip}

\pagebreak
During the first tests with \textit{fftwignerh} it became evident that it is necessary to get rid of the high Fourier modes which grow during time evolution.
If the integration steps are too large, the high Fourier modes start growing exponentially due to numerical instability.
Even if the step size is small enough, high Fourier modes will grow due to propagation terms in Fourier space.
After some time the bounds of the reciprocal grid are reached and the evolution breaks down, an example can be seen in \Fig\ref{fig:fftwignerb_multi_1}a).
In principle it is possible to compensate by using a large enough grid, but this consumes lots of memory and computation time.
Even though a number of example calculations were successful using this approach and an example is given in \Fig\ref{fig:fftwignerb_multi_1}b), it was found to be computationally too expensive.

Another idea was to add a linear damping term to the equation of motion
\begin{align*}
  \partial_t W \to \partial_t W - \mathrm{FT}^{-1}\left[u_s\left(\frac{|\vv{s}|}{s_\mx}\right)\cdot \mathrm{FT}[W(\vv{p}')](\vv{s})\right](\vv{p})\,,
\end{align*}
resulting in an exponential decay of any population of the high modes.
This term should of course only act on the higher modes, such that the lower modes evolve as the equation of motion dictates.
The prefactor was defined as
\begin{align*}
  u_s(a) = \frac{d_s}{2}\bigl( \tanh\left( h_s (a-x_s) \right) +1\bigr),
\end{align*}
where $d_s$ would define the damping strength for the high modes, $x_s$ defines the point where the damping is half of that value and $h_s$ defines the steepness of the function $u_s(a)$.
Using this damping for the high Fourier modes improved the numerical results, but artifacts still remained in the final spectra, an example can be seen in \Fig\ref{fig:fftwignerb_multi_2}a).
To prevent further numerical problems arising from those artifacts, the same damping term was added to the equation of motion in momentum space
\begin{align*}
  \partial_t W \to \partial_t W - u_p\left(\frac{|\vv{p}|}{p_\mx}\right)\cdot W\,
\end{align*}
introducing a second set of damping parameters $d_p$, $x_p$ and $h_p$.
The resulting spectra have much smoother artifacts that are less likely to produce numerical problems, an example can be seen in \Fig\ref{fig:fftwignerb_multi_2}b).
\begin{figure}
 \centering
 \includegraphics{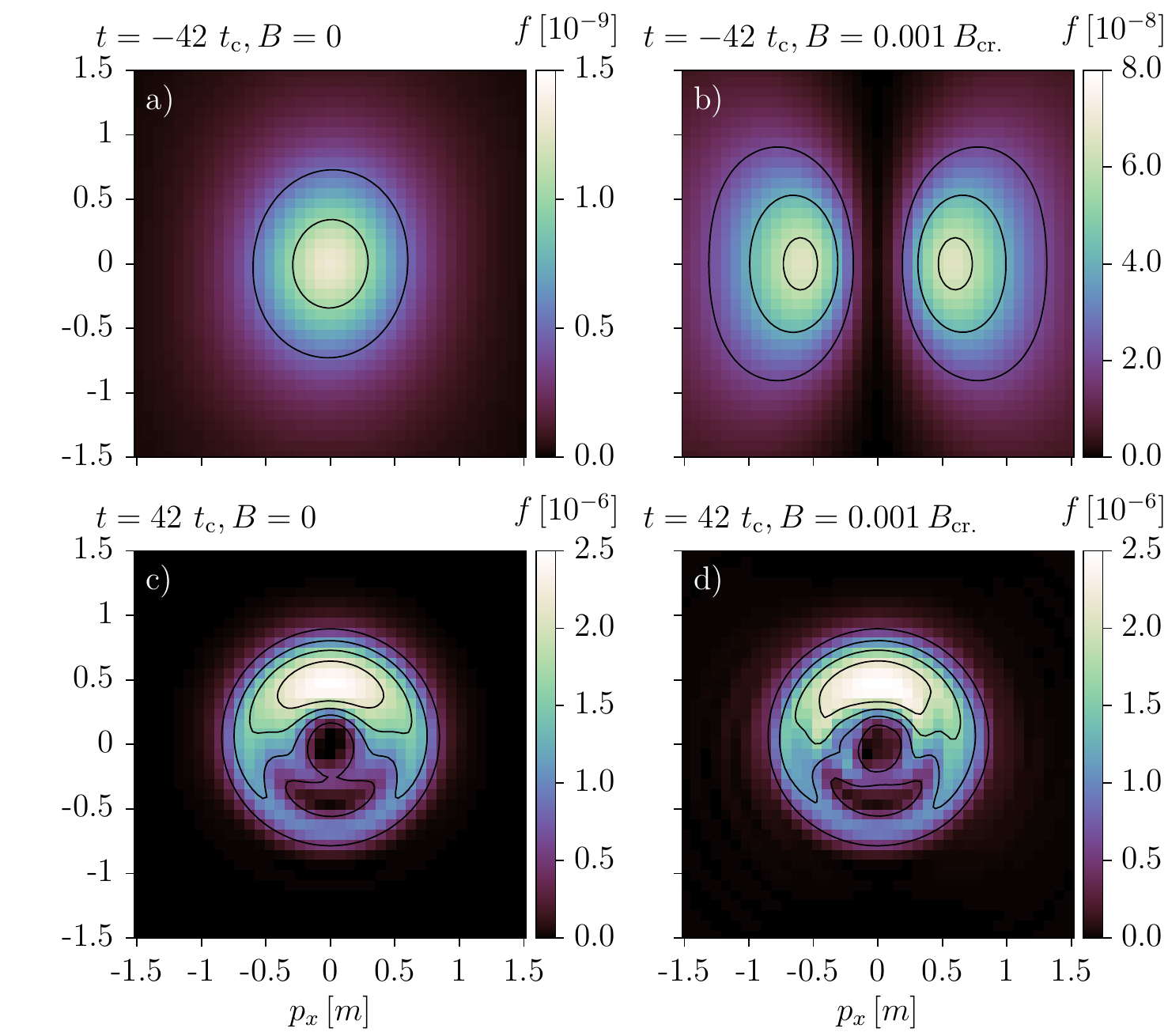}
 \caption[Comparison of the time evolution with and without magnetic field]{
   Comparison of the time evolution with and without magnetic field.
   The electric field is given by the rotating Sauter pulse \Eqref{eqn:puls-sauter-rot} with parameters $\tau=10\,\tc$, $\Omega=0.6\,m$.
   The magnetic field is given by \Eqref{eqn:magnetic_field_switched} with a switching parameter of $\Delta t=100\,\tc$.
   The levels of the contour lines are indicated by the marks of the color box.
   Panel a) and b): Intermediate stages of the time evolution before the peak of the electric field pulse without magnetic field or while the magnetic field is switched on.
   Panel c): An intermediate stage of the time evolution without magnetic field after the peak of the electric field pulse.
   The distribution has almost arrived at its final shape.
   Panel d): An intermediate stage of the time evolution while the magnetic field is being switched off after the peak of the electric field pulse.
   Some differences to Panel c) are visible that vanish completely once the magnetic field is completely switched off.
 }
 \label{fig:fftwignerb_multi_3}
\end{figure}

\pagebreak
\begin{figure}[ht]
 \centering
 \includegraphics{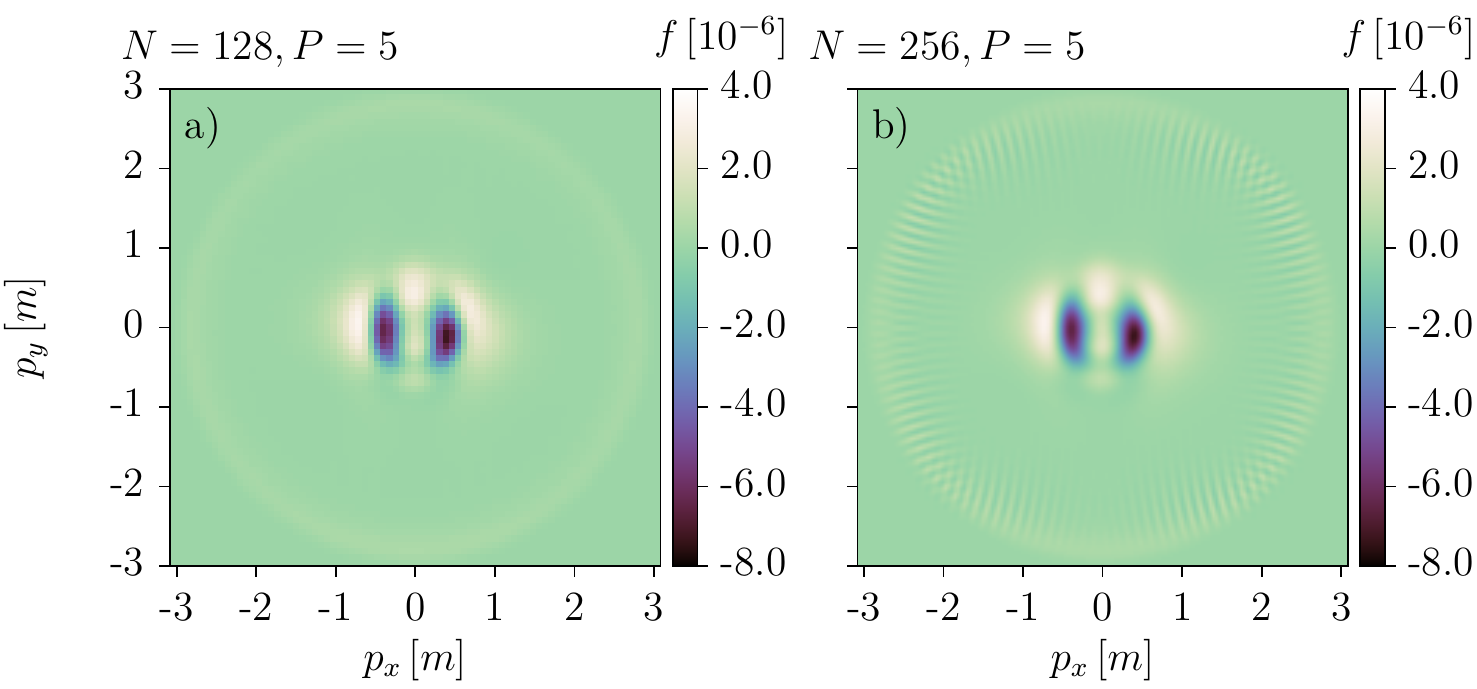}
 \caption[Pair production spectra for $B=0.01\,B_\crit$]{
   Comparison of the results for $B=0.01\,B_\crit$ with different grids.
   Electric field given by the rotating Sauter pulse \Eqref{eqn:puls-sauter-rot} with parameters $\tau=10\,\tc$, $\Omega=0.6\,m$.
   The result seems to be stable w.\,r.\,t. the grid size.
 }
 \label{fig:fftwignerb_multi_4}
\end{figure}
\section{Preliminary Results}
At that stage the magnetic field was added as defined in \Eqref{eqn:magnetic_field_switched}.
The values for $B$ were chosen to be $B=0.001\,B_\crit$ and later $B=0.01\,B_\crit$, both with a switching parameter $\Delta t=100\,\tc$.

In case of the weak magnetic field $B=0.001\,B_\crit$ it turns out that the final pair production spectrum is the same as without magnetic field.
At intermediate times, especially during the switching of the magnetic field, the states are different.
Some examples can be seen in \Fig\ref{fig:fftwignerb_multi_3}.

In case of the stronger magnetic field $B=0.01B_\crit$ the resulting pair production spectra contain negative values for $f$, an example can be seen in \Fig\ref{fig:fftwignerb_multi_4}.
This should not be possible and clearly indicates that our approach poses open questions.
Increasing the grid size does not change the result, so we conclude that it is numerically stable.
One possible conclusion might be that our assumptions for the approximation are not right and that our approach might not be useful.
But we need to stress that these are very early results and further research needs to be done.

  \chapter{Conclusion}
We have investigated electron positron pair production in rotating electric fields.
To this end we constructed the modified quantum kinetic equation \Eqref{eqn:Wigner-EoM-ModQKT} and implemented a numerical solver.
We also developed a \textit{C++} code based on the semiclassical method to calculate pair production in the rotating Sauter pulse.
Computation of the pair production spectra for long pulse durations is substantially easier using the semiclassical method.

We studied pair production in the rotating Sauter pulse with the sub-critical field strength $E=0.1\,E_\crit$ in dependency of the parameters pulse duration $\tau$ and rotation frequency $\Omega$.
Doing so we interpreted the spectra with respect their rich phenomenology in terms of the vector potential, Schwinger pair production and multiphoton pair production.
We demonstrated that the total particle yield per unit spacetime volume is dependent only on a combined Keldysh parameter, apart from multiphoton resonances, see \Fig\ref{fig:total_yield_unified}.
Multiphoton pair production in the rotating Sauter pulse has some measure of asymmetry regarding the magnetic moment.

We considered other kinds of configurations for a spatially homogeneous electric field.
Studying pair production in elliptically polarized Sauter pulses disclosed parameter sets which lead to the production of collimated electron bunches.
If a chirp is added to the rotating Sauter pulse the resulting spectra show quite chaotic shapes, which could be relevant for high\hyp{}power lasers using chirped pulse amplification.

Bichromatic fields have been studied in the form of superposed linear oscillating and rotating Sauter pulses.
We compared the resulting spectra in the rotating case to experimental results from atomic ionization.
In the linear case another variant of experimentally useful features is discovered.

We worked on an approach to include magnetic fields in a spatially homogeneous setup and discussed our preliminary results.

While clearly some of the results presented in this work should be complemented by additional research, the results at hand are already useful to gain an overall picture of pair production in rotating fields.

  \cleardoublepage %
  \pagestyle{scrheadings}
  \setcounter{page}{1}
  \pagenumbering{Roman}

  \appendix
  \chapter{Appendix}
\section{Matrices for the Equation of Motion Including Homogeneous Magnetic Fields}
\label{app:c-matrices}
The equation of motion for the inclusion of homogeneous magnetic fields in Chapter~\ref{chap:magnetic}, \Eqref{eqn:Wigner-EoM-Fierz-Hom-GeneralForm}, was given in a general form.
The matrices are still to be defined, which will be done via block matrices.
Let us first define some vectors and matrices used to carry out the various vector operations that occur in the equation of motion.
Evaluating a cross product $\vv{B}\times\vv{v}$ can be expressed using a matrix $b$ defined by
\begin{align*}
  b\vv{v} \definedby \vv{B}\times\vv{v}\,.
\end{align*}
The matrix $b$ is then given by
\begin{align*}
  b = \begin{pmatrix}
                             0 & -B_z & B_y \\
                             B_z & 0 & -B_x \\
                             -B_y & B_x & 0
                          \end{pmatrix}
                          =: \begin{pmatrix}
                            \vv{b_x} & \vv{b_y} & \vv{b_z}
                          \end{pmatrix},
\end{align*}
where the columns of $b$ are designated $\vv{b_x} , \vv{b_y} , \vv{b_z}$.
The two types of terms containing $\vv{B}\times\nablap$ can then be expressed using these vectors as
\begin{align*}
  \left( \vv{B}\times\nablap \right)a &= \sum_{i=x,y,z}\vv{b_i}\left( \partial_{p_i}a \right) \\
  \left( \vv{B}\times\nablap \right)\cdot\vv{v} &= \sum_{i=x,y,z}\vv{b_i}\cdot\left( \partial_{p_i}\vv{v} \right)\,.
\end{align*}
Terms with a triple cross product $\left( \vv{B}\times\nablap \right)\times \vv{t}$ need to be evaluated separately resulting in
\begin{align*}
 \left( \vv{B}\times\nablap \right)\times \vv{t} &= \sum_{i=x,y,z} T_i \left( \partial_{p_i}\vv{t} \right)
\end{align*}
with matrices $T_i$ defined by
\begin{align*}
  T_x \definedby \begin{pmatrix}
          0 & B_y & B_z \\
          -B_y & 0 & 0 \\
          - B_z & 0 & 0
        \end{pmatrix},
  &
  T_y \definedby \begin{pmatrix}
          0 & -B_x &  0 \\
          B_x & 0 & B_z \\
          0 & -B_z & 0
        \end{pmatrix},
  &
  T_z \definedby \begin{pmatrix}
          0 & 0 & -B_x \\
          0 & 0 & -B_y \\
          B_x & B_y & 0
        \end{pmatrix}\,.
\end{align*}
To define the matrices $C_i$ and $M$ for \Eqref{eqn:Wigner-EoM-Fierz-Hom-GeneralForm} block matrix notation is used.
They are given by
{\allowdisplaybreaks
\begin{align}
  \label{eqn:definition_c_matrices}
  C_i \definedby e\begin{pmatrix}
 0 & 0 & 0        & 0                  & 0        & 0                  & 0   & 0   \\
 0 & 0 & 0        & 0                  & 0        & 0                  & 0   & 0   \\
 0 & 0 & 0        & \vv{b_i}^\intercal & 0        & 0                  & 0   & 0   \\
 0 & 0 & \vv{b_i} & 0                  & 0        & 0                  & 0   & 0   \\
 0 & 0 & 0        & 0                  & 0        & \vv{b_i}^\intercal & 0   & 0   \\
 0 & 0 & 0        & 0                  & \vv{b_i} & 0                  & 0   & 0   \\
 0 & 0 & 0        & 0                  & 0        & 0                  & 0   & T_i \\
 0 & 0 & 0        & 0                  & 0        & 0                  & T_i & 0
\end{pmatrix}
-eE_i\mathbbm{1}_{16}
\\
\nonumber
&=e\begin{pmatrix}
 -eE_i & 0     & 0        & 0                  & 0        & 0                  & 0                  & 0                  \\
 0     & -eE_i & 0        & 0                  & 0        & 0                  & 0                  & 0                  \\
 0     & 0     & -eE_i    & \vv{b_i}^\intercal & 0        & 0                  & 0                  & 0                  \\
 0     & 0     & \vv{b_i} & -eE_i\mathbbm{1}_3 & 0        & 0                  & 0                  & 0                  \\
 0     & 0     & 0        & 0                  & -eE_i    & \vv{b_i}^\intercal & 0                  & 0                  \\
 0     & 0     & 0        & 0                  & \vv{b_i} & -eE_i\mathbbm{1}_3 & 0                  & 0                  \\
 0     & 0     & 0        & 0                  & 0        & 0                  & -eE_i\mathbbm{1}_3 & T_i                \\
 0     & 0     & 0        & 0                  & 0        & 0                  & T_i                & -eE_i\mathbbm{1}_3
\end{pmatrix}
\end{align}
and
\begin{align}
  \label{eqn:definition_m_matrix}
  M \definedby \begin{pmatrix}
 0        & 0       & 0 & 0                   & 0      & 0     & 2\vv{p}^\intercal   & 0                  \\
 0        & 0       & 0 & 0                   & -2\, m & 0     & 0                   & -2\vv{p}^\intercal \\
 0        & 0       & 0 & 0                   & 0      & 0     & 0                   & 0                  \\
 0        & 0       & 0 & 0                   & 0      & -2\,P & 2\cdot\mathbbm{1}_3 & 0                  \\
 0        & 2\,m    & 0 & 0                   & 0      & 0     & 0                   & 0                  \\
 0        & 0       & 0 & -2\,P               & 0      & 0     & 0                   & 0                  \\
 -2\vv{p} & 0       & 0 & 2\cdot\mathbbm{1}_3 & 0      & 0     & 0                   & 0                  \\
 0        & 2\vv{p} & 0 & 0                   & 0      & 0     & 0                   & 0
               \end{pmatrix}
\end{align}}
with
\begin{align*}
  P\vv{v} \definedby \vv{p}\times\vv{v} \\
  \Rightarrow P\definedby \begin{pmatrix}
                             0 & -p_z & p_y \\
                             p_z & 0 & -p_x \\
                             -p_y & p_x & 0
                          \end{pmatrix}\,.
\end{align*}

\section{Semiclassical Results for the Constant Rotating Pulse}\label{app:constrot}
The integrals of \Eqs\eqref{eq:Kint} and \eqref{eq:K_xy} at the turning points for the constant rotating field have been computed beforehand in \cite{Strobel:2014tha}. Here we reproduce them for the sake of completeness. The turning points and the integrals can be found to be
\begin{align}
 \Omega t_k^\pm=\arcsin\left(\frac{\cm_x}{\cm_\parallel}\right)\pm\i\,\operatorname{arcosh}\left(\frac{\left( \cm_\parallel^2+\epsilon_\perp^2\right)\left(\frac{\Omega}{m\varepsilon}\right)^2+m^2}{2\left(\frac{\Omega}{m\varepsilon}\right)\,  \cm_\parallel\, m}\right)+2\pi k\,. \label{eq:rotatingcompltp}
\end{align}
The integrals $K(t_k)$ and $K_{xy}(t_k)$ are given by
\begin{align}
\begin{split}
\label{eq:Krot}
 K(t_k)%
 &=\frac{4\epsilon_\perp}{\Omega}\sqrt{1-y_+^2}
      \left[
            2k \E\left(\sqrt{\frac{y_-^2-y_+^2}{1-y_+^2}}\right)
      \right]
      +\Phi
    \\
    &\qquad+\i\frac{4\epsilon_\perp}{\Omega}\sqrt{1-y_+^2}
      \left[
            \E\left(\sqrt{\frac{1-y_-^2}{1-y_+^2}}\right)
            -\K\left(\sqrt{\frac{1-y_-^2}{1-y_+^2}}\right)
      \right]\,,
\end{split}
 \\
\begin{split}
\label{eq:K_xyrot}
 K_{xy}(t_k)%
 &=2k\frac{g ({y_-^2-1})}{\sqrt{1-y_+^2}}
         \frac{y_+}{y_-}\PiI\left(\frac{1}{y_-^2}\frac{y_-^2-y_+^2}{1-y_+^2},\sqrt{\frac{y_-^2-y_+^2}{1-y_+^2}}\right)
      \\
    &\qquad+2k\frac{g\left(1-y_-y_+\right)}{\sqrt{1-y_+^2}}
          \K\left(\sqrt{\frac{y_-^2-y_+^2}{1-y_+^2}}\right)
      +\Phi_{xy}
    \\
    &\qquad +\frac{\i g(y_-y_+)}{\sqrt{1-y_+^2}}
          \PiI\left(1-y_-^2,\sqrt{\frac{1-y_-^2}{1-y_+^2}}\right)
    \\
    &\qquad -\frac{\i g}{\sqrt{1-y_+^2}}
          \K\left(\sqrt{\frac{1-y_-^2}{1-y_+^2}}\right)
      \,,
 \end{split}
\end{align}
with
\begin{align*}
 y_{\pm}:=\i\frac{\Omega\,\cm_\parallel\pm e \varepsilon E_\crit}{\Omega\,\epsilon_\perp}\,,&&
\cm_\parallel:=\sqrt{\cm_x^2+\cm_y^2}
\end{align*}
and the elliptic integrals \autocite{gradshteyn2007}
\begin{align*}
  \K(x)\definedby\int_0^{\nicefrac{\pi}{2}}\dd \Theta\frac{1}{\sqrt{1-x^2\sin^2\Theta}}
  \\
  \E(x)\definedby\int_0^{\nicefrac{\pi}{2}}\dd \Theta\sqrt{1-x^2\sin^2\Theta}
  \\
  \PiI(n,x)\definedby\int_0^{\nicefrac{\pi}{2}}\dd \Theta\frac{1}{\left( 1-n\sin^2\Theta \right)\sqrt{1-x^2\sin^2\Theta}}\,.
\end{align*}
The quantities \(\Phi\) and \(\Phi_{xy}\) are physically irrelevant global phases.

\section{Calculation of the Momentum Spectrum for the Sauter Pulse} \label{app:Sauter}
In this appendix we want to calculate the integral \(K_0(t)\) for the non-rotating Sauter pulse which is given by \Eqs\eqref{eqn:puls-sauter-rot} for \(\Omega=0\).
We start from the potential
\begin{align*}
 A_\mu(x)&=(0,A(t),0,0),\\
 A(t)&=eE_0\tau \left[1+\tanh\left(\frac{t}{\tau}\right)\right].
\end{align*}
The turning points, as defined in \Eqref{eq:turningpoints}, are found to be
\begin{align}
 t_j^\pm&=\tau \artanh\left[\frac{\cm_x\pm\ii\tilde{\epsilon}_\perp}{e E_0 \tau}-1\right]+\ii \pi j \tau\,, \label{eq:TPSauter}
\end{align}
for \(j\in\mathbb{N}\) with
\begin{align*}
\tilde{\epsilon}_\perp^2\definedby \epsilon_\perp^2+\cm_y^2\,.
\end{align*}
This means we find an infinite number of turning points which all have the same real part
\begin{align*}
 s_j=\Re(t_j^\pm)=\frac{1}{4}\log\left(\frac{\cm_x^2+\tilde{\epsilon}_\perp^2}{(\cm_x-2eE_0\tau)^2+\tilde{\epsilon}_\perp^2}\right).
\end{align*}
The integral from \Eqref{eq:Kint} gives
\begin{align}
\begin{split}
 K_0(t)&=-\tau\frac{2}{\gamma}\log\left[\frac{\gamma}{m}(\omega(t)+\cm_x)+\tanh\left(\frac{t}{\tau}\right)\right]
 \\&\hspace{1cm}
    -\tau\sum_{l=\pm1}l\cm_l\Bigg(\log\left[1-l\tanh\left(\frac{t}{\tau}\right)\right]
    \\&\hspace{2cm}
    -\log\biggl[\left(\frac{\gamma}{m}\cm_x+l\right)\left(\frac{\gamma}{m}\cm_x+\tanh\left(\frac{t}{\tau}\right)\right)
    \\&\hspace{3cm}
    +\frac{\gamma^2}{m^2}\left(\cm_l\omega(t)+\tilde{\epsilon}_\perp^2\right)\biggr]\Bigg)
        \\&\hspace{1cm}+\tilde{\Phi}\,,
\end{split}\label{eq:Kint3}
\end{align}
where \(\tilde{\Phi}\) is a physically irrelevant global phase and we introduced the Keldysh parameter for the pulse length \(\tau\) which is defined as
\begin{align*}
 \gamma:=\frac{m }{e E_0\tau}
\end{align*}
and we also defined
\begin{align*}
 \cm_\pm\definedby \sqrt{\left(  \cm_x-\frac{ms}{\gamma}(1\pm1)\right)^2+\tilde{\epsilon}_\perp^2}\,.
\end{align*}
Using the explicit form of the turning points of \Eqref{eq:TPSauter} and assuming \(E_0>0\) we find that the imaginary part of the integral from \Eqref{eq:Kint} at the turning points is given by
\begin{align*}
\Im[K_0(t_j^\pm)]=&\mp\frac{\pi}{2} \tau \frac{1}{\gamma}\left(\gamma \cm_++\gamma \cm_--2m\right)\,.
\end{align*}
According to the condition in \Eqref{eq:constraint} only the turning points for which the imaginary part is negative contribute.
That still leaves an infinite number of turning points which will give the same contribution to the sum in \Eqref{eq:MomentumSpectrum}.
However \Eqref{eq:MomentumSpectrum} only holds if the turning points have a different real part.
This is connected to how the contour is deformed to extract the contributions of the poles.
We chose the contour such that it follows the real axis up to \(s_p\) and then approaches the turning point in a line parallel to the imaginary axis.
If turning points have the same real part it is sufficient to take one integral which encircles all of the turning points.
Using \Eqref{eq:Kint3} we find
\begin{align*}
 \int_{t_j^\pm}^{t_{j+1}^\pm}\omega(t')dt'=0\,.
\end{align*}
This means that only the integral from \(s_p\) to \(t_p^+\) contributes for the Sauter pulse, since the contributions of the other ones vanish due to the periodic form of \(\omega(t) \).
Accordingly the semiclassical momentum spectrum defined in \Eqref{eq:MomentumSpectrum} takes the form
\begin{align*}
 W^{\text{SC}}_s(\vec{\cm})=\exp\left(-\frac{\pi}{\epsilon}\frac{1}{\gamma^2}\left( \frac{\gamma \cm_+}{m}+\frac{ \gamma  \cm_-}{m}-2\right)\right)\,.
\end{align*}
This can be compared to the exact result, which for instance can be obtained in the real-time DHW formalism and is \cite{Hebenstreit:2010vz,Hebenstreit:2011pm}
\begin{align*}
 W_s(\vec{\cm})=\frac{\sinh\left(\frac{1}{2}\frac{\pi}{\epsilon}\frac{1}{\gamma^2}\left[2+\frac{\gamma \cm_+}{m}-\frac{\gamma \cm_-}{m}\right]\right)\sinh\left(\frac{1}{2}\frac{\pi}{\epsilon}\frac{1}{\gamma^2}\left[2 -\frac{\gamma \cm_+}{m }+\frac{\gamma \cm_-}{m }\right]\right)}{\sinh\left(\frac{\pi}{\epsilon}\frac{1}{\gamma} \frac{\cm_+}{m }\right)\sinh\left(\frac{\pi}{\epsilon}\frac{1}{\gamma} \frac{\cm_-}{m }\right)}\,.
\end{align*}
Using the fact that
 \( \sinh(x)\approx\frac12 \exp(x) \)
for large $x$ we find for \(\epsilon\gamma\sim \nicefrac1{\tau m}\ll1\)
\begin{align*}
 W_s(\vec{\cm})\overset{\tau m\gg1}\approx W^{\text{SC}}_s(\vec{\cm})
\end{align*}
such that the semiclassical result is approximating the exact one well for long enough pulses.
As described in Sec.~\ref{sec:cmp} for shorter pulses the turning points get too close in the complex plane and the approximation in \Eqref{eq:APPROX} breaks down.

    {
      \emergencystretch=2.5em
      \hbadness=10000
      \printbibliography
    }

    \listoffigures

  \chapter{List of Symbols and Abbreviations}
\section*{Symbols}
\markboth{List of symbols and abbreviations}{Symbols}
\begin{list}{}{
    \setlength{\labelwidth}{3cm}
    \setlength{\labelsep}{0.3cm}
    \setlength{\leftmargin}{\labelwidth+\labelsep}
}
  \item[$\mu,\nu,\alpha,\beta$] Greek indices running from 0 to 3 over spatial \& time indices
  \item[$i,j,k,l$] Latin indices running from 1 to 3 over spatial indices
  \item[$\vv{v}=v^i\vv{e_i}$] Cartesian unit vectors $\vv{e_i}$ and Einstein's sum convention
  \item[$a,b$] spinor indices
  \item[$\hat{\mathcal{L}}$] operators in a Hilbert space carry a hat
  \item[$(P\hat{\Psi})_a=P_{ab}\hat{\Psi}_b$] spinor indices are also summed if they are on the same level
  \item[$\bbW_{,p_i}=\partial_{p_i} \bbW$] short-hand comma notation for derivatives
  \item[$\id$] identity matrix
  \item[$\epsilon_{ijk},\epsilon_{\mu\nu\alpha\beta}$] Levi-Civita pseudo tensor
  \item[$\e^{\ii\cdot\pi}+1=0$] Euler's constant $\e$ and the imaginary unit $\ii$
  \item[$\eta^{\mu\nu}$] Minkowski metric, $\eta^{\mu\nu}=\operatorname{diag}(1,-1,-1,-1)$
  \item[$\gamma^\mu$] Dirac matrices, $\anticommutator{\gamma^\mu}{\gamma^\nu}=2\eta^{\mu\nu}$
  \item[$\sigma^i$] Pauli matrices
  \item[{$\dd\Gamma=\dd[3]{\vv{x}}\frac{\dd[3]{\vv{p}}}{\left( 2\pi \right)^3}$}] phase space measure $\dd\Gamma$
  \item[$\vv{E}, \vv{B}, \vv{A}$] electric field, magnetic field and vector potential (\Eq\eqref{eqn:temporalgauge})
  \item[$\vv{q}, q_{x,y,z}$] canonical momentum
  \item[$\vv{p}, p_{x,y,z}$] kinetic momentum, $\vv{p}=\vv{q}-e\vv{A}$
  \item[$\Dt, \vDx, \vv{P}$] pseudo-differential operators (\Eq\eqref{eqn:diff-ops})
  \item[$\mathcal{W}$] Wigner function, 4 by 4 matrix, transforms as a Dirac matrix
  \item[$\bbs,\bbp,\bbv_\mu,\bba_\mu,\vv{\bbt}^{1,2}$] Wigner function components (\Eqs\eqref{eqn:wigner_fierz} and \eqref{eqn:wigner_comp_t})
  \item[$\vv{\bbt}$] in the spatially homogeneous case $\vv{\bbt}=\vv{\bbt}^1$
  \item[$\bbw$] vector containing the 10 components of the spatially homogeneous Wigner function in the purely electric case\newline $\bbw = \begin{pmatrix}
    \bbs & \vv{\bbv} & \vv{\bba} & \vv{\bbt}
  \end{pmatrix}^\intercal$
  \item[$\bbW$] vector containing the 16 components of the Wigner function\newline $\bbW = \begin{pmatrix}
    \bbs & \bbp & \bbv^0 & \bba^0 & \vv{\bbv} & \vv{\bba} & \vv{\bbt}^1 & \vv{\bbt}^2
  \end{pmatrix}^\intercal$
  \item[$\epsilon$] phase space energy density of the Dirac field (\Eq\eqref{eqn:energdens})
  \item[$f$] one-particle distribution function (\Eq\eqref{eqn:general_f})
  \item[$\mathcal{N}$] total particle yield (\Eq\eqref{eqn:numberofparticles})
  \item[$\mathcal{N}_{xy}$] particle yield for $z=0$ plane (\Eq\eqref{eqn:numberofparticles2d})
  \item[$c$] speed of light, here $c\equiv1$
  \item[$e$] elementary charge, $e\approx 1.6\cdot10^{-19}\,\mathrm{C}$
  \item[$\hbar$] Planck's constant, here $\hbar\equiv1$
  \item[$m$] mass of the electron, unit of mass and due to $c=\hbar=1$ unit of momentum in the numerical calculations
  \item[$\tc$] Compton time $\tc=\frac{\hbar}{c^2 m}=\frac{1}{m}\approx1.29\cdot10^{-21}\,\mathrm{s}$, unit of time and due to $c=1$ unit of length in numerical calculations
  \item[$E_\crit$] critical field strength $E_\crit=\frac{m^2c^3}{e\hbar}\approx1.3\cdot10^{18}\,\unitfrac{V}{m}$
  \item[$\varepsilon$] amplitude of electric field in units of the critical field strength
  \item[$\Omega$] frequency of electric field pulse
  \item[$\omega$] total energy of a particle with momentum $\vv{p}$ according to $\omega^2=m^2+\vv{p}^2$
  \item[$\tau$] pulse duration
  \item[$\gamma_\Omega$] Keldysh adiabaticity parameter regarding oscillation/rotation frequency $\Omega$, $\gamma_\Omega=mc\frac{\Omega}{eE}=\frac{\Omega}{\varepsilon m}$ (\Eqref{eq:keldysh_omega})
  \item[$\gamma_\tau$] Keldysh adiabaticity parameter for Sauter pulse, $\gamma_\tau=mc\frac{1}{\tau\, eE}=\frac{1}{\tau\varepsilon m}$ (\Eqref{eq:keldysh_tau})
  \item[$\gamma^*$] combined Keldysh adiabaticity parameter for rotating fields, $\gamma^*=\frac{1}{\varepsilon m}\sqrt{\left( \frac{\pi}{2}\frac{1}{\tau}  \right)^2 + \Omega^2}$ (\Eqref{eq:keldysh_combined})
  \item[$\Re\left( \cdot \right), \Im\left( \cdot \right)$] taking the real or imaginary part of a complex number
  \item[$\phantom{x}+\mathrm{c.c.}$] plus complex conjugated, $z+\mathrm{c.c.}:=z+z^*=2\Re\left( z \right)$.
\end{list}

\section*{Abbreviations}
\markright{Abbreviations}
\begin{list}{}{
    \setlength{\labelwidth}{3cm}
    \setlength{\labelsep}{0.3cm}
    \setlength{\leftmargin}{\labelwidth+\labelsep}
}
  \item[DHW] Dirac-Heisenberg-Wigner (formalism)
  \item[QED] Quantum electrodynamics
  \item[QKT] Quantum kinetic theory
  \item[WKB] Wentzel-Kramers-Brillouin (approach)
  \item[ODE] ordinary differential equation
  \item[PDE] partial differential equation
\end{list}

    \begin{otherlanguage}{ngerman}
\setlength{\parskip}{1ex} %
\setlength{\parindent}{0pt}
\chapter{Danksagungen}
An dieser Stelle möchte ich allen danken, die mich bei der Erstellung dieser Dissertation unterstützt haben.
Eine besondere Stellung nimmt hier mein Betreuer Holger Gies ein, dem ich die Möglichkeit verdanke, an diesem spannenden Thema zu arbeiten.
Ebenfalls erwähnt werden muss ebenfalls die Finanzierung durch die DFG in den Bereichen SFB/TR-18 und GRK 1523.

Ohne die inspirierenden Gespräche mit den Kollegen wären viele Ideen so sicher nicht entstanden.
Daher möchte ich meinen Dank insbesondere Julia Borchardt, Tom Dörffel, Michael Kahlisch, Steven Krause, Christian Kohlfürst, Julian Leiber, Stefan Lippoldt, David Schinkel und Nico Seegert aussprechen.

Nicht zuletzt möchte ich auch meiner Familie für ihre Unterstützung danken und meinen Freunden, die für meine wunderbare Zeit in Jena während des Studiums und der Promotion gesorgt haben. Danke Luja.

\end{otherlanguage}

    \begin{otherlanguage}{ngerman}
\setlength{\parskip}{1ex} %
\setlength{\parindent}{0pt}
\chapter{Ehrenwörtliche Erklärung}
Ich erkläre hiermit ehrenwörtlich, dass ich die vorliegende Arbeit selbstständig, ohne unzulässige Hilfe Dritter und ohne Benutzung anderer als der angegebenen Hilfsmittel und Literatur angefertig habe.
Die aus anderen Quellen direkt oder indirekt übernommenen Daten und Konzepte sind unter Angabe der Quelle gekennzeichnet.

Bei der Auswahl und Auswertung des Materials haben mir die nachstehend aufgeführten Personen %
unentgeltlich geholfen:

\begin{enumerate}
  \item Prof. Dr. Holger Gies
  \item Dr. Christian Kohlfürst
\end{enumerate}

Weitere Personen waren an der inhaltlich-materiellen Erstellung der vorliegenden Arbeit nicht beteiligt.
Insbesondere habe ich hierfür nicht die entgeltliche Hilfe von Vermittlungs- bzw. Beratungsdiensten (Promotionsberater oder andere Personen) in Anspruch genommen.
Niemand hat von mir unmittelbar oder mittelbar geldwerte Leistungen für Arbeiten erhalten, die im Zusammenhang mit dem Inhalt der vorgelegten Dissertation stehen.

Die Arbeit wurde bisher weder im In- noch im Ausland in gleicher oder ähnlicher Form einer anderen Prüfungsbehörte vorgelegt.

Die geltende Promotionsordnung der Physikalisch-Astronomischen Fakultät ist mir bekannt.

Ich versichere ehrenwörtlich, dass ich nach bestem Wissen die reine Wahrheit gesagt und nichts verschwiegen habe.

\vspace{1cm}
\noindent Ort, Datum\hspace{3cm} Unterschrift d. Verfassers
\end{otherlanguage}

\end{document}